%% file: main.tex
\documentclass[twocolumn]{aastex701}

\usepackage{etoolbox}
\makeatletter
\let\original@label\label
\makeatother

\usepackage{multirow}
\usepackage{subcaption}
\usepackage{siunitx}

\usepackage{dcolumn}

\usepackage{amsmath}
\usepackage{enumitem}
\usepackage{todonotes}
\usepackage{makecell}
\usepackage{comment}
\usepackage{gensymb}
\usepackage{mathtools}

\usepackage{todonotes}

\usepackage{hyperref}

\hypersetup{
    colorlinks=true,
    linkcolor=blue,
    filecolor=magenta,      
    urlcolor=cyan,
    pdftitle={Improving Pulsar Timing Precision with Single Pulse Fluence Clustering},
    pdfpagemode=FullScreen,
    }

\begin{document}

\title{The NANOGrav 15 yr Dataset: Modified Functional Forms for Profile-Evolution Time Delays in Pulsar Timing Models}

\input{authorsLIFD.tex}




\begin{abstract}
Radio pulsars exhibit frequency-dependent variations in the shape of their pulse profiles that introduce systematic delays in the observed pulse times of arrival (TOA) across an observing band. The North American Nanohertz Observatory for Gravitational Waves (NANOGrav) models these delays using a frequency-dependent (FD) log-polynomial function whose coefficients are fit to TOA measurements at different radio frequencies. However, FD parameters are prone to absorbing power from other chromatic effects, such as interstellar dispersion, leading to fitted FD values that overestimate profile evolution and vary significantly across datasets. To separate these effects, we introduce two new mathematical formulations of profile evolution: a monomial expansion (IFD) and a Legendre polynomial expansion (LIFD). We test these models on observations of 11 pulsars in the NANOGrav 15-year dataset and on simulated data, and also compare their performance to that of frequency-resolved pulse templates. We find that Legendre-based models generally predict timing delays that more closely replicate the true profile evolution, reducing the residual offset relative to FD by as much as $85\%$ for the simulated data. Additionally, we find that the choice of profile evolution model can bias the fitted dispersion model, with FD-based models overestimating the injected DM by $0.013~\mathrm{pc~cm^{-3}}$ compared to $0.0002~\mathrm{pc~cm^{-3}}$ for LIFD. These results support the use of Legendre bases to mitigate covariances between chromatic effects, directly benefiting targeted timing studies of these effects and the development of advanced noise-modeling techniques for nanohertz gravitational-wave detection.
\end{abstract}

\keywords{Pulsar timing --- Interstellar Medium --- Pulsars --- Radio astronomy}


\section{Introduction} \label{sec:intro}

\subsection{Profile Evolution With Frequency}\label{subsec:prof_evolution_theory}

Pulsars are rapidly rotating neutron stars that emit highly regular pulsed radio emission. In particular, millisecond pulsars (MSPs) are extremely stable rotators, whose pulsed emission has been measured with sub-microsecond precision \cite[e.g.,][]{NG15}. By comparing the observed times of arrival (TOAs) of these pulses with the predictions from sophisticated mathematical models known as \textit{timing models}, we can detect timing offsets generated by a myriad of astrophysical phenomena and probe their underlying physical properties. Notably, long-term monitoring of an array of the most stable MSPs has led to strong evidence for a signal consistent with nanohertz-frequency gravitational radiation from a population of supermassive black hole binaries \citep[e.g.,][]{NG15_GWB, antoniadis2023second, reardon2023search, 2023RAA....23g5024X}.

High-precision pulsar timing experiments require careful modeling of noise sources that change the TOAs and obscure underlying timing signatures. Some of these are sources extrinsic to the pulsar, including the interstellar medium \citep[ISM; e.g.,][]{10.1093/mnras/sts486, 2025ApJ...986..191G}, Solar wind \citep[e.g.,][]{2019ApJ...872..150M, 2021A&A...647A..84T, 10.1093/mnras/stae2727}, and Solar System ephemeris \citep[e.g.,][]{Lazio:2017fos, Vallisneri_2020, Dai_2024}. Other noise sources are intrinsic to the pulsar emission processes, such as pulse \textit{jitter} \citep[e.g.,][]{10.1111/j.1365-2966.2011.19578.x, 10.1093/mnras/stu1213, Lam_2019}, mode changes \citep[e.g.,][]{1982ApJ...258..776B, 2022MNRAS.510.5908M}, and spin noise \citep[e.g.,][]{Shannon_2010, Lam_2017} including \textit{glitches} \citep[e.g.,][]{2024A&A...689A.191Z}.

\begin{figure*}[htp]
\centering
\begin{subfigure}{0.4\linewidth}
    \includegraphics[width=\linewidth]{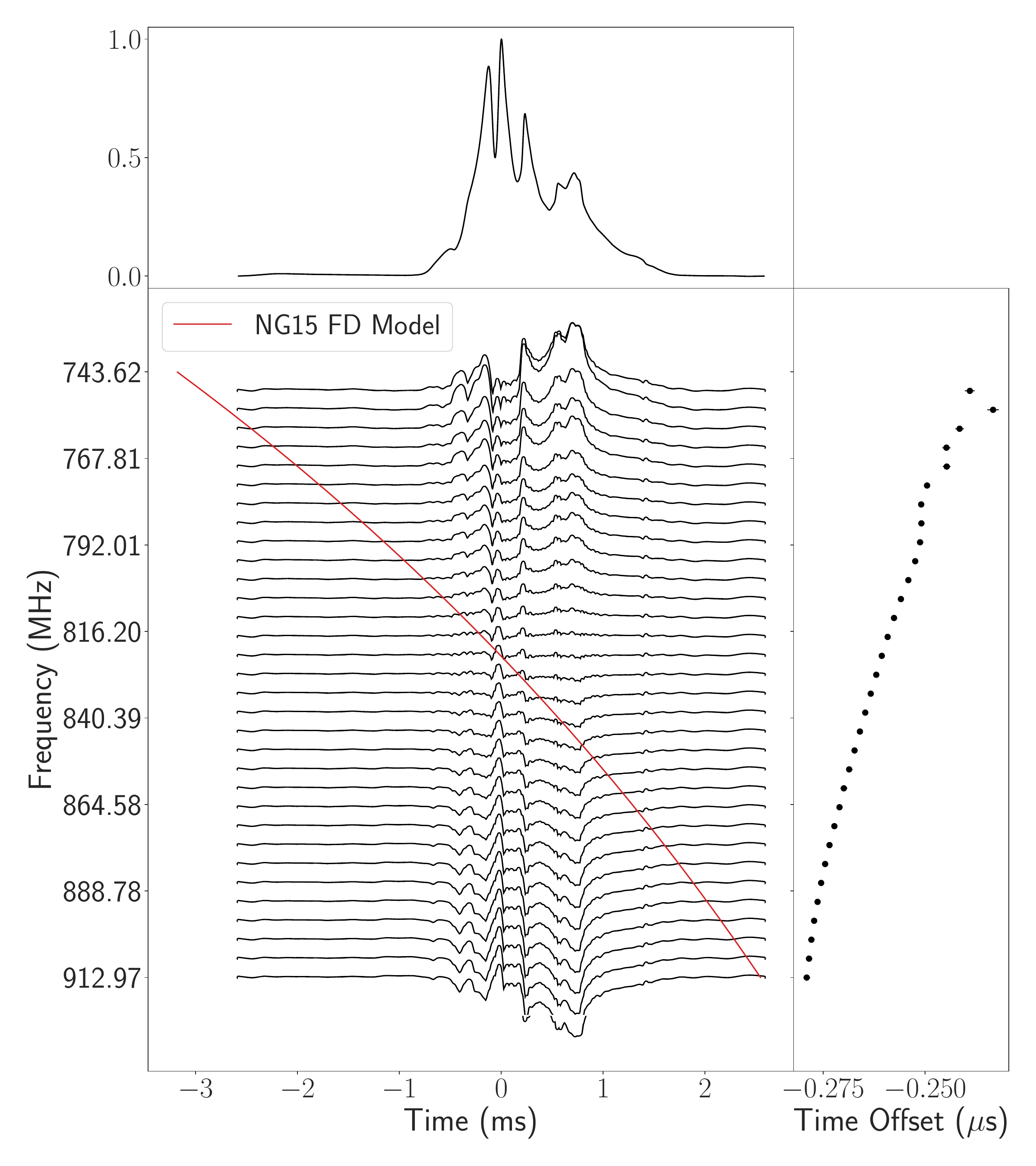}\label{fig:J1024_portrait}
    \caption{J1024$-$0719 with the 820 MHz receiver.}
\end{subfigure}
\
\centering
\begin{subfigure}{0.4\linewidth}
\includegraphics[width=\linewidth]{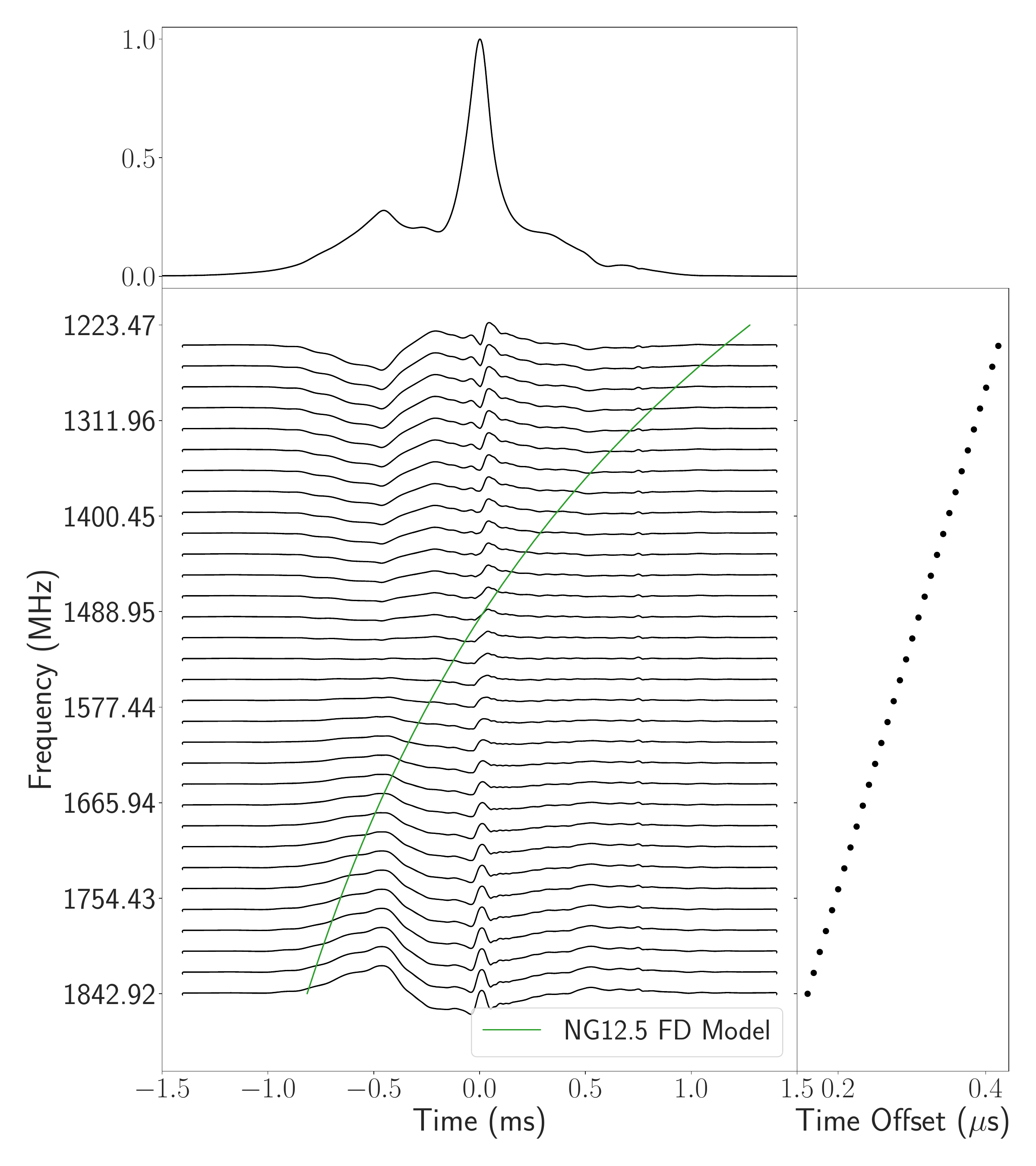}\label{fig:J1918_portrait}
\caption{J1918$-$0642 with the L-band receiver.}
\end{subfigure}

\caption{Pulse profile evolution of PSR J1024$-$0719 in the 820-MHz band (left) and PSR J1643$-$1224 in the L-band (right). Bottom left: pulse profiles at different frequencies, after de-dispersing them using the pulsar's nominal DM value reported in \cite{NG15} and subtracting the pulse profile at the middle of the frequency band. The red and green lines show the PE-induced timing delay expected at each frequency as predicted by the FD models reported in the NANOGrav 15- and 12.5-year datasets, respectively. Bottom right: TOAs obtained from cross-correlating the pulse profiles at different frequencies with the pulsar's profile template. Top: frequency-averaged pulse profile.}\label{fig:FD_evolution}
\end{figure*}


A known intrinsic pulsar phenomenon that can contribute to timing noise is the evolution of the observed pulse-profile shape with radio frequency. Most pulsars exhibit a systematic increase in pulse width and a separation of pulse-profile components at lower frequencies, unrelated to ISM effects \citep[e.g.,][]{Kramer_1998}, as exemplified in Fig.~\ref{fig:FD_evolution}. This effect is often more pronounced in young pulsars than in MSPs, which show less pronounced evolution in pulse width and component separation with frequency \citep[e.g.,][]{1999ApJ...526..957K}.

Different physical mechanisms have been proposed to explain pulse profile evolution (PE) with frequency. \cite{1970Natur.225..612K} first suggested that lower-frequency emission originates higher in the magnetosphere, where dipolar field lines subtend a larger angular region, causing the emission beam to sweep across the observer’s line of sight for a longer fraction of the pulsar’s rotation and thus producing wider pulses. This idea, known as \textit{radius-to-frequency mapping} (RFM), is theoretically grounded in the polar cap model of \cite{1975ApJ...196...51R} and was first proposed for canonical pulsars by \cite{1978ApJ...222.1006C}. \cite{2017MNRAS.468.4389S} proposed that these frequency-dependent emission heights arise because radio emission at different frequencies is produced most efficiently in distinct regions of the magnetosphere. In contrast, \cite{10.1093/mnras/stu2262} argued that spectral gradients within different emission beam geometries (cones or streams) can change how the line of sight samples the emission at different frequencies, producing component evolution without requiring RFM's emission at different altitudes. Other explanations include intrinsic magnetospheric effects, including refraction and coupling between orthogonally polarized modes \citep{1968Natur.220..676C, 10.1093/mnras/stad070}, as well as extrinsic contributions from propagation through the interstellar medium \citep{2025A&A...704A.214J}. More wideband studies are needed to test and constrain the underlying radiation mechanisms \citep[e.g.,][]{Xu_2021}.

Since observations for high-precision timing typically span one or more broad frequency bands, PE across that band can impact TOA calculation in different ways:

\begin{itemize}
    \item In narrow-band timing, each band is divided into multiple sub-bands \citep{2016MNRAS.458.3341D} or frequency channels \citep[e.g.,][]{NG15}. A separate TOA is measured for each sub-band or channel by estimating the time shift of its pulse profile relative to an idealized noise-free pulse shape, known as a ``template'' \citep{10.1098/rsta.1992.0088}. However, a frequency-constant template will not correctly capture PE-induced effects and will thus deviate from the true pulse shape at different frequencies. These mismatches between the data and the template result in small but systematic frequency-dependent TOA offsets \citep[e.g.,][]{Brook_2018}.
    \item Some studies frequency-average the data across each observing band and compare the resulting integrated profile to a template to produce a single TOA for the full receiver band \citep[e.g.,][]{2013PASA...30...17M}. However, if the bandwidth is sufficiently large, PE introduces frequency-dependent timing offsets across frequency channels, which, when averaged, smear the integrated pulse profile and reduce TOA precision.
    \item In wideband timing, a single TOA is produced for the entire bandwidth by fitting all frequencies simultaneously using a two-dimensional template, known as a ``pulse portrait,'' which encodes pulse evolution as a sum of either Gaussian components \citep{Pennucci_2014}, channelized Fourier transforms \citep{10.1093/mnras/stu1420}, or low-degree piecewise polynomial spline functions \citep{Pennucci_2019}.  
    Even then, residual frequency-dependent structure has been reported in the timing data from wideband timing \citep[e.g.,][]{Curylo_2023, 2021ApJS..252....5A}, similarly to that seen in narrowband analyses with frequency-averaged templates.    
\end{itemize}

\noindent In either case, mis-modeled PE results in reduced TOA precision and reduced sensitivity to effects targeted in high-precision pulsar timing array (PTA) experiments, including gravitational-wave (GW) detection \citep[][]{10.1111/j.1365-2966.2011.19578.x, 10.1111/j.1365-2966.2011.19452.x, 10.1093/mnras/stx580}.



 \subsection{Accounting for Profile Evolution}





The European Pulsar Timing Array typically splits wideband data into discrete frequency bands within certain bandwidths, such as 140 or 240 MHz for the L-band (1.4 GHz), 300 MHz for S-band (2.2 GHz), and 500 MHz for C-band (4.8 GHz) \citep{2023A&A...678A..48E}. A template is independently created and used to match the data within each band. Because the TOAs are measured relative to this local template, PE-induced shifts are minimized at the TOA-generation level. To align the data from different observing systems, an arbitrary time-constant offset, also known as a ``jump,'' is fitted for each sub-band \citep{2016MNRAS.458.3341D}.

Similarly, the Parkes Pulsar Timing Array averages narrowband observations over the full bandwidth (64 MHz, 256MHz, or 1024MHz) of each of its three receivers. Then, they use different template profiles to measure a TOA for each band and fit ``jumps'' between bands \citep{2013PASA...30...17M, 2020PASA...37...20K, 2023PASA...40...49Z}. With the Ultra-Wideband Low receiver \citep{2020PASA...37...12H}, the bandwidth is large enough that the pulse profile can evolve significantly across this frequency range, so they use pulse portraits to model PE continuously across the band \citep{Curylo_2023}.

The North American Nanohertz Observatory for Gravitational Waves (NANOGrav) in its 5-year dataset \citep{2013ApJ...762...94D} included a ``jump'' for each frequency channel (rather than each frequency band) in the observations, which is then adjusted as a free parameter in the timing fit. 
However, starting in their 9-year dataset \citep{NG9}, they replaced this scheme with FD parameterization. In this approach, band-specific templates are created for each frequency receiver (e.g., the 820- and L- bands at the Green Bank Telescope), and additional timing corrections are incorporated into the timing models using a log-polynomial function of frequency to remove PE frequency structure. This function can be written as:

\begin{equation}\label{eq:old_FD_formulation}
    t_{\mathrm{PE}, \nu} = \sum_{k=1}^{n}{\mathrm{FD}_k \left[ \ln{ \left( \frac{\nu}{1~\mathrm{GHz}} \right) } \right]^k},
\end{equation}

\noindent The coefficients of this polynomial (known as FD parameters) are estimated post-TOA generation by including them as free parameters in the timing fit. The number of FD terms for each pulsar is determined via an $F$-test \citep[e.g.,][]{NG15}, but usually ranges from $n = 2$ to $5$. There is no $k = 0$ term because it would be a constant phase offset that is removed when the mean is subtracted from the timing residuals.





\subsection{FD Mis-estimations}\label{subsec:FD_mis_estimations}

\begin{figure}[ht]
\centering
    \includegraphics[width=\linewidth]{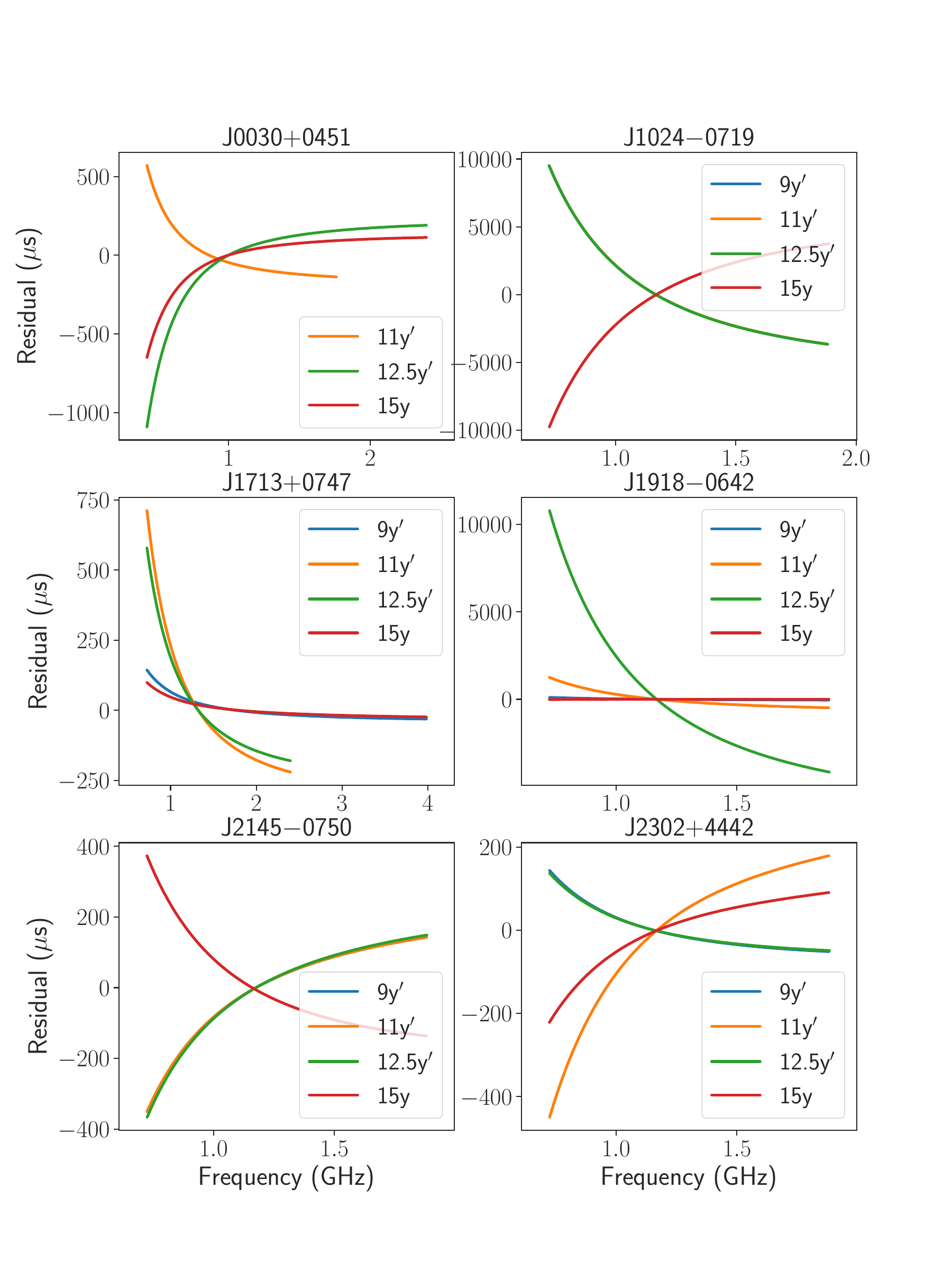}
\vspace{-6mm}
\caption{PE timing delays as a function of frequency according to the FD models reported in the NANOGrav 9- \citep{NG9}, 11- \citep{NG11}, 12.5- \citep{NG12}, or 15-year \citep{NG15} datasets. Since the datasets use different nominal DM values, we shifted the 9-, 11- and 12.5-year curves by $-K(\mathrm{DM}^\mathrm{(x)}+\mathrm{DMX}_0^\mathrm{(x)}-\mathrm{DM}^\mathrm{(NG15)})/\nu^2$, where $K$ is the dispersion constant, $\mathrm{DM}^\mathrm{(x)}$ is the nominal DM value of the NG dataset in question, $\mathrm{DMX}_0^\mathrm{(x)}$ is its first $\mathrm{DMX}$ parameter, and $\mathrm{DM}^\mathrm{(NG15)}$ is the nominal DM of the 15-year dataset. For some MSPs, the delay curve amplitudes are larger than the pulsar's rotational period and flip concavity across datasets, which is inconsistent with the behavior expected from PE.}\label{fig:FD_curves}
\end{figure}

When we use FD models to account for PE and marginalize over the chromatic terms of the timing model, the resulting predictions have been accurate enough to preserve sensitivity to achromatic signals, including multiple classes of GW signals \citep[e.g.,][]{NG15_NB}. However, when we look at the FD term's individual contribution to the overall timing model across the NANOGrav datasets (see Fig.~\ref{fig:FD_curves}), we find that the FD terms usually predict PE-induced timing delay amplitudes that are difficult to reconcile with the amplitudes observed in the PE portraits (see Fig.~\ref{fig:FD_curves}). For example, the 12.5-year FD model for PSR J1918$-$0642 predicts a delay exceeding 12.5 ms; for this 7.65-ms pulsar, this corresponds to a delay larger than one full rotational phase and is substantially larger than the empirically observed PE, as illustrated in Fig.~\ref{fig:FD_evolution}. 




We also observe in Fig.~\ref{fig:FD_curves} that the FD curve for a given pulsar can vary substantially depending on the dataset used to fit the FD parameters. This includes changes in concavity (e.g., J0030$+$0451 and J2145$-$0750) and in the delay amplitude (e.g., J1713$+$0747 and J1918$-$0642). While a few pulsars are known to exhibit temporal variations in pulse profile shape \citep[e.g.,][]{2000Natur.406..484S}, these are long-term and only observed in canonical pulsars. Furthermore, since the NANOGrav datasets are not independent (earlier datasets are subsets of later ones), any temporal profile changes should be correlated and continuous across datasets. Therefore, these pronounced, non-monotonic variations suggest that the FD models do not robustly isolate PE from dataset-dependent effects.

A possible explanation for these shortcomings is that FD parameters are estimated through a global least-squares fit of the timing model to minimize the timing residuals; however, chromatic effects other than PE can introduce frequency-dependent delays, such as dispersion and scattering, that also contribute to the TOAs \citep[e.g.,][]{Brook_2018}. Because these chromatic effects are all fitted simultaneously to the observed TOA frequency dependence, without explicitly separating the physical origin of those TOAs, the resulting fit is susceptible to parameter covariances. In other words, we do not know a priori what the timing shifts due to profile evolution and due to other effects are; we can only fit them empirically a posteriori.

While it is generally assumed that the largest component of the frequency-dependent time delay accounted for by FD parameters is due to intrinsic PE \citep{Zhu_2015}, due to these covariances, power from other imperfectly modeled chromatic effects can be redistributed and absorbed into the FD term during the timing fit, changing the inferred $\mathrm{FD}_k$ parameters away from values that reflect intrinsic PE alone \citep[e.g.,][]{2013MNRAS.428.1147V}. The combined noise model may adequately describe the global chromatic noise, but individual models associated with specific physical processes may be imperfectly constrained, complicating the physical interpretation of individual chromatic terms.

The most significant chromatic effect other than PE is dispersion. FD model describe PE using a polynomial in $\mathrm{log}(\nu)$, while dispersive delays scale as $\nu^{-2}$. These functional forms are not orthogonal, so the two effects have partial overlap. Similarly, FD parameters will also account for the average scattering broadening over the course of a data set \citep[e.g.,][]{2021ApJ...909..219S}. As a result, the timing fit can partially redistribute dispersion and scattering delays into the FD coefficients.


In this paper, we aim to develop a formulation of PE that minimizes covariances with other chromatic effects during the timing fit. In particular, we study PE polynomial formulations in monomial and Legendre bases. In Sec.~\ref{sec:theoretical_background}, we describe the chromatic effects affecting pulsar TOA measurements and how NANOGrav accounts for them, and we introduce our own PE formulations. In Sec.~\ref{sec:results}, we fit our new PE models to NANOGrav's 15-year dataset and to simulated observations, and compare our results to current FD predictions. Finally, in Sec.~\ref{sec:conclusions}, we discuss the implications of this work, especially concerning the accuracy of our noise models. Processed data products presented here are publicly available as of the date this work is published.

\section{Theoretical Background}\label{sec:theoretical_background}

In this section, we describe the chromatic effects on the TOAs, and how they are accounted for in NANOGrav's timing models. Then we lay out the mathematical formalism for our novel PE models.

\subsection{Frequency-Dependent Effects}\label{subsec:freq_dependent_effects}

Pulsar observations are never monochromatic, but rather performed over a finite bandwidth comprised of several frequencies. To obtain a TOA, these multi-frequency measurements are usually averaged across the observing bandwidth. However, TOAs exhibit frequency dependence arising from a variety of astrophysical sources. These include effects intrinsic to the pulsar (e.g., pulse jitter and PE), the interstellar medium (e.g., dispersion and scattering), and the Solar System (solar activity and ionospheric delays). As a result, pulses with frequency $\nu$ will arrive at Earth at a time $t_\nu$ that is delayed with respect to the expected arrival time $t_\infty$ for a signal of infinite frequency (i.e., in the absence of chromatic effects). Pulsar timing applications must then account for these chromatic delays.


\subsubsection{Dispersion}

As the pulsar's radio signal propagates through the interstellar medium (ISM), which is an ionized plasma, it encounters a frequency-dependent index of refraction. For a cold, homogeneous, unmagnetized plasma, the cold-plasma dispersion law predicts that a pulse emitted at frequency $\nu$ is delayed relative to an infinite-frequency signal by a time $t_\mathrm{DM}=K\times\mathrm{DM}/\nu^2$. Here, the dispersion measure $\mathrm{DM}=\int_0^d{n_e(l)dl}$ represents the integral of free electron density $n_e$ along the line-of-sight (LOS) $l$ to a pulsar at distance $d$, and $K=e^2/2\pi m_ec \simeq4.149~\mathrm{ms~GHz^2~pc^{-1}~cm^3}$ is the dispersion constant.

In practice, the pulsar LOS samples electron-density fluctuations due to turbulent and bulk motions in the ISM and the solar wind. Relative motion between the pulsar, Earth, and the scattering medium converts these spatial fluctuations into temporal variations \citep[e.g.,][]{1991ApJ...382L..27P, Cordes_1998, Lam2016}. These processes result in DM variations on timescales from hours to years. NANOGrav models this variability using a stepwise-constant representation in which the DM has independent values on fixed time intervals, ranging from 0.5 to 15 days. The offsets from the globally fixed fiducial $\mathrm{DM}$ value are parameterized by the epoch-dependent $\mathrm{DMX}$ terms \citep[e.g.,][]{Jones_2017, 2021ApJ...909..219S}. This introduces a time-dependent correction to the dispersive delay as:

\begin{equation}\label{eq:dispersive_delay}
    t_{\mathrm{DM}, {i}} (\nu)=\frac{K \times (\mathrm{DM}+\mathrm{DMX}_i)}{\nu^2}
\end{equation}

\noindent where $\mathrm{DMX}_{i}$ is the DM offset at observing epoch $i$.



\subsubsection{Refractive Effects}

As pulsar radiation encounters density gradients on scales larger than the Fresnel scale, the coherent ray deviates from straight-line propagation, bending toward regions of higher refractive index (just like light bending at a water/air interface, but applied to a large, extended, slowly-varying gradient). This has two consequences:

\vspace{2mm}

\noindent \textbf{Angle-of-arrival delays}: The change in the angle of arrival of the ray at the Earth shifts the apparent direction to the pulsar. Since the timing model assumes a single fixed direction to calculate the Roemer delay, this causes a chromatic delay that varies sinusoidally with an annual period \citep{1990ApJ...364..123F}. Since the bending angle, $\theta_r$, is a first-order effect in the refractive index perturbation ($\propto n_e/\nu^2$), the induced timing delay is $\propto \theta_r \propto \nu^{-2}$ \citep{Lam_2016b}.

\vspace{2mm}

\noindent \textbf{Geometric delays}: A ray deviated from straight-line propagation has an increased path length and suffers a geometric delay, arriving later than photons traveling in a straight path. Because the extra path length for a small bend is a quadratic effect in $\theta_r \propto \nu^{-2}$, the geometric timing delay scales as $\left( \nu^{-2} \right)^2 = \nu^{-4}$.

\vspace{2mm}

Additionally, refractive ray bending produces weak lensing, modulating the observed flux on long timescales (days to weeks). This is called \textit{refractive scintillation}, but is usually not explicitly modeled in PTA pipelines.

\subsubsection{Diffractive Effects}

When a ray encounters a density structure smaller than the Fresnel scale, it is no longer described by a single deterministic bent ray. Instead, different patches of the medium send a coherent sub-ray to the observer, resulting in multiple, simultaneous, stationary-phase rays \citep{1986ApJ...310..737C}. This results in two effects:

\vspace{2mm}

\noindent \textbf{Scattering}: Photons deflected at small angles travel longer paths and arrive at later rotational phases. The resulting pulse profile is the convolution of the intrinsic pulse and an exponential \textit{scattering tail}, the timescale of which is $\tau_\mathrm{scat} \propto \nu^{-\alpha}$ \citep[e.g.,][]{10.1093/mnras/stt989, Bansal_2019}. This produces a deterministic, chromatic timing delay that is commonly modeled using a thin-screen approximation \citep[][]{2017MNRAS.464.2075S}, assuming an isotropic, turbulent medium with a Gaussian \citep[$\alpha=4$;][]{1971ApJ...164..249L} or Kolmogorov \citep[$\alpha=4.4$;][]{10.1093/mnras/220.1.19} distribution of inhomogeneities. 

Most MSPs used in PTAs are at relatively high Galactic latitudes, so their lines of sight pass through less turbulent ISM and their $\tau_\mathrm{scat}$ at observing frequencies ($\sim$1–2 GHz) is sub-microsecond or sub-nanosecond \citep[e.g.,][]{Turner_2021}. Therefore, scattering does not have dedicated deterministic parameters in NANOGrav's timing model, and it is instead included in the noise models as a time-variable stochastic chromatic process with a free spectral index \citep{2025arXiv251122597H}.

\vspace{2mm}

\noindent \textbf{Diffractive scintillation}: When photons scattered through different angles recombine at the observer, they will no longer be in phase and will thus interfere with one another, generating stochastic intensity variations across frequency and time known as \textit{scintles}. For a Kolmogorov medium, the bandwidth and timescale of these variations are $\Delta \nu_\mathrm{d}\propto\nu^{4.4}$ and $\Delta t_\mathrm{d}\propto\nu^{1.2}$ \citep{1985ApJ...288..221C}, although studies have empirically found shallower power indices \citep[][]{2022A&A...664A.116L, Brown2025}.

When averaged over the frequency band, the intensity modulation can cause epoch-to-epoch TOA variations, rather than a deterministic timing delay. Therefore, scintillation is treated as a white‑noise contribution that is empirically determined per pulsar (using the white-noise parameters EFAC, EQUAD, and ECORR) rather than following a fixed power law \citep{Lam2016}.

\subsubsection{Total Chromatic Model}

Combining the previous effects, we model the total chromatic delay at a frequency $\nu$ and epoch $i$ as:

\begin{equation}\label{eq:full_chromatic_delays}
    t_{\nu, i} = t_{\infty, i} + \frac{K ~ (\mathrm{DM}+\mathrm{DMX}_i)}{\nu^2} + t_\mathrm{PE, \nu} + t_{\mathrm{C},\nu,i} + \epsilon_\nu
\end{equation}

\noindent where $t_\infty$ is the infinite-frequency arrival time, $\epsilon_\nu$ are measurement errors, and

\begin{equation}\label{eq:additional_chromatic}
    t_{\mathrm{C},\nu, i} = \sum_{k=1}^{N_\mathrm{c}} C_{k,i} \nu^{-\alpha_k}
\end{equation}

\noindent models additional TOA variations due to $ N_\mathrm{c}$ non-PE and non-DM chromatic effects with power-law spectral scalings $\alpha_k$ and power spectrum amplitudes $C_{k,i}$.


\noindent The parameters in this model are refined by including them as free parameters when fitting a timing model to the TOAs observed at different frequencies. The differences between observed and modeled TOAs are quantified by the residuals. Systematic structures in the residuals with non-random deviations from zero are indicative of parameter mis-estimations. In that case, the timing model is refined by iteratively adjusting its parameters using least-squares fitting. This process continues until the residuals resemble a white noise distribution, resulting in more precise parameter estimates. This fit is usually carried out using software like TEMPO2 \citep{2006MNRAS.372.1549E} or PINT \citep{Luo_2021}.

\subsection{Polynomial PE Model in a Monomial Basis}

As explained in Sec.~\ref{subsec:FD_mis_estimations}, FD mis-estimations can be explained through leakage of spectral power from propagation-induced chromatic effects into the $\mathrm{FD}_k$ parameters in the timing fit. In particular, consider a timing model that accounts for dispersion using a $\mathrm{DM}$ with an error $\Delta \mathrm{DM}$ from the real dispersion measure. On a given epoch, the delay at a frequency $\nu$ is (see Eq.~\ref{eq:full_chromatic_delays}):

\begin{align}\label{eq:naive_tm}
    t_{\nu} &= t_{\infty} + \frac{K ~ (\mathrm{DM}+\Delta \mathrm{DM})}{\nu^2} + t_\mathrm{PE, \nu} + t_{\mathbf{C},\nu} \\ \nonumber
    &= t_{\infty} + \frac{K ~ \mathrm{DM}}{\nu^2} + t_{\nu}^\mathrm{excess}
\end{align}

\noindent where $t_{\nu}^\mathrm{excess}$ is the chromatic delay excess that the dispersion model does not capture, comprising all the remaining frequency-dependent components, including PE, scattering, refraction, and the residual dispersive delay due to $\Delta \mathrm{DM}$. If we assume that refraction and scattering scale as $\nu^{-4}$ (see Sec.~\ref{subsec:limitations} for a discussion on the limits of this assumption), we can write it as:

\begin{equation}\label{eq:new_polynomial_2}
	t_{\nu}^\mathrm{excess} = K ~ \Delta \mathrm{DM}~{\nu_0}^2 \left(\tfrac{\nu}{\nu_0}\right)^{-2} + t_C \left(\tfrac{\nu}{\nu_0}\right)^{-4} + t_{\mathrm{PE}, \nu}
\end{equation}

\noindent We take $\nu_0 = 1~\mathrm{GHz}$ as a reference frequency, for numerical stability and given typical bands used in timing.

We will now assume that the PE-induced timing delay at frequency $\nu$, $t_{\mathrm{PE}, \nu}$, can be completely expressed as a polynomial in inverse frequency in a monomial basis:

\begin{equation}\label{eq:IFD_polynomial}
	t_{\mathrm{PE}, \nu} = \sum_{k=0} a_k \left(\frac{\nu}{\nu_0}\right)^{-k},
\end{equation}

\noindent All the $a_k$ coefficients have time units, and the power-law indices are integers. 
Substituting Eq.~\ref{eq:IFD_polynomial} into Eq.~\ref{eq:new_polynomial_2}:

\begin{equation}\label{eq:monomial_basis}
    \begin{split}
        t_{\nu}^\mathrm{excess} = a_1 \left(\tfrac{\nu}{\nu_0}\right)^{-1} & + \left( K~\Delta \mathrm{DM}~\nu_0^2 + a_2 \right) \left(\tfrac{\nu}{\nu_0}\right)^{-2} \\ + a_3 \left(\tfrac{\nu}{\nu_0}\right)^{-3}& + (t_C + a_4) \left(\tfrac{\nu}{\nu_0}\right)^{-4} + a_5 \left(\tfrac{\nu}{\nu_0}\right)^{-5} + \ldots        
    \end{split}
\end{equation}

\noindent We see that $\Delta \mathrm{DM}$ is isolated in the $\nu^{-2}$ term and only has the same frequency dependence as the $a_2$ term of PE. We then define $a_2' \equiv K \mathrm{DM} \nu_0^2 + a_2$ as the misestimated DM that absorbs the $a_2$ coefficient. Similarly, refraction and scattering are isolated in the $\nu^{-4}$ term with only the $a_4$ contribution from PE, so we define $a_4'\equiv t_C+a_4$.

We will now express the radio-frequency powers in terms of radio wavelengths. Defining $\lambda_0$ as the wavelength equivalent to the reference frequency $\nu_0$ (for $\nu_0=1~\mathrm{GHz}$, $\lambda_0 = 1~\mathrm{ns})$, we can write

\begin{equation}\label{eq:lambda_transformation}
    \frac{\lambda}{\lambda_0} = \frac{c/\nu}{c/\nu_0} = \left( \frac{\nu}{\nu_0} \right)^{-1},
\end{equation}

\noindent Under this variable transformation, Eq.~\ref{eq:monomial_basis} becomes:

\begin{equation}\label{eq:monomial_basis_lambdas}
\begin{split}
    t_\lambda^\mathrm{excess} &  = a_0 + a_1 \left( \tfrac{\lambda}{\lambda_0} \right)^{1} + a_2' \left( \tfrac{\lambda}{\lambda_0} \right)^{2} + a_3 \left( \tfrac{\lambda}{\lambda_0} \right)^{3} + \\ & a_4' \left( \tfrac{\lambda}{\lambda_0} \right)^{4} + a_5 \left( \tfrac{\lambda}{\lambda_0} \right)^{5} + \ldots ,
\end{split}
\end{equation}

\noindent The timing model in Eq.~\ref{eq:naive_tm} with $t_{\nu}^\mathrm{excess}$ replaced with Eq.~\ref{eq:monomial_basis_lambdas} would then model PE-induced chromatic delays using a monomial-basis polynomial on inverse frequency. We will refer to this as the Inverse-Frequency Dependent (IFD) model and to the $\vec{a}=\{ a_0, a_1, \ldots \}$ polynomial coefficients as the $\mathrm{IFD}$ parameters, \{$\mathrm{IFD_0}$, $\mathrm{IFD_1}$, $\ldots$\}.

\subsubsection{Advantages and Limitations of a Monomial Basis}\label{sec:advantages_monomial}

The IFD formulation presents a major advantage over FD parameters: an excess dispersive $K~\Delta \mathrm{DM}~\nu^{-2}$ delay is not exactly representable by a finite log-polynomial basis (as in Eq.~\ref{eq:old_FD_formulation}), so it cannot be isolated into specific PE terms. Instead, modeling it requires a linear combination of several logarithmic terms. As a result, a misestimated $\mathrm{DM}$ can result in DM-induced timing delays not captured by the dispersion model, being projected onto multiple FD logarithmic terms, biasing the fitted values of multiple $\mathrm{FD}_k$ parameters.


In contrast, when PE is modeled in a monomial basis, a dispersive delay is represented exactly by a single basis element, $\lambda^{2}$, so it is only covariant with $\mathrm{IFD_2}$. Similarly, under the assumptions above that the refraction/scattering contributions scale as $\lambda^{4}$, those effects couple only to $\mathrm{IFD}_4$. All remaining coefficients describe a chromatic structure distinct from propagation effects, and we can attribute it to real PE. Therefore, IFD models isolate dispersion- and scattering-induced covariances to components geometrically distinguishable and physically interpretable in the chromatic model.



Nevertheless, this formulation still has a major limitation: the monomial basis functions, $\phi_k(\lambda)=(\lambda/\lambda_0)^{k}$, are not orthogonal over a finite observing band. As a result, the associated design matrix can be ill-conditioned, and the fitted coefficients become statistically correlated. In the presence of realistic chromatic structure beyond pure dispersion, this correlation means that uncertainty in one parameter is statistically coupled to uncertainty in others, thereby reducing the precision with which individual coefficients can be determined. Furthermore, in the presence of noise, the projections of small noise components onto strongly correlated basis elements can be amplified, leading to larger fluctuations in the estimated parameter. Although this does not introduce systematic biases, it degrades the numerical stability and overall precision of the timing fit.

From a geometric perspective, we can think of the monomial basis elements as vectors that are not fully orthogonal. An excess dispersive delay is a vector that lies exactly in the direction of the basis element proportional to $\lambda^2$, so including this signal will only change the fitted value of $a_2$. However, because the basis vectors are not orthogonal, small perturbations from the $\lambda^2$ direction can project onto multiple basis vectors and cause large variations in coefficients other than $a_2$.

\subsection{Polynomial PE Model in a Legendre Basis}\label{sec:LIFD}



 To reduce parameter correlations, we will model $t_{\nu}^\mathrm{excess}$ as a polynomial in an orthogonal basis, i.e., a set of functions $[p_n(x)]$ defined over a domain $x\in[a,b]$ that satisfy the orthogonality relation
 
\begin{equation}\label{eq:orthogonal}
    \langle p_m, p_n \rangle \coloneq \int_a^b w(x) p_m(x) p_n(x) dx = \delta_{mn}c_n,
\end{equation}

\noindent where $w(x)$ is a weighting function, $\delta_{mn}$ is the Kronecker delta, and $c_n$ is a normalization constant \citep[e.g.,][]{AbramowitzStegun1965}. This condition ensures that the basis functions are uncorrelated in $[a,b]$ under the inner product defined by $w(x)$.

Geometrically, the elements of such a basis can be visualized as mutually orthogonal vectors. A delay proportional to $\nu^{-2}$ will not, in general, align exactly with a single basis vector and must be expressed as a sum of projections into multiple basis directions. However, because the basis vectors are orthogonal, these projections do not overlap geometrically, and variations along one basis direction do not induce variations along another. For a linear least-squares fitting with uncorrelated noise and weighting consistent with the chosen inner product, this choice of basis produces a diagonal covariance matrix and, therefore, uncorrelated coefficient estimates.

Orthogonal bases are widely used in data analysis. For example, in linear principal component analysis, the data are transformed into a new coordinate system defined by an orthogonal set of vectors aligned in the directions of the largest data variance \citep[e.g.,][]{10.1145/3447755}. Because these directions are orthogonal, the transformed variables are linearly uncorrelated, which helps minimize model-fitting error and leads to a more accurate representation of the data.

\cite{2023AAS...24120907S} first explored using an orthogonal basis to model chromatic timing delays through Hermite polynomials, $H_n(\lambda)$. These polynomials are defined on $(-\infty,\infty)$, preserving the wavelength domain. 
However, the weighting function used for Hermite polynomials is $w_\mathrm{H}(x)=e^{-x^2}$, which introduces two potential problems:

\begin{enumerate}
    \item When fitting a timing model, the effective inner product is a sum over observations weighted by their inverse variance, i.e., $w(\lambda) \propto 1/\sigma^2(\lambda)$. As a result, even if the Hermite polynomials are mathematically orthogonal under the inner product defined by the $e^{-x^2}$ weight, they are not with respect to the inner product used in timing fits. 
    \item The weighting function $e^{-\lambda^2}$ assigns more weight to the region near an arbitrary $\lambda=0$, while points farther away get exponentially downweighted. In real pulsar observations, many important chromatic effects are more prominent at the edges of the frequency band, particularly at low frequencies where dispersion and PE are strongest, and can vary nonlinearly across the frequency range.
\end{enumerate}

In view of these shortcomings, we introduce the use of Legendre polynomials, $L_n(x)$, which are orthogonal under an inner product that uses a flat weighting function, $w_\mathrm{L}(x)=1$. While the data weight $w(\lambda) \propto 1/\sigma^2(\lambda)$ is not exactly flat, it is typically closer to flat \citep[to zeroth order;][]{Lam_2018, NG15} for modern timing uncertainties than the Gaussian weighting used by Hermite polynomials. As a result, Legendre polynomials remain closer to orthogonal under the inner product used in timing, reducing parameter covariance.

Unlike Hermite polynomials, Legendre polynomials are defined over the interval $[-1, 1]$, which does not coincide with the wavelength domain used in pulsar observations. Therefore, we define a bijective mapping to scale our inverse frequencies $\lambda$ into $x \in[-1, 1]$:

\begin{equation}\label{eq:mapping}
    x = 2 \frac{\lambda-\lambda_\mathrm{min}}{\lambda_\mathrm{max} - \lambda_\mathrm{min}} - 1
\end{equation}

\noindent where $\lambda_\mathrm{min}$ and $\lambda_\mathrm{max}$ are some minimum and maximum wavelengths. The inverse transformation is:

\begin{equation}\label{eq:inverse_mapping}
    \begin{split}
     \frac{\lambda}{\lambda_0} = \frac{\Delta}{2 \lambda_0} x + \frac{S}{2 \lambda_0}
    \end{split}
\end{equation}

\noindent where $S \coloneq \lambda_{\mathrm{max}}+\lambda_{\mathrm{min}}$ and $\Delta \coloneq \lambda_{\mathrm{max}}-\lambda_{\mathrm{min}}$.

Substituting Eq.~\ref{eq:inverse_mapping} into Eq.~\ref{eq:monomial_basis_lambdas}, we obtain a polynomial expansion on $x$, which we rearrange by collecting together terms involving the same powers of $x$ to obtain:

\begin{equation}\label{eq:monomial_basis_x}
    t_x = b_0 (1) + b_1 x + b_2 x^2 + b_3 x^3 + \ldots 
\end{equation}

\noindent The coefficients $b_k$ of the polynomial in the normalized variable $x$ are, in principle, functions of the coefficients $a_k$ of the polynomial in the wavelength $\lambda$ (Eq.~\ref{eq:monomial_basis_lambdas}). For a fifth-degree polynomial, the functions are:

\begin{align}\label{eq:as_to_bs}
b_0 &= \tfrac{1}{32}\left(a_5 S^{5} + 2 a_4 S^{4} + 4 a_3 S^{3} + 8 a_2 S^{2} + 16 a_1 S + 32 a_0  \right) \nonumber \\
b_1 &= \tfrac{\Delta}{32}\left(5 a_5 S^{4} + 8 a_4 S^{3} + 12 a_3 S^{2} + 16 a_2 S + 16 a_1 \right) \nonumber \\
b_2 &= \tfrac{\Delta^{2}}{16}\left(5 a_5 S^{3} + 6 a_4 S^{2} + 6 a_3 S + 4 a_2 \right) \\
b_3 &= \tfrac{\Delta^{3}}{16}\left(5 a_5 S^{2} + 4 a_4 S + 2 a_3 \right) \nonumber \\
b_4 &= \tfrac{\Delta^{4}}{32}\left(5 a_5 S + 2 a_4 \right) \nonumber \\
b_5 &= \tfrac{\Delta^{5}}{32}a_5 \nonumber
\end{align}

Eq.~\ref{eq:monomial_basis_x} represents a polynomial in canonical form in the monomial basis, $\{1, x, x^2, \ldots \}$, whose independent variable, $x$, is in $[-1, 1]$. In this domain, we can perform a change of basis to the Legendre basis, $\{L_0(x), L_1(x), L_2(x), \ldots \}$, by rewriting the monomials $x^n$ as linear combinations of the Legendre polynomials and then rearranging them to the canonical form:

\begin{equation}\label{eq:legendre_basis_x}
\begin{split}
    t_x &= b_0 L_0 + b_1 L_1 + b_2 \left( \tfrac{2}{3} L_2 + \tfrac{1}{3} L_0 \right) + b_3 \left( \tfrac{2}{5} L_3 + \tfrac{3}{5} L_1 \right) + \ldots \\
    & \coloneq c_0 L_0(x) + c_1 L_1(x) + c_2 L_2(x) + c_3 L_3(x) + \ldots \\
\end{split}
\end{equation}

\noindent Eq.~\ref{eq:legendre_basis_x} models PE using a Legendre-basis polynomial on the normalized inverse frequency $x$. We will now refer to this as the Legendre Inverse-Frequency Dependent (LIFD) model and to the $\vec{c}=\{ c_0, c_1, \ldots \}$ coefficients as the $\mathrm{LIFD}$ parameters, \{$\mathrm{LIFD_0}$, $\mathrm{LIFD_1}$, $\ldots$\}.

As seen in Eq.~\ref{eq:legendre_basis_x}, the Legendre $c_k$ coefficients are functions of the monomial $b_k$ coefficients, and the exact functional forms will depend on the polynomial degree. For a fifth-degree polynomial, these functions are:

\begin{equation}\label{eq:bs_to_cs}
\begin{split}
    c_0 &= b_0 + \tfrac{1}{3} b_2 + \tfrac{1}{5} b_4 \\
    c_1 &= b_1 + \tfrac{3}{5} b_3 + \tfrac{3}{7} b_5 \\
    c_2 &= \tfrac{2}{3} b_2 + \tfrac{4}{7} b_4 \\
    c_3 &= \tfrac{2}{5} b_3 + \tfrac{4}{9} b_5 \\
    c_4 &= \tfrac{8}{35} b_4 \\
    c_5 &= \tfrac{8}{63} b_5
\end{split}
\end{equation}


We can combine the transformations in Eqs.~\ref{eq:as_to_bs} and \ref{eq:bs_to_cs} to write the coefficients $\vec{a}=(a_0 \ldots a_5)^\mathrm{T}$ of the polynomial in the monomial ``physical'' basis (where we can attribute different physical effects to different terms) as functions of the coefficients $\vec{c}=(c_0\ldots c_5)^\mathrm{T}$ in the Legendre ``orthogonal'' basis (where the timing fit is performed to minimize parameter correlations). If we do that, we will obtain a linear transformation $\vec{a} = \mathbf{T} \vec{c}$, where $\mathbf{T}$ is the upper-diagonal matrix given by:

\begin{widetext}
\begin{equation}\label{eq:transformation_matrix}
    \mathbf{T} = 
    \begin{pmatrix}
    
    1 & -\frac{2S}{\Delta} & \frac{3S^{2}-\frac{\Delta^{2}}{3}}{\Delta^{2}} & \frac{-4 S^{3}+S\Delta^{2}}{\Delta^{3}} &  \frac{5 S^{4}-2 S^{2}\Delta^{2}+\frac{\Delta^{4}}{5}}{\Delta^{4}} & \frac{-6 S^{5}+3 S^{3}\Delta^{2}-S \frac{\Delta^{4}}{2}}{\Delta^{5}} \\
    
    0 & \frac{2}{\Delta} & -\frac{12 S}{\Delta^{2}} & \frac{40 S^{2}-4\Delta^{2}}{\Delta^{3}} & \frac{-70 S^{3}+14 S\Delta^{2}}{\Delta^{4}} & \frac{252 S^{4}-84 S^{2}\Delta^{2}+3\Delta^{4}}{\Delta^{5}} \\
    
    0 & 0 & \frac{6}{\Delta^{2}} & -\frac{60 S}{\Delta^{3}} & \frac{210 S^{2}-30\Delta^{2}}{\Delta^{4}} & \frac{-1260 S^{3}+210 S\Delta^{2}}{\Delta^{5}} \\
    
    0 &0 &0 & \frac{20}{\Delta^{3}} & -\frac{140 S}{\Delta^{4}} & \frac{1680 S^{2}-140\Delta^{2}}{\Delta^{5}} \\
    
    0 & 0 & 0 & 0 & \frac{70}{\Delta^{4}} & -\frac{630 S}{\Delta^{5}} \\
    
    0 & 0 & 0 & 0 &0 & \frac{252}{\Delta^{5}}
    
    \end{pmatrix}
\end{equation}
\end{widetext}

\subsubsection{Advantages and Limitations of a Legendre basis}\label{sec:advantages_Legendre}

Since the Legendre weighting function, $w_\mathrm{L}(x)=1$, is a close approximation to the weighting used in timing fits, $1/\sigma^2(\nu)$, we expect the fitted LIFD coefficients to be nearly uncorrelated in the timing fit. Therefore, their covariance matrix will be approximately diagonal, i.e., $\mathrm{Cov}(\vec{c}) \approx \mathrm{diag}(\sigma_{c_1}^2, \ldots, \sigma_{c_n}^2)$. However, the physical interpretation of the fitted model requires transforming to the monomial basis, where the covariance matrix is:

\begin{equation}
    \mathrm{Cov}(\vec{a})=\mathbf{T}\mathrm{Cov}(\vec{c})\mathbf{T}^{T}
\end{equation}

\noindent Since $\mathbf{T}$ is not diagonal, $\mathrm{Cov}(\vec{a})$ will, in general, also be non-diagonal. Therefore, transforming from the Legendre basis back to the monomial basis reintroduces covariance among the polynomial coefficients $\vec{a}$, even if the Legendre coefficients $\vec{c}$ are uncorrelated.

Nevertheless, orthogonal bases preserve a major structural advantage. Consider the impact of a contaminating dispersive delay, $t_\mathrm{DM}\propto \lambda^2$, within the Legendre formulation. According to Eq.~\ref{eq:inverse_mapping}, it can be written as:

\begin{equation}
    t_\mathrm{DM}\propto\lambda^2 = \left(\tfrac{S}{2} + \tfrac{\Delta}{2} x \right)^{2} =  \tfrac{S^2}{4} + \tfrac{S}{2} \Delta x + \tfrac{\Delta^2}{4} x^2
\end{equation}

\noindent In turn, we can rewrite the monomial basis $\{1, x, x^2\}$ in term of Legendre polynomials to obtain:

\begin{equation}
    \begin{split}
    t_\mathrm{DF} \propto& \tfrac{S^2}{4} L_0 + \tfrac{S}{2} \Delta L_1 + \tfrac{\Delta^2}{2} \left( \tfrac{2}{3} L_2 + \tfrac{1}{3} L_0 \right) \\ =& \underbrace{\left( \tfrac{S^2}{4} + \tfrac{\Delta^2}{12} \right)}_{\coloneq c_0} L_0 + \underbrace{\left( \tfrac{S}{2} \Delta \right)}_{\coloneq c_1} L_1 + \underbrace{\tfrac{\Delta^2}{6}}_{\coloneq c_2} L_2
    \end{split}
\end{equation}

\noindent We see that the Legendre expansion of a contaminating dispersive delay is finite and confined to the first three modes. Therefore, it only affects the fits of the Legendre coefficients $c_0$, $c_1$, and $c_2$, and does not populate higher-order Legendre modes. In other words, dispersion should not produce spectral power leakage into higher-order Legendre coefficients. 

When we transform the fitted Legendre coefficients to the monomial basis using $\vec{a}=\mathbf{T}\vec{c}$, the matrix $\mathbf{T}$ is upper diagonal (see Eq.~\ref{eq:transformation_matrix}), giving a hierarchical structure:

\begin{itemize}
    \item $a_5$ is uniquely determined by $c_5$,
    \item $a_4$ depends only on $\{c_4, c_5\}$, and%
    \item $a_3$ depends only on $\{c_3, c_4, c_5\}$.
\end{itemize}

\noindent Since a purely dispersive delay only affects the fitted values of $\{c_0, c_1, c_2\}$, it follows that $a_3$, $a_4$, and $a_5$ receive no deterministic contributions from dispersion. In contrast, $a_2$, $a_1$, and $a_0$ depend on, at least, $c_2$ and are therefore affected by contributions from dispersion.

In conclusion, because the transformation matrix to the monomial basis is upper triangular, higher-order monomial coefficients $a_i$ depend only on equal or higher-order Legendre coefficients $c_i$. As a result, coefficients of order higher than quadratic exhibit a simpler variance structure and receive no contamination from dispersive effects. Fitting in an orthogonal basis, therefore, separates low-order chromatic behavior from higher-order profile evolution and should provide a more accurate description of PE than the FE log-polynomial modeling.

\section{Implementation}\label{sec:implementation}

In this section, we will describe our numerical implementations of the PE models described in Sec.~\ref{sec:theoretical_background}.

\subsection{Data Selection}

\begin{table}
    \centering
    \begin{tabular}{ccccc}
        \hline
         \multirow{2}{*}{PSR}& $\mathrm{DM}^\mathrm{(NG15)}$ & $\mathrm{DM}^\mathrm{(new)}$& Percentual $\mathrm{DM}$\\
 & [$\mathrm{pc}/\mathrm{cm}^3$] & [$\mathrm{pc}/\mathrm{cm}^3$]& difference (\%)\\
         \hline
         B1937+21&  71.1001 & 71.0192 & $-$0.1138 \\
         J0610$-$2100& 61.277 & 60.6805 & $-$0.9734 \\
         J1012+5307& 8.924 & 9.0177 & 1.05 \\
         J1022+1001&  9.436 & 10.3126 & 9.29 \\
         J1024$-$0719& 8.4181 & 6.456 & $-$23.3081 \\
         J1643$-$1224&  62.2996 & 62.3918 & 0.148 \\
         J1713+0747& 15.9743 & 15.9914 & 0.107 \\
         J1909$-$3744& 10.3912 & 10.3914 & 0.0019 \\
         J1918$-$0642& 26.5904 & 26.5899 & $-$0.0019 \\
         J2145$-$0750& 8.9295 & 8.9998 & 0.7873 \\
         J2302+4442 & 13.7632 & 13.7265 & $-$0.2667\\
         \hline
    \end{tabular}
    \caption{NG15 pulsars used to test PE models. $\mathrm{DM}^\mathrm{(NG15)}$ is the published dispersion measure, the $\mathrm{DM}^\mathrm{(new)}$ is the value that we obtained after applying the algorithm in Sec.~\ref{subsec:DM_refitting}, and $\mathrm{DM}$ difference presents the relative difference between these two values as a percentage of the $\mathrm{DM}^\mathrm{(NG15)}$ value, i.e., $(\mathrm{DM}^\mathrm{(new)} - \mathrm{DM}^\mathrm{(NG15)})/\mathrm{DM}^\mathrm{(NG15)} \times 100$.}
    \label{tab:PSRs_and_DMs}
\end{table}

We tested our PE models on a sample of 11 pulsars included in the NANOGrav 15-year dataset \citep[NG15;][]{NG15}, selected to represent a wide variety of pulse morphologies, PE behaviors, dispersion measures, and observing baselines. This ranges from MSPs with multi-component pulse structure that changes substantially with frequency, such as PSR B1937+21 \citep{Alam_2021}, PSR J1022+1001 \citep{2025ApJ...984..139F}, PSR J1643$-$1224 \citep{2018ApJ...868..122B}, and PSR J2145$-$0750 \citep{2018ApJ...868..122B}, to well-timed MSPs with simple pulse profiles and relatively small PE across NANOGrav bands, such as PSR J1012+5307 and PSR J1909-3744.

For each pulsar, the NG15 dataset provided: 1) a set of TOA measurements at different frequencies and epochs (contained in a \texttt{.tim} text file), 2) a timing model that includes values of the DM, DMX, and FD parameters (in a \texttt{.par} text file), and 3) a set of time-averaged but phase- and frequency-resolved pulse profiles, one for each observing epoch (in \texttt{.ff} PSRFITS-format files).

\subsection{Fiducial DM Re-Fitting}\label{subsec:DM_refitting}

NANOGrav's timing models account for interstellar dispersion by fitting a fiducial $\mathrm{DM}$ value and $\mathrm{DMX}$ corrections (see Sec.~\ref{subsec:freq_dependent_effects}). However, correlations between dispersion and PE terms can cause the least-squares fit to redistribute spectral power between these two components (see Sec.~\ref{subsec:FD_mis_estimations}). Furthermore, estimating the $\mathrm{DM}$ by fitting a timing model to TOA measurements over a finite frequency band can yield biased estimates due to incomplete frequency sampling \citep{Sosa_Fiscella_2024}. In either case, the fitted $\mathrm{DM}$ value can be shifted away from the true dispersion measure.

Since both the $a_2'$ IFD coefficient and the lower-order ${c_0, c_1, c_2}$ LIFD coefficients are still covariant with $\nu^2$-delays (including dispersion), performing a timing fit while using a strongly misestimated initial $\mathrm{DM}$ value can produce excess residual structure that may be partially redistributed between $\mathrm{DM/DMX}$ and these PE parameters. To reduce the magnitude of the $\mathrm{DMX}$ corrections from the fiducial $\mathrm{DM}$, and the excess residual amplitude that can be redistributed to the PE parameters, we fitted more robust $\mathrm{DM}$ values independently from the timing fit. To that end, for each pulsar in our sample:



\begin{enumerate}
    \item We took an interval ${[\mathrm{DM}^{0}-\Delta \mathrm{DM}, \mathrm{DM}^{0}+\Delta \mathrm{DM}]}$ around the NG15 nominal dispersion measure $\mathrm{DM}^{0}$, where we took $\Delta \mathrm{DM}$ as $25\%$ of $\mathrm{DM}^{0}$.
    \item We selected a set of trial $\mathrm{DM}$ values within this range. However, instead of blindly testing $\mathrm{DM}$ values, we used the Brent optimization algorithm \citep{brent1973algorithm} from the \texttt{SciPy} \texttt{optimize} module \citep{2020SciPy-NMeth}. In this approach, trial $\mathrm{DM}$ values are updated iteratively by successively narrowing the range of values on the interval.
    \item For each trial $\mathrm{DM}$ value, we used \texttt{PyPulse} \citep{2017ascl.soft06011L} to de-disperse and frequency-average each of the pulsar's profiles, thus obtaining a time- and frequency-integrated pulse profile per observing epoch, which we cross-correlated to the pulsar's NG15 template (at the observation's frequency band) to obtain its signal-to-noise (S/N) ratio.
    \item Lastly, we averaged the S/N of all the observations de-dispersed with a given trial $\mathrm{DM}$ value.
\end{enumerate}

\noindent Since $\langle \mathrm{S/N} \rangle$ is computable for all the $\mathrm{DM}$ within ${[\mathrm{DM}^{0}-\Delta \mathrm{DM}, \mathrm{DM}^{0}+\Delta \mathrm{DM}]}$, the Brent optimization algorithm should converge to the $\mathrm{DM}$ value that maximizes $\langle \mathrm{S/N} \rangle$ \citep{brent1973algorithm}.

This physically motivated approach takes advantage of the fact that using incorrect $\mathrm{DM}$ estimates to de-disperse frequency-resolved data results in a smearing of the integrated pulse profile, reducing the peak amplitude and, therefore, the resulting S/N. Therefore, maximizing the S/N effectively finds the $\mathrm{DM}$ that best aligns the pulses across frequency. This provides a DM estimate tied directly to the profile morphology and independent of timing model assumptions.

The original and new $\mathrm{DM}$ values are presented in Table ~\ref{tab:PSRs_and_DMs}. The new estimates, $\mathrm{DM}^\mathrm{(new)}$, are generally within $\sim1$\% of the original value, $\mathrm{DM}^\mathrm{(NG15)}$. The only notable exception is PSR J1024$-$0719, whose adjusted dispersion measure value, $\mathrm{DM}^\mathrm{(new)}=6.456 ~\mathrm{pc/cm^3}$, represents a $23.3$\% reduction from $\mathrm{DM}^\mathrm{(NG15)}=8.4181~\mathrm{pc/cm^3}$. Coincidentally, this $\mathrm{DM}^\mathrm{(new)}$ value is consistent with the DM value fitted in datasets previous to NG15. Therefore, we consider the NG15 value to be an overstimation of the real DM, which, when coupled to covariances with the FD parameters, would also explain the flip in the concavity of the NG15 FD delay curve in Fig.~\ref{fig:FD_curves}.


\subsection{Model Implementation and Significance Testing}\label{subsec:model_fitting}

After changing the nominal $\mathrm{DM}$ value for each pulsar to the value found in Sec.~\ref{subsec:DM_refitting}, we performed timing fits of its TOAs to timing models that account for PE using either an FD (Eq.~\ref{eq:old_FD_formulation}), IFD (Eq.~\ref{eq:IFD_polynomial}), or LIFD (Eq.~\ref{eq:legendre_basis_x}) model. \texttt{PINT} implements the FD model in its \texttt{FD} class. For the other two models, we implemented new \texttt{PINT} classes named \texttt{IFD} and \texttt{LIFD} that can be attached to a timing model in replacement of the \texttt{FD} class (see Appx.~\ref{sec:appendix}). In brief, these new classes operate as follows:

\begin{enumerate}
    \item Given a set of frequency-resolved TOA measurements, the \texttt{IFD}/\texttt{LIFD} classes transform the radio frequencies $\nu$ to inverse $\lambda$ or normalized-inverse frequencies $x$, respectively. For the \texttt{LIFD} class, we have defined two new timing parameters, \texttt{LIFD\_FMIN} and \texttt{LIFD\_FMAX}, representing the maximum and minimum frequencies of the dataset, and which can be readily converted to the maximum and minimum wavelengths defining the $\lambda \rightarrow x$ mapping (see Eq.~\ref{eq:mapping}), so the model is reproducible and portable across TOA subsets.
    \item PE is modeled as a new frequency-dependent delay term, given either by Eq.~\ref{eq:IFD_polynomial} or Eq.~\ref{eq:legendre_basis_x}, included in the timing model's global TOA estimation.
    \item Finally, the polynomial coefficients are included as free parameters when using a weighted least-squares algorithm to fit the model to the pulsar's TOA measurements \citep[e.g.,][]{Luo_2021}. 
\end{enumerate}

\begin{figure*}[ht]
\centering

\begin{subfigure}{\textwidth}
    \begin{subfigure}{.5\textwidth}
        \centering
        \includegraphics[width=0.9\linewidth]{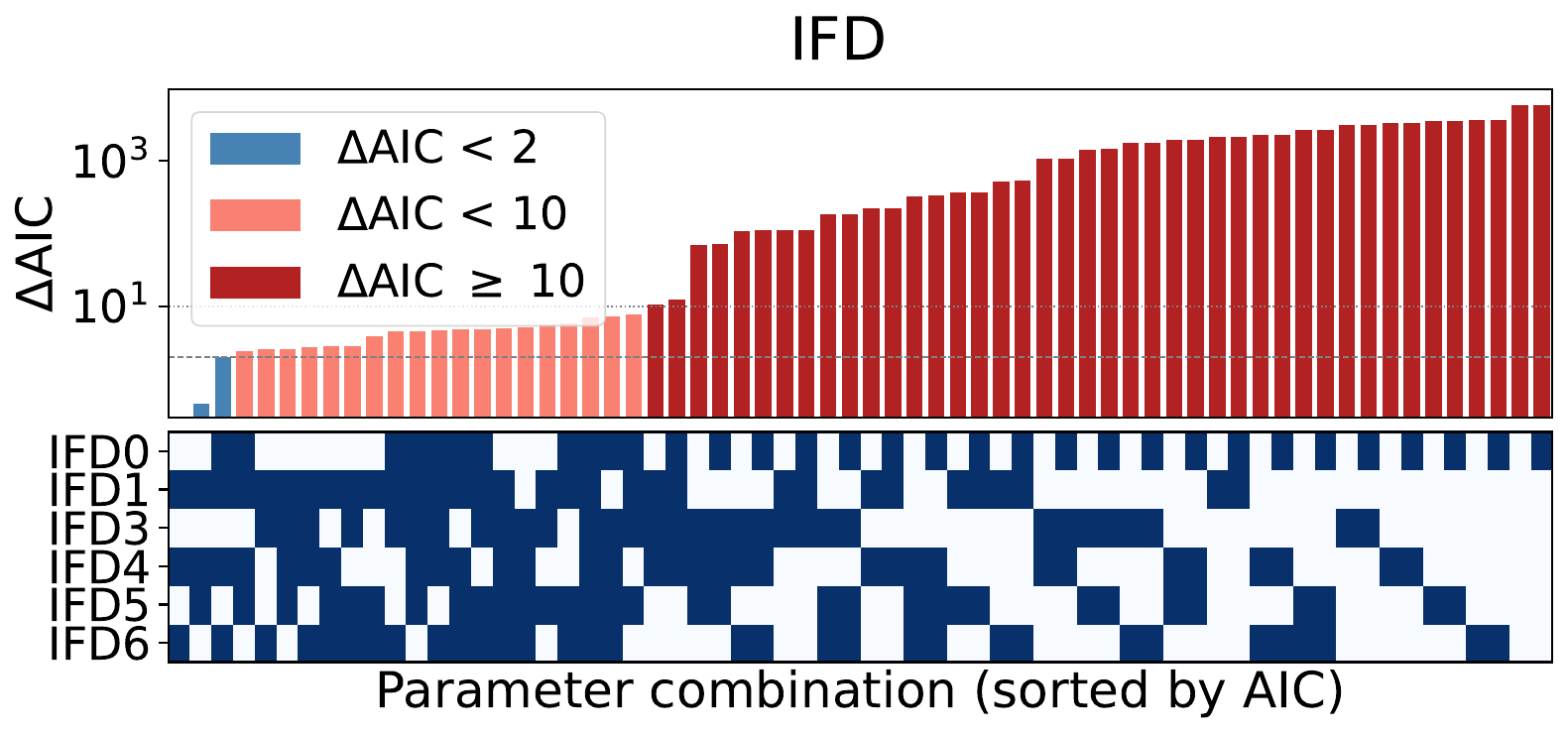}
    \end{subfigure}%
    \begin{subfigure}{.5\textwidth}
        \centering
        \includegraphics[width=0.9\linewidth]{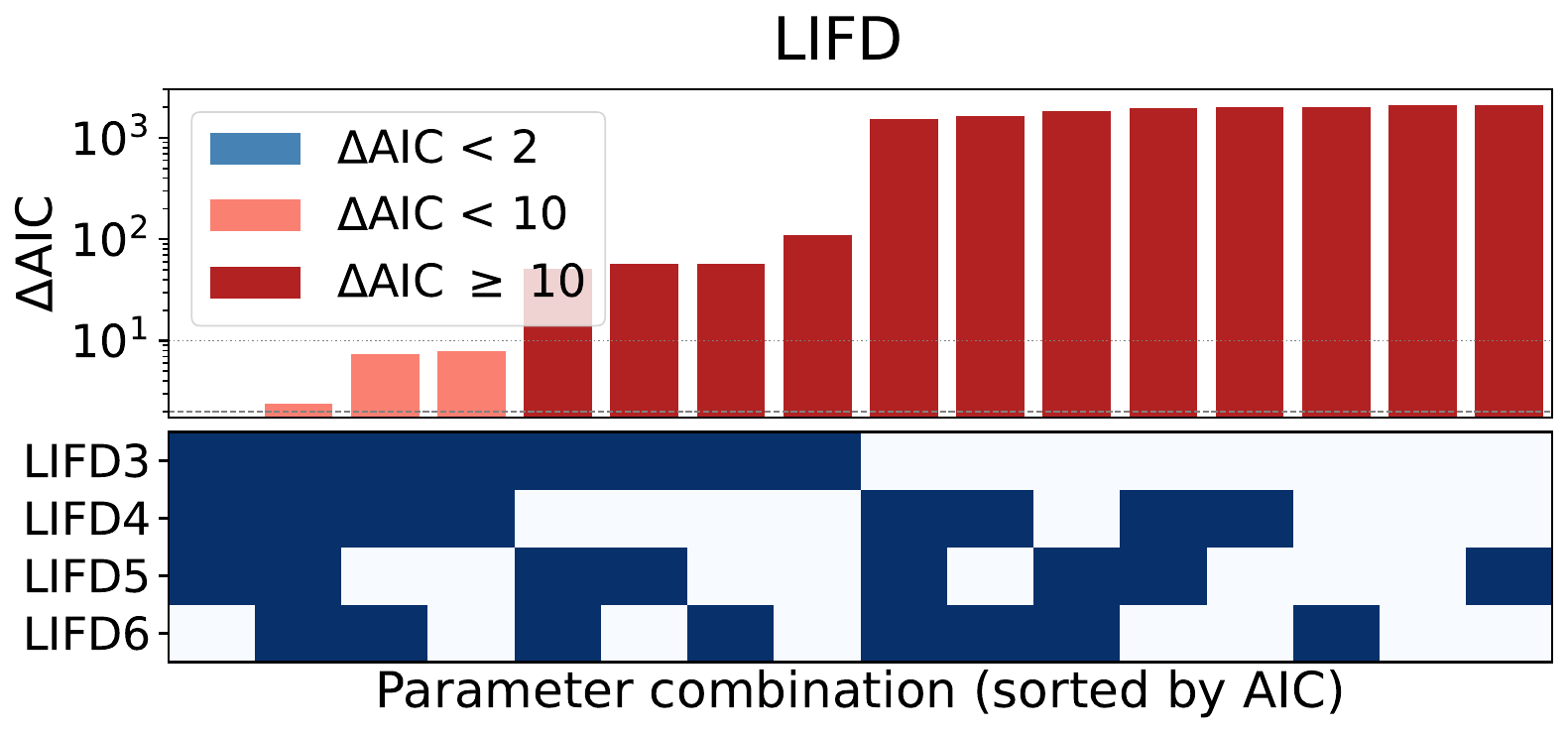}
    \end{subfigure}
        \subcaption{PSR J1643$-$1224}
        \label{subfig:J1643_pair}
\end{subfigure}

\begin{subfigure}{\textwidth}
    \begin{subfigure}{.5\textwidth}
        \centering
        \includegraphics[width=0.9\linewidth]{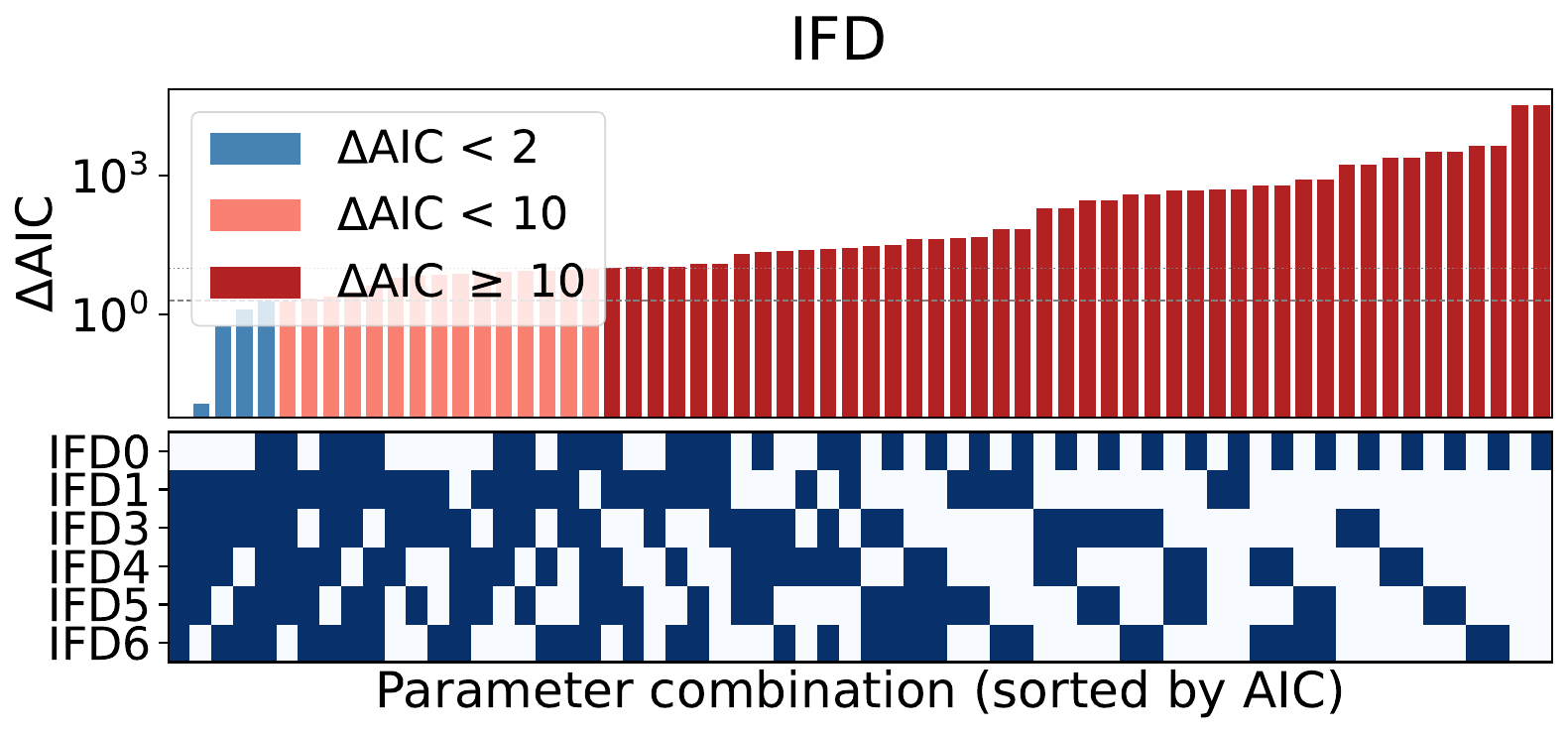}
    \end{subfigure}%
    \begin{subfigure}{.5\textwidth}
        \centering
        \includegraphics[width=0.9\linewidth]{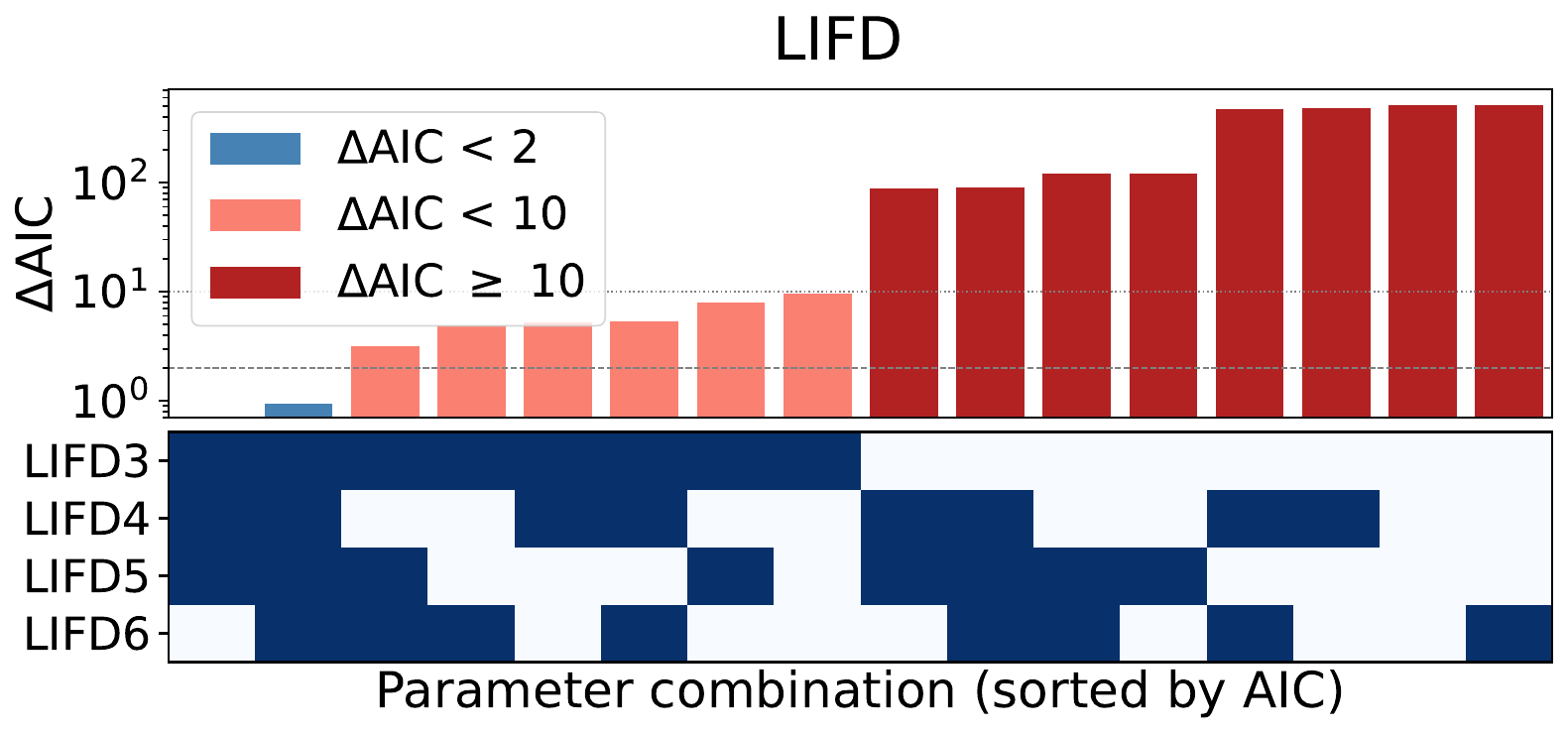}
    \end{subfigure}
        \subcaption{PSR J2145$-$0750}
        \label{subfig:J2145_pair}
\end{subfigure}

\caption{Top: Bar chart of $\Delta \mathrm{AIC} = \mathrm{AIC} - \mathrm{AIC}_\mathrm{min}$ for each combination of IFD/LIFD parameters, sorted by $\Delta \mathrm{AIC}$ (best model on the left). The conventional thresholds ($\Delta \mathrm{AIC} < 2$: comparable support; $\Delta \mathrm{AIC} > 10$: strongly disfavored) are shown as horizontal dashed lines. Bottom: A binary heatmap aligned with the bars above, where rows are the free parameters and columns are the model combinations. A filled box means the parameter was included, and an empty box means it was removed.}
\label{fig:aic_comparison}

\end{figure*}

These \texttt{PINT} classes can be set up with an arbitrary polynomial order. However, fitting an arbitrarily high order increases the number of free parameters, which is both computationally inefficient and can lead to overfitting. Instead, an optimal number of parameters should reflect the significance of PE for each pulsar.

In the NANOGrav data releases, the optimal number of FD parameters is found iteratively using an $F$-statistic test \citep{NG9}. This test is valid when one model is a subset of another, so the data releases have required that all FD parameters up to the highest-order significant FD parameter be included in the fit, even if the lower-order parameters are not found to be significant. For example, if FD3 is significant but FD2 is not, then FD1, FD2, and FD3 are included \citep{NG12}.

However, IFD and LIFD models isolate chromatic effects into specific terms, so we decided to allow non-contiguous combinations of parameters. Therefore, to find the optimal number of IFD/LIFD parameters, we used the Akaike information criterion (AIC) since, unlike the $F$-test, it does not require the models to be nested. Given a model with best-fit parameters, the AIC is defined as $\mathrm{AIC} = 2k - 2\ln(L)$, where $k$ is the number of free parameters in the model and $L$ is the model likelihood \citep{aic}.

Since PE varies smoothly over our observing bandwidth, higher-order polynomial terms are not expected to contribute significantly, as they would capture fine-scale frequency structure that is not physically consistent with PE and also exceeds our frequency resolution, making such terms poorly constrained. We therefore restricted the AIC test to IFD/LIFD polynomials of order $N\leq6$ as an upper bound, and considered all possible combinations of its 7 terms. The number of combinations is simplified by considering that dispersion is covariant with $\mathrm{IFD2}$ (see Sec.~\ref{sec:advantages_monomial}) and with $\mathrm{LIFD0}$, $\mathrm{LIFD1}$, and $\mathrm{LIFD2}$ (see Sec.~\ref{sec:advantages_Legendre}), so these terms can always be assumed to be relevant.

For each of the parameter combinations, we performed a timing fit of the IFD/LIFD model to the pulsar's NG15 TOAs, and then applied the \texttt{akaike\_information\_criterion} function from the \texttt{pint.utils} module to compute the model's AIC value. The preferred model is the one that minimizes the AIC. 
Generally speaking, we found that:
\begin{itemize}
    \item The favored LIFD models preserve lower-order ($\mathrm{LIFD3}$ and $\mathrm{LIFD4}$) but remove higher-order coefficients ($\mathrm{LIFD6}$ and, in some cases, $\mathrm{LIFD5}$).
    \item IFD models including $\mathrm{IFD0}$ are usually disfavored. This can be interpreted as $\mathrm{IFD0}$ being a frequency-independent timing offset, which is completely degenerate with other parameters in the timing model (e.g., the pulsar's phase or period).
\end{itemize}

We exemplify these findings for two representative pulsars in Fig.~\ref{fig:aic_comparison}, but we obtained similar AIC results for all MSPs we analyzed. Therefore, in all the subsequent timing fits, we adopted LIFD polynomials with all terms up to the $5^\mathrm{th}$ order, and IFD polynomials of degree $5$ that remove the degenerate $\mathrm{IFD0}$ term.

\section{Results}\label{sec:results}

In this section, we present the results of fitting each PE model to the NG15 TOAs of our sampled MSPs, including the impact on the interstellar dispersion model.

\subsection{Profile Evolution Curves Across Models}\label{subsec:evolution_across_models}

\begin{figure*}[ht]
\centering

\begin{subfigure}{0.3\linewidth}
    \centering
    \includegraphics[width=\linewidth]{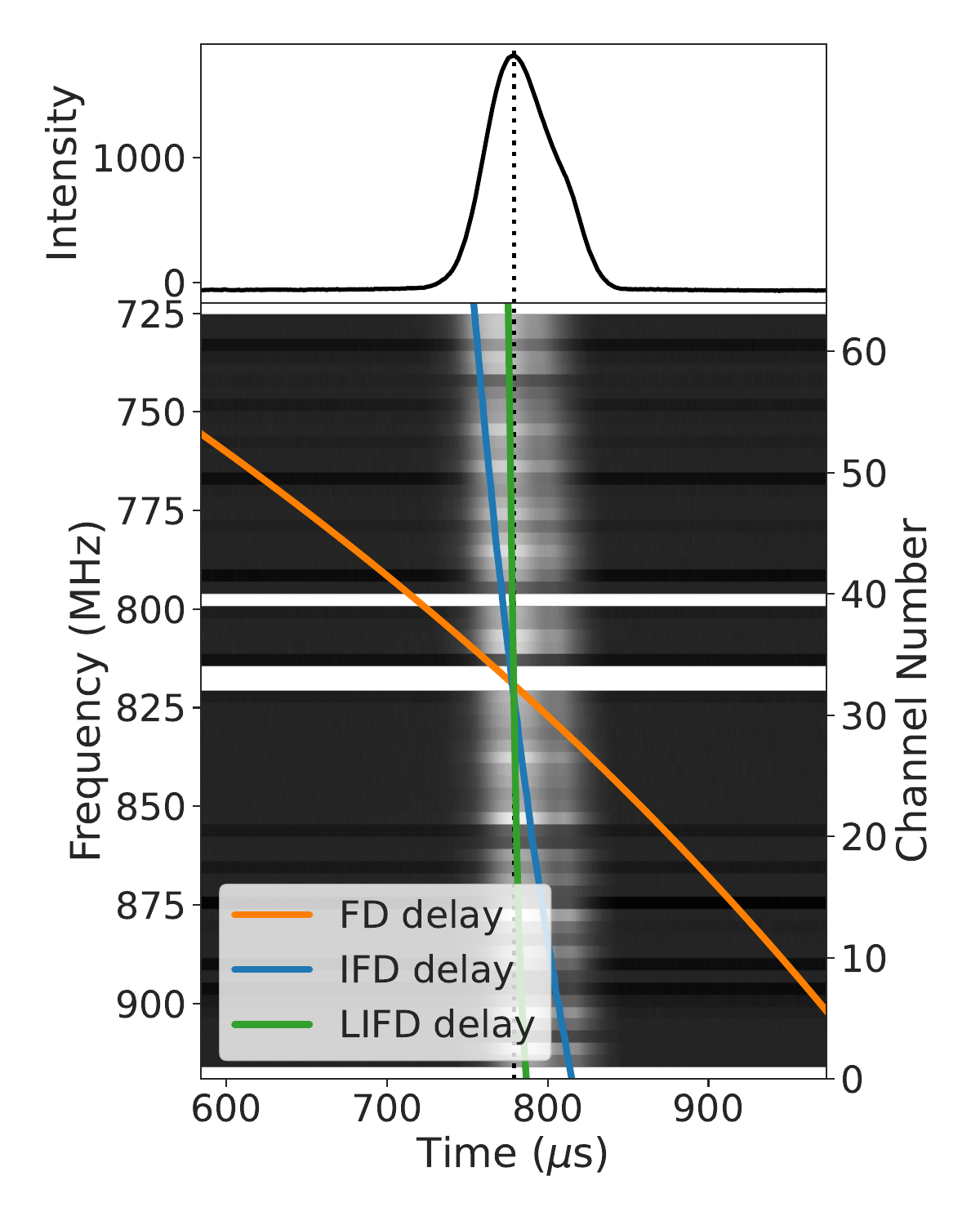}
    \subcaption{B1937+21}
    \label{subfig:B1937+21_waterfall}
\end{subfigure}%
\begin{subfigure}{0.3\linewidth}
    \centering
    \includegraphics[width=\linewidth]{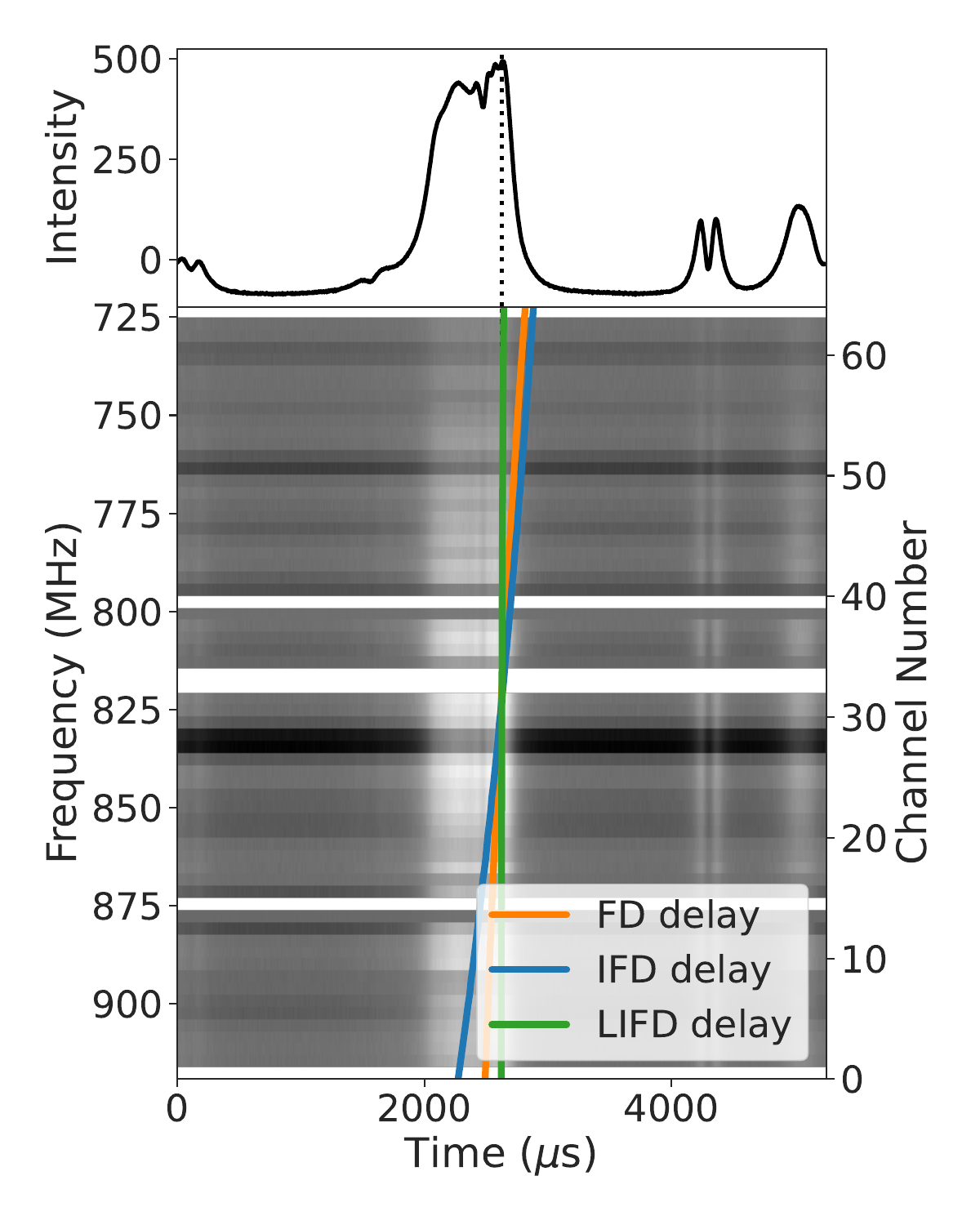}
    \subcaption{J1012+5307}
    \label{subfig:J1012+5307_waterfall}
\end{subfigure}
\begin{subfigure}{0.3\linewidth}
    \centering
    \includegraphics[width=\linewidth]{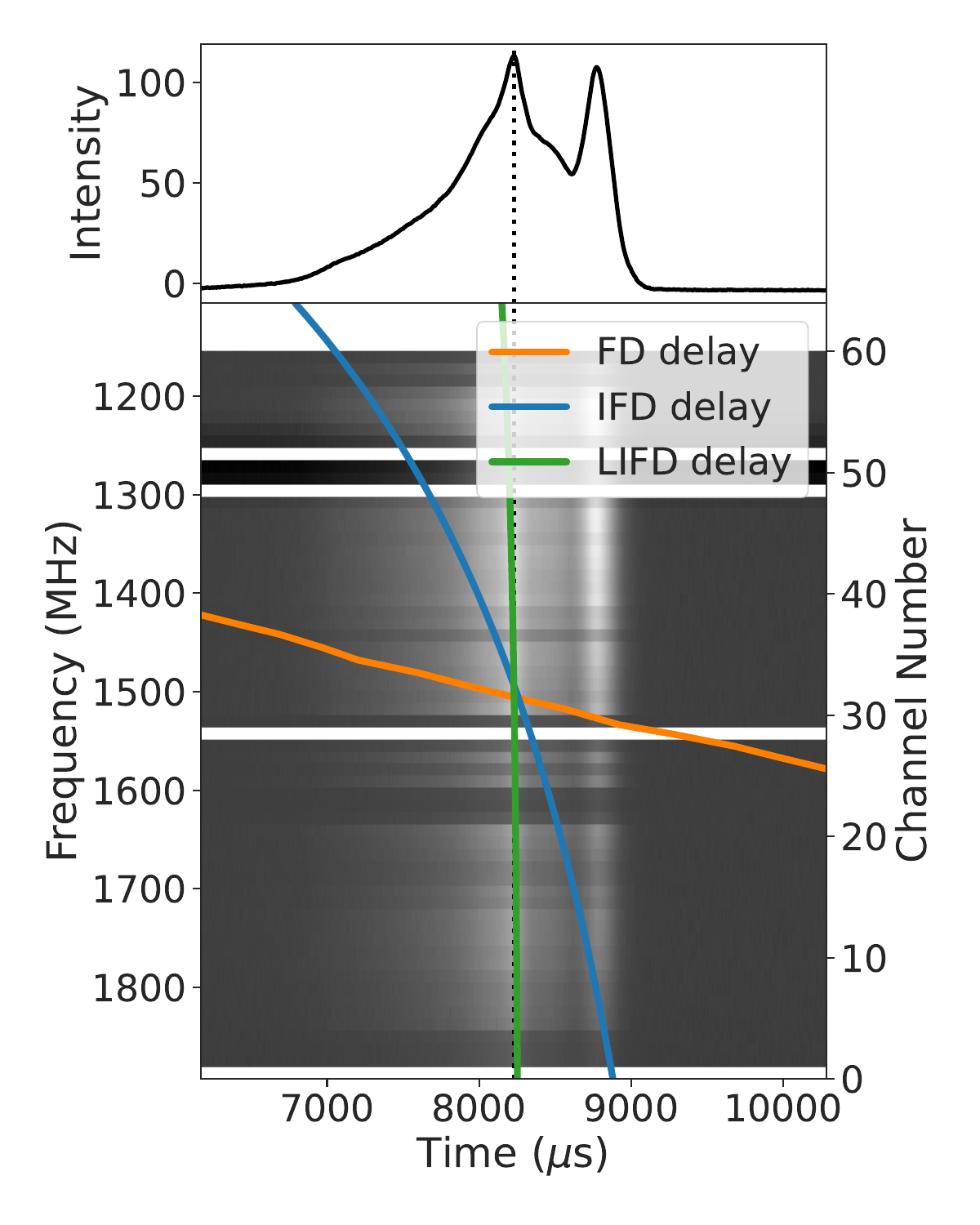}
    \subcaption{J1022+1001}
    \label{subfig:J1022+1001_waterfall}
\end{subfigure}%

\begin{subfigure}{0.3\linewidth}
    \centering
    \includegraphics[width=\linewidth]{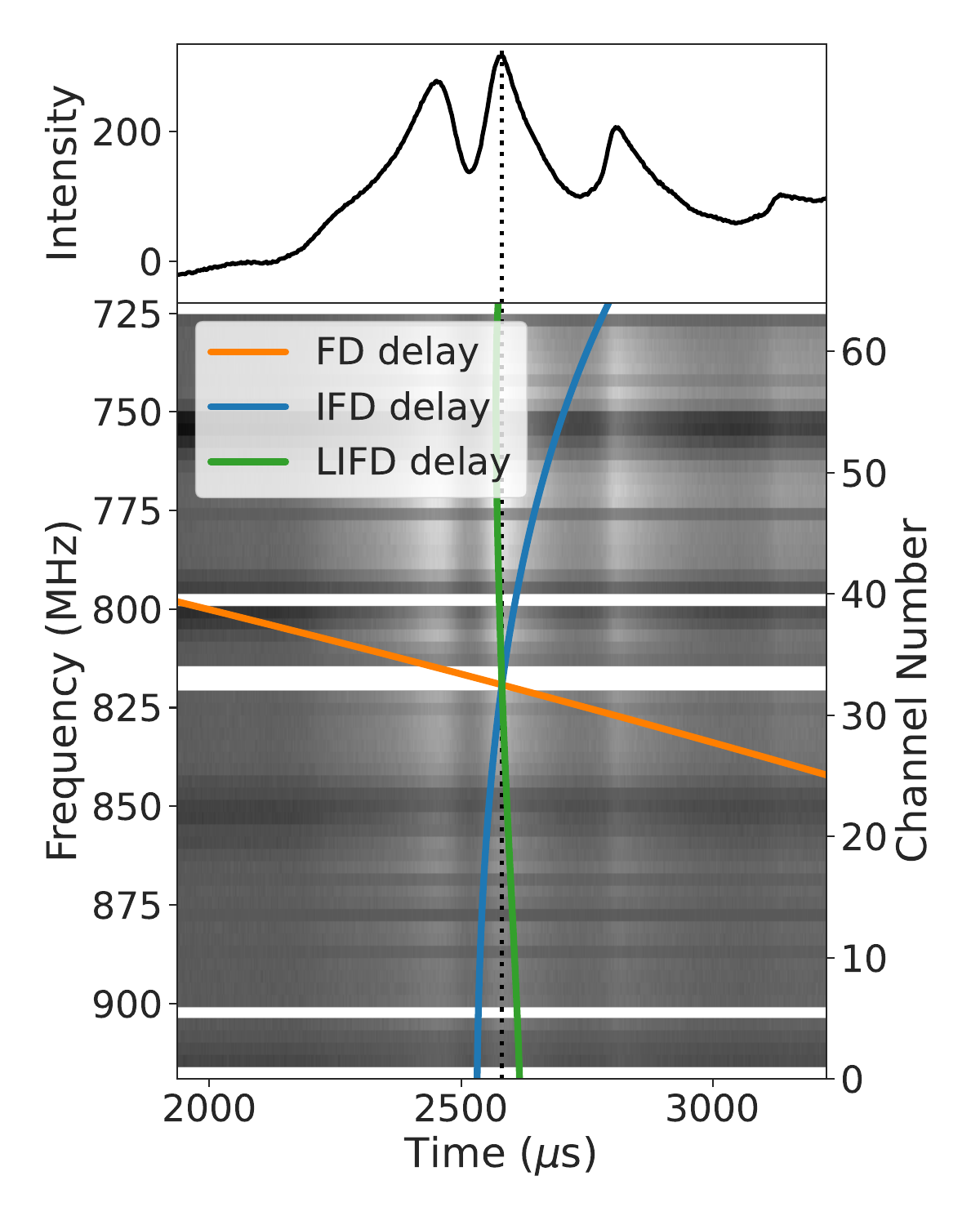}
    \subcaption{J1024$-$0719}
    \label{subfig:J1024-0719_waterfall}
\end{subfigure}
\begin{subfigure}{0.3\linewidth}
    \centering
    \includegraphics[width=\linewidth]{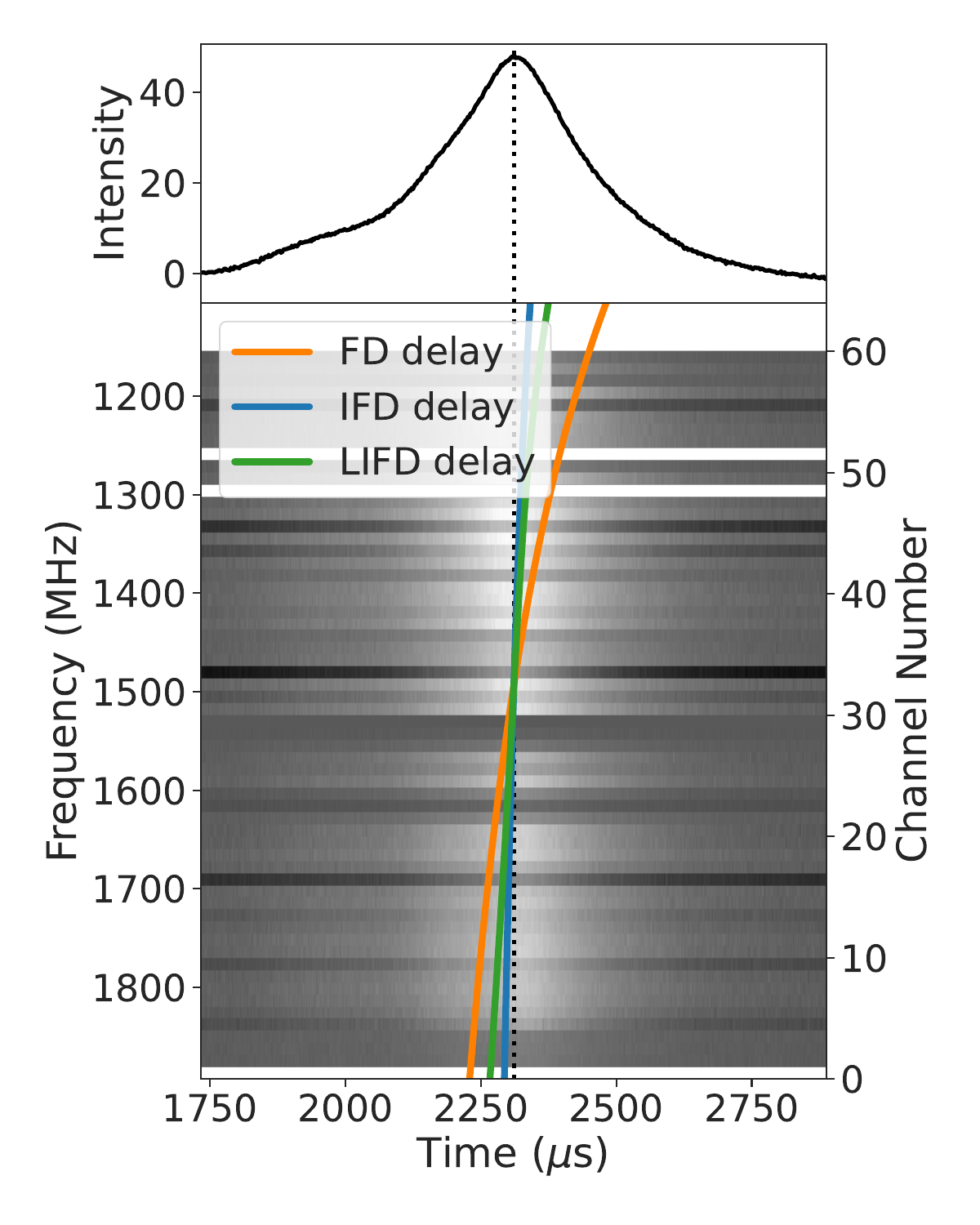}
    \subcaption{J1643$-$1224}
    \label{subfig:J1643-1224_waterfall}
\end{subfigure}
\begin{subfigure}{0.3\linewidth}
    \centering
    \includegraphics[width=\linewidth]{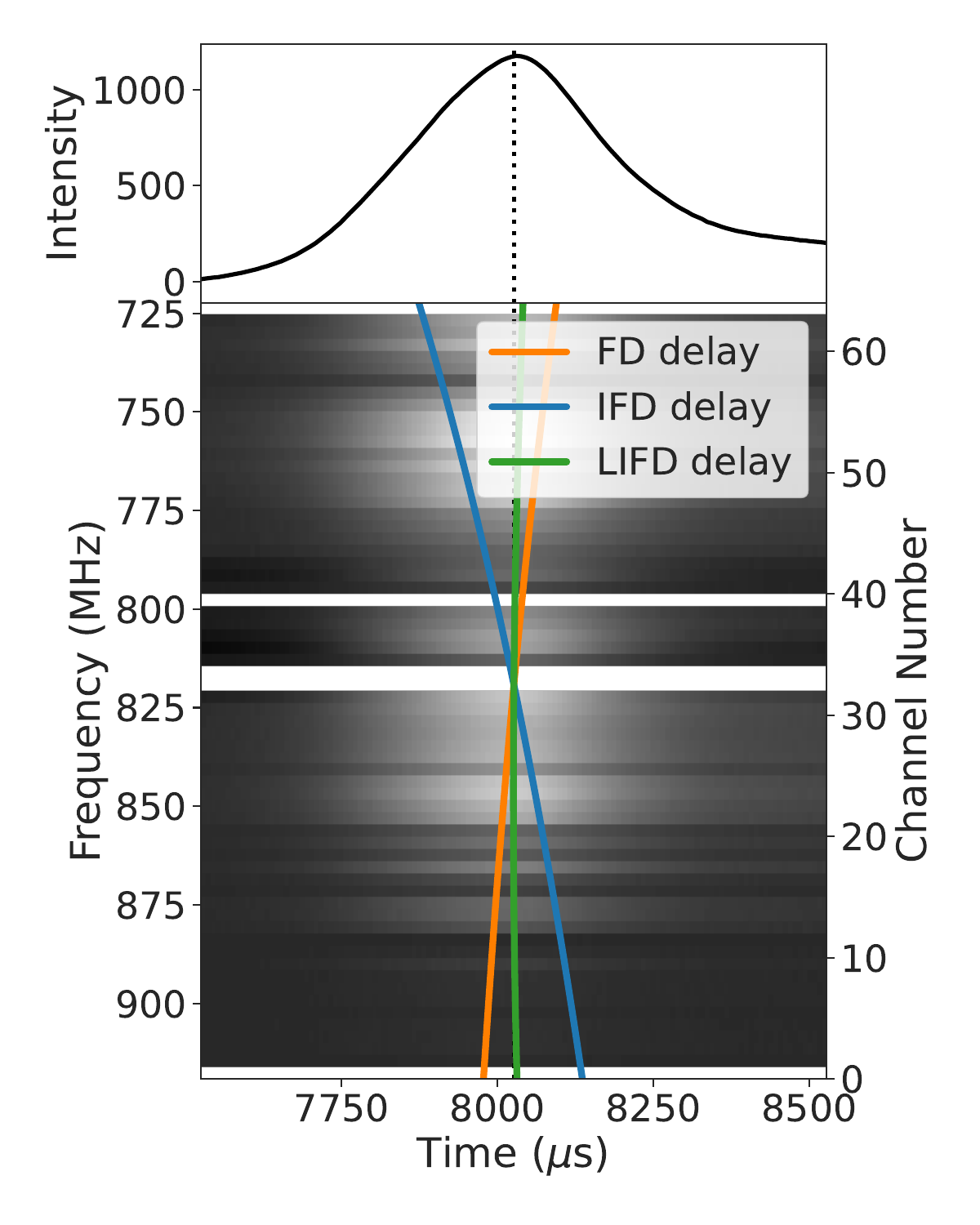}
    \subcaption{J2145$-$0750}
    \label{subfig:J2145-0750_waterfall}
\end{subfigure}

\caption{In each subfigure, the color map shows the intensity of the (de-dispersed) pulse profile as a function of frequency and rotational phase for the pulsar's highest-S/N observation. We overlay the PE delay curves as fitted by the FD (orange), IFD (green), or LIFD (blue) model. 
The top plot shows the corresponding frequency-integrated pulse profile. We see that FD models can result in delay curves that significantly overestimate the sweep of the pulse profile across frequency, while the LIFD-derived delay curves follow this sweep more closely across all MSPs.}
\label{fig:profile_evolution_curves}
\end{figure*}

After implementing and fitting the PE models described in Sec.~\ref{sec:implementation} to the NG15 TOAs of each MSP in our sample, we obtained a set of FD/IFD/LIFD parameters describing the pulsar's profile evolution across the NANOGrav frequency bands. We summarize the results for a representative subset of MSPs in Fig.~\ref{fig:profile_evolution_curves}. For each pulsar, we present the profile frequency portrait from its highest-S/N observation, after de-dispersing it using the pulsar's NG15 DM value. We overlap PE delay curves obtained by creating a polynomial in a logarithmic (FD), monomial (IFD), or Legendre (LIFD) basis, whose coefficients are the FD/IFD/LIFD parameters resulting from the timing fit, and evaluating it at the observation's frequencies $\nu$ (FD), inverse frequencies $\lambda$ (IFD), or normalized inverse frequencies $x$ (LIFD).

Our fitted FD models tend to overestimate the empirical PE delay amplitude. This effect is more significant in pulsars with multi-component pulse structure and substantial PE, such as PSRs B1937+21, J1024$-$0719, and, J1022+1001; however, it becomes less pronounced in pulsars with simpler pulse structures and less significant PE, such as PSRs J1643$-$1224 and J1012+5307. Interestingly, both B1937+21 and J1643$-$1224 are high-DM pulsars (see Table~\ref{tab:PSRs_and_DMs}), but the PE overestimation is significantly larger for the first pulsar than for the second. Conversely, both J1024$-$0719 and J1012+5307 are relatively low-DM pulsars, but they also exhibit markedly different degrees of PE mis-estimation. We found similar results for other MSPs in our sample. Therefore, PE mis-estimations seem more strongly correlated with the MSP's pulse structure and evolution than with its DM.

For high-evolution pulsars such as B1937+21 and J1024$-$0719, the IFD models yield delay amplitudes that are significantly closer to the observed pulse profiles than those predicted by the FD models. However, for most low-evolution pulsars, the FD and IFD delay amplitudes are comparable. In some cases, the IFD curves depart more significantly from the empirical PE than the FD curves (e.g., J1012+5307). Therefore, the IFD formulation is still susceptible to mis-estimations, and its advantages over FD models are only significant for high-evolution pulsars.

Across our sample of MSPs, the most robust approximations to the empirical PE were derived using LIFD models. The improvements are most prominent for high-evolution MSPs such as B1937+21 and J1024$-$0719. 
However, even for low-evolution MSPs where the IFD models had overestimated PE, such as J2145$-$0750, we find that the LIFD-derived delay curves also closely follow the pulse profiles. LIFD models are still susceptible to spectral power leakage and, in some cases, perform comparably to FD and IFD models (e.g., J1643$-$1224 and J1022+1001). Even so, we find that LIFD generally provides more accurate and consistent PE delay curves across our pulsar sample than either FD or IFD models.

\subsection{Dispersion Response}

In Sec.~\ref{subsec:FD_mis_estimations}, we conjectured that covariances between different chromatic effects could lead to spectral power being inadvertently moved from the dispersion to the PE terms of the timing model during the timing fit, thus overestimating the FD contribution while underestimating that of the DM. Now, LIFD resulting in smaller PE delay amplitudes than FD models could be indicative of LIFD models better constraining cross-term covariances and limiting spectral power leakage into the PE term. In that case, we would expect the dispersion model to also change in response to this spectral power excess no longer being absorbed into the PE term. We will then verify this assertion by comparing the dispersion models fitted when using either FD, IFD, or LIFD.

\begin{figure*}[ht]
\centering

\begin{subfigure}{\linewidth}
  
    \begin{subfigure}{0.3\linewidth}
        \centering
        \includegraphics[width=\linewidth]{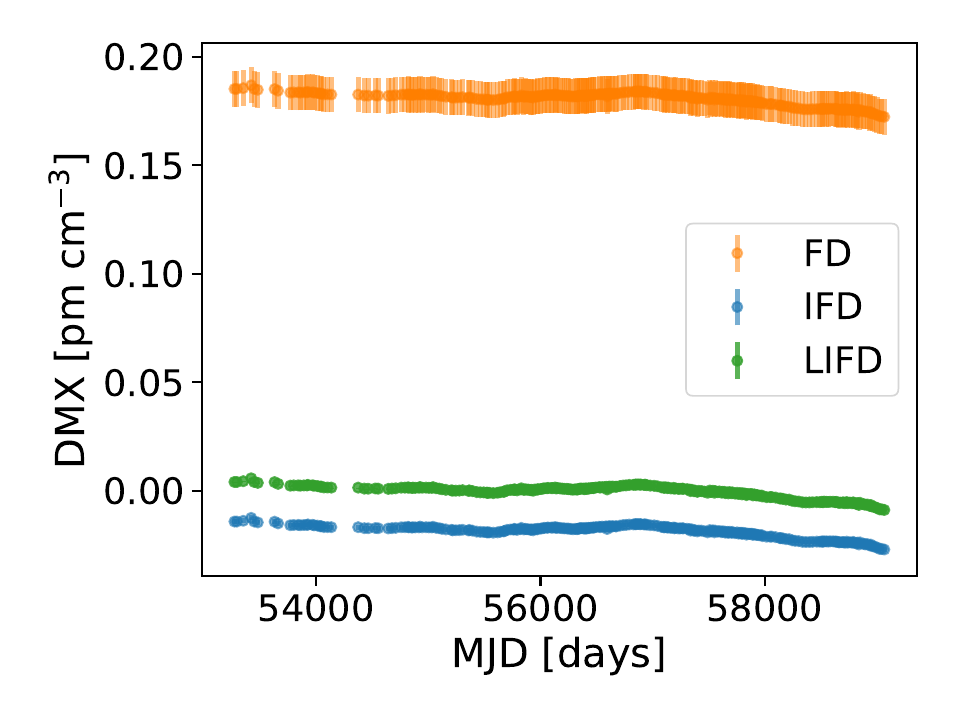}
        \subcaption{B1937+21}\label{subfig:DMX_B1937}
    \end{subfigure}
    \begin{subfigure}{0.3\linewidth}
        \centering
        \includegraphics[width=\linewidth]{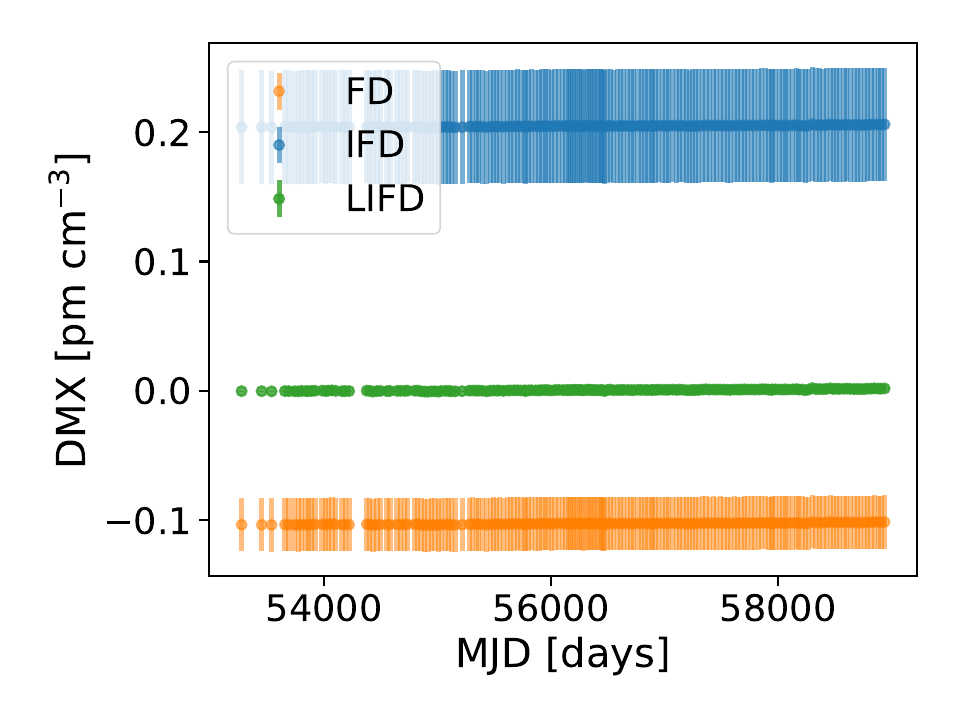}
        \subcaption{J1012+5307}
    \end{subfigure}    
        \centering
        \begin{subfigure}{0.3\linewidth}
        \includegraphics[width=\linewidth]{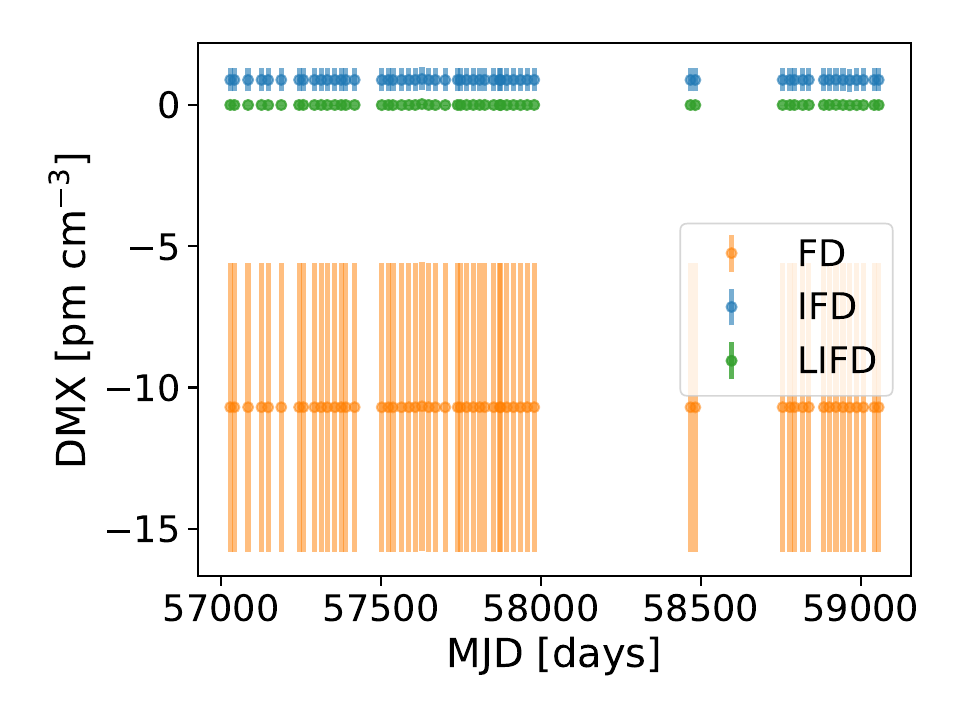}
        \subcaption{J1022+1001}
    \end{subfigure}
\end{subfigure}

\begin{subfigure}{\linewidth}
    \centering
    \begin{subfigure}{0.3\linewidth}
        \centering
        \includegraphics[width=\linewidth]{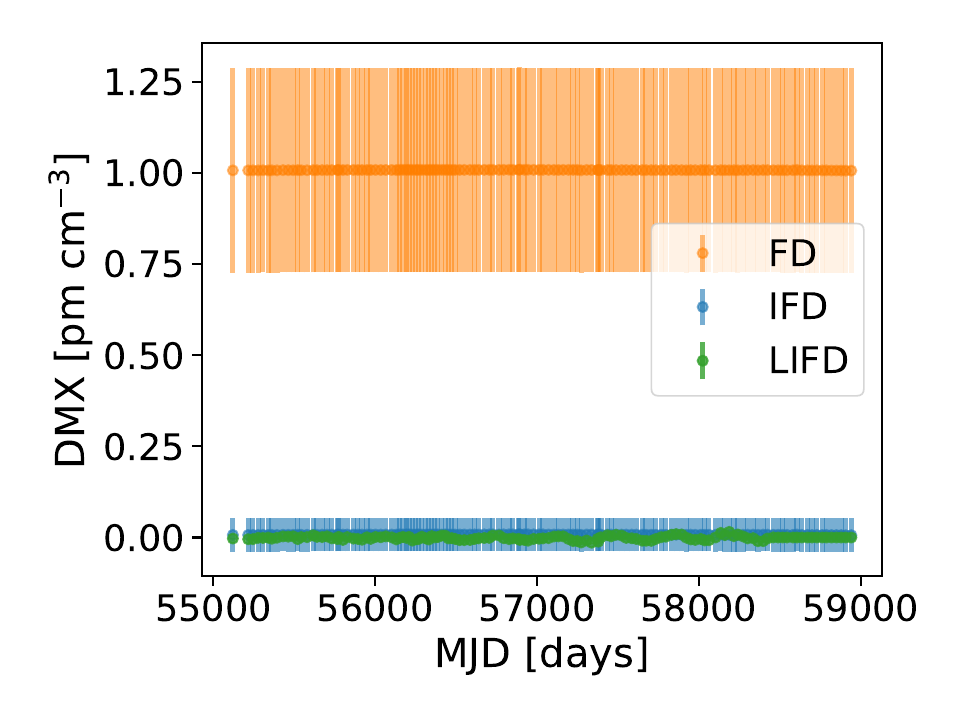}
        \subcaption{J1024$-$0719}
    \end{subfigure}
    \begin{subfigure}{0.3\linewidth}
        \centering
        \includegraphics[width=\linewidth]{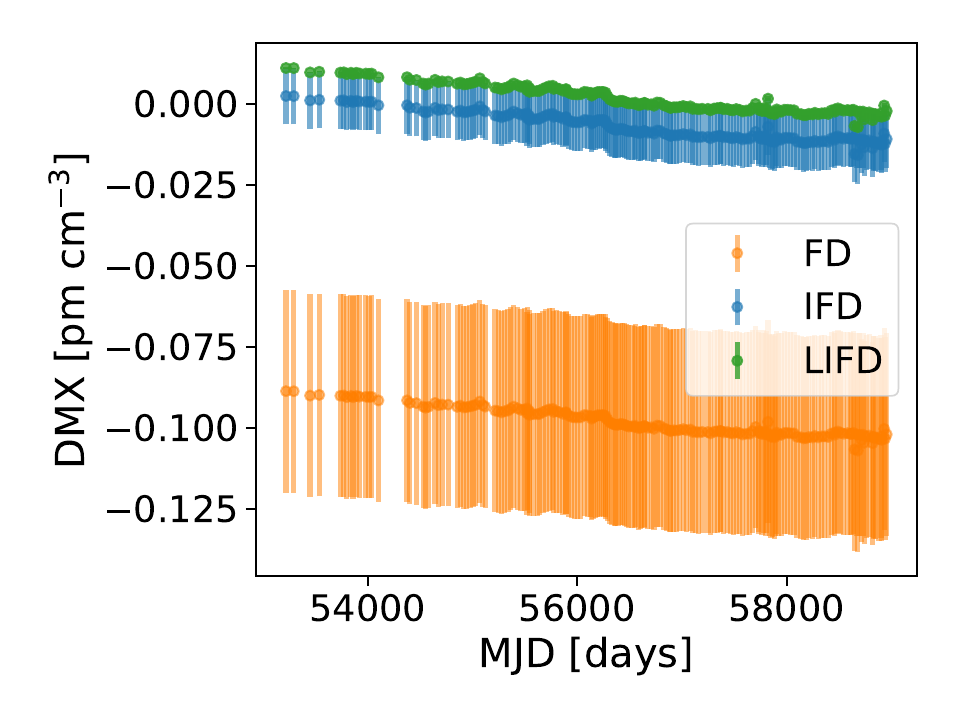}
        \subcaption{J1643$-$1224}
    \end{subfigure}
    \begin{subfigure}{0.3\linewidth}
        \centering
        \includegraphics[width=\linewidth]{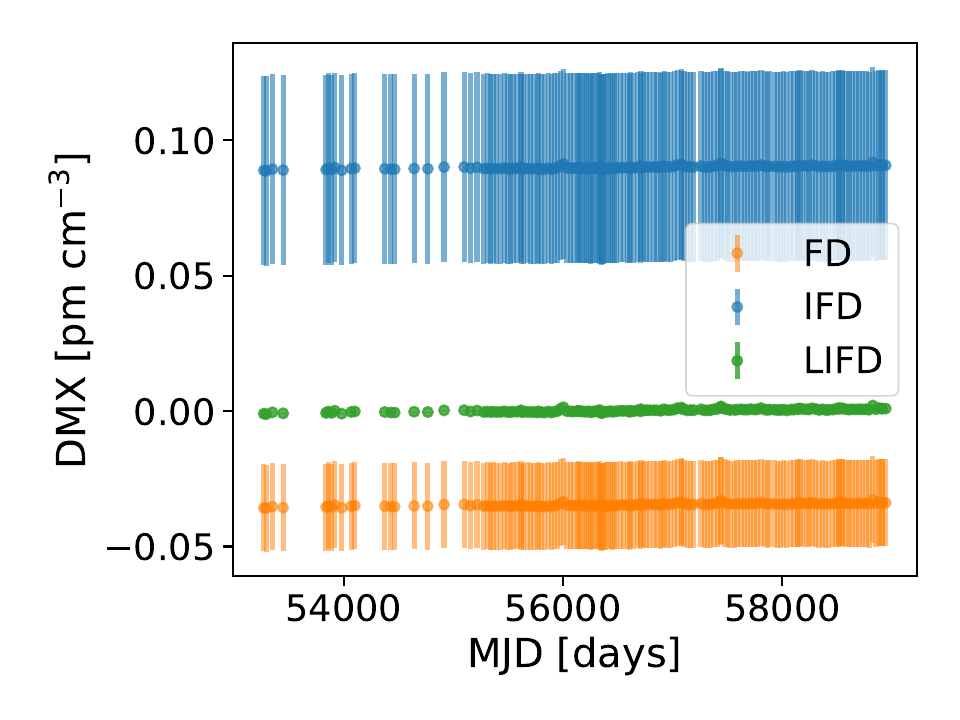}
        \subcaption{J2145$-$0750}
    \end{subfigure}
\end{subfigure}

\caption{Response of the dispersion model to using different PE models. For each pulsar, we present the epoch-dependent DMX correction as a function of the center of the corresponding time window. The DMX parameters were obtained by fitting a timing model that used an FD (orange), IFD (blue), or LIFD (green) model to account for PE. While the DM temporal trends are the same, FD and IFD models result in recovering larger DMX corrections and with larger uncertainties than LIFD models.}
\label{fig:DMX_response}
\end{figure*}

When fitting the timing models in Sec.~\ref{subsec:evolution_across_models}, we set each pulsar's nominal DM to the value found in Sec.~\ref{subsec:DM_refitting} and left the piecewise DMX corrections as free parameters during the timing fit. Therefore, any response in the dispersive model should manifest through these fitted DMX parameters. Therefore, we decided to analyze how the fitted DMX values change when using the same fiducial DM and set of TOAs, and only changing the PE model. We present the results for a subset of representative MSPs in Fig~\ref{fig:DMX_response}.

We see that the recovered DM time series changes when using an FD, IFD, or LIFD model, with FD and IFD resulting in larger DMX errors than LIFD. This agrees with our original conjecture: different PE models absorb different amounts of chromatic power, and DMX responds by also shifting the predicted dispersive delay. Moreover, because the FD and IFD bases are not orthogonal, they are susceptible to correlations and error propagation. That same ill-conditioning propagates into the DMX fit, since the fitter has to simultaneously solve for DMX and FD/IFD parameters, which are all covariant with each other. Conversely, LIFD uses basis elements that are orthogonal to each other and partially orthogonal to a dispersive $\nu^{-2}$-delay, leading to better-constrained LIFD and DMX parameters.

Nevertheless, we also find that the LIFD formulation, which predicts smaller delays due to PE, can also result in smaller $\mathrm{DMX}$ values (e.g., Fig.~\ref{subfig:DMX_B1937}). This contradicts our initial conjecture, since we expected that smaller PE-induced delays would force the excess spectral power that had leaked into the PE term to be reabsorbed into the dispersion term, thus fitting larger DMX values, to keep total chromatic delay constant.


One possible explanation is that the PE and dispersion model always have some degree of degeneracy in the $\nu^{-2}$ direction, so the timing fit can approximate the same observed chromatic delay with different combinations of PE and DMX parameters. In particular:

\begin{itemize}
    \item Since the FD covariance matrix is ill-conditioned, one such possible combination is a large and positive FD delay compensated by fitting a large and negative dispersive delay, or vice versa. The modeled FD + DM chromatic delay remains consistent with the data, but the individual components are inflated in opposite directions.
    \item In contrast, since LIFD models have more terms orthogonal to a dispersive $\nu^{-2}$ delay, they have lower degeneracy with the DMX parameters. As a result, the resulting LIFD and DM terms operate more independently and do not need to cancel each other, thus yielding more realistic predictions for both DM- and PE-induced timing delays.
\end{itemize}


\begin{figure*}
\centering

\begin{subfigure}{\linewidth}
    \centering
    \begin{subfigure}{0.325\linewidth}
        \includegraphics[width=\linewidth]{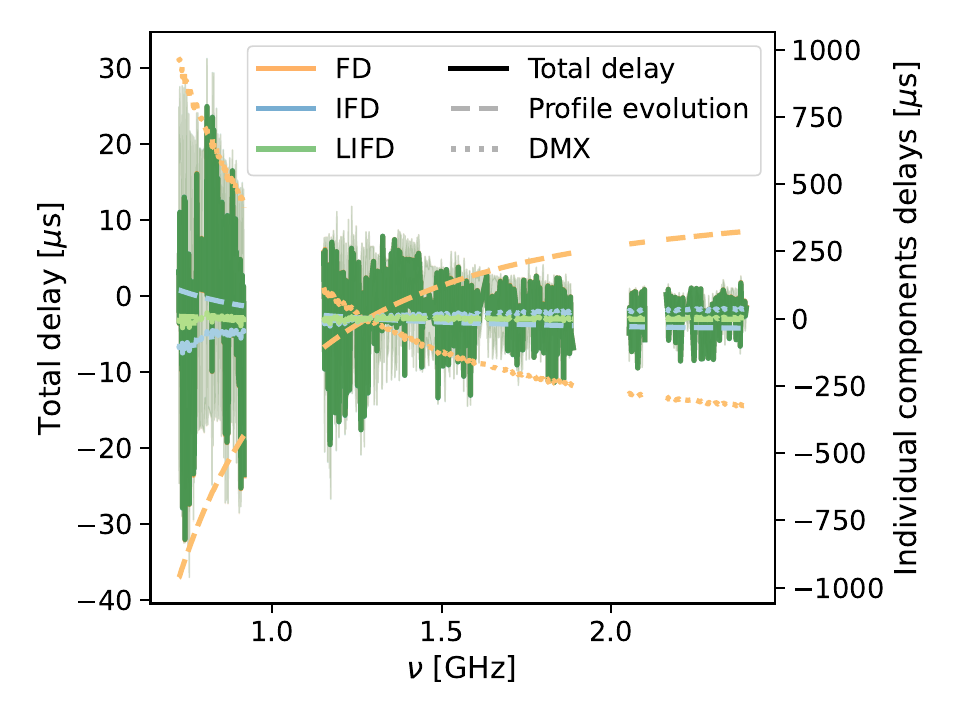}
        \subcaption{B1937+21}
    \end{subfigure}
    \hfill
    \begin{subfigure}{0.325\linewidth}
        \includegraphics[width=\linewidth]{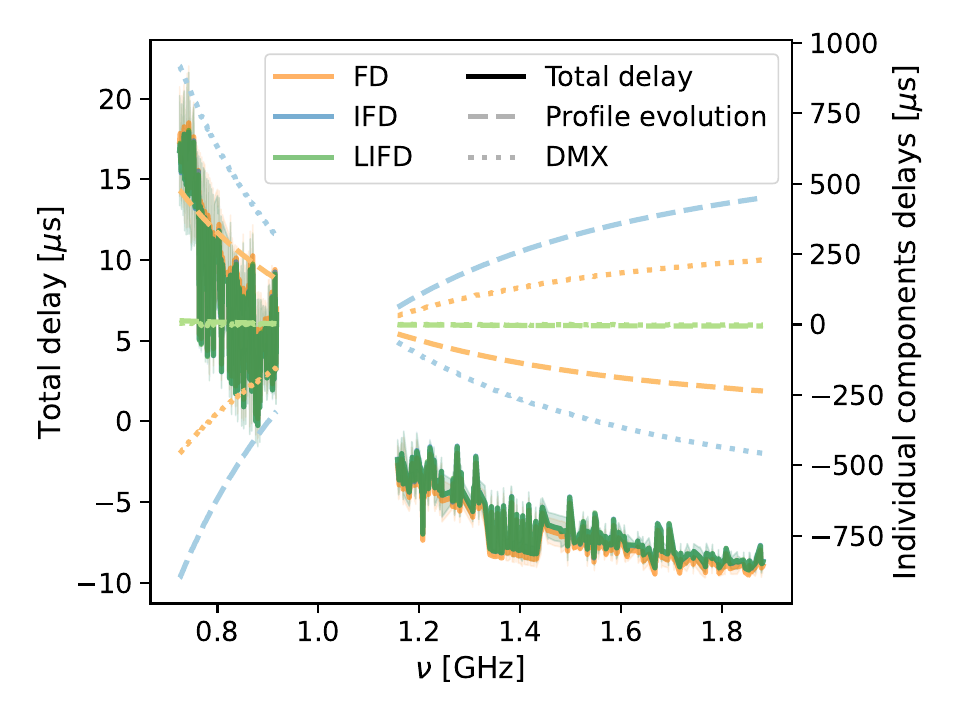}
        \subcaption{J1012+5307}
    \end{subfigure}    
    \hfill
        \begin{subfigure}{0.325\linewidth}
        \includegraphics[width=\linewidth]{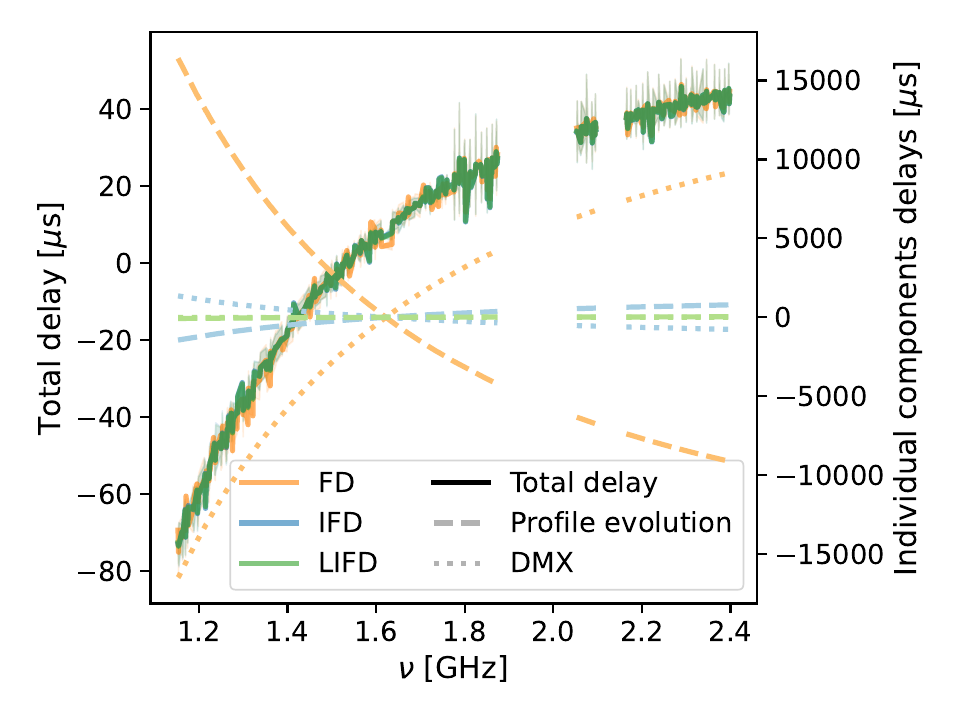}
        \subcaption{J1022+1001}
    \end{subfigure}
\end{subfigure}

\begin{subfigure}{\linewidth}
    \centering
    \hfill
    \begin{subfigure}{0.325\linewidth}
        \includegraphics[width=\linewidth]{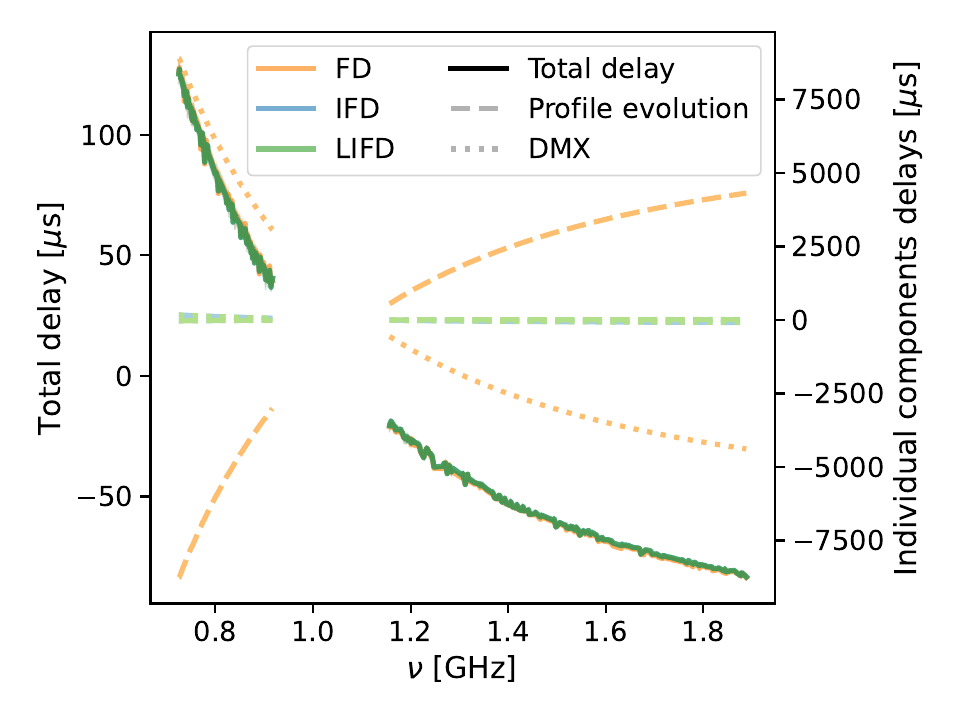}
        \subcaption{J1024$-$0719}
    \end{subfigure}
    \hfill
    \begin{subfigure}{0.325\linewidth}
        \includegraphics[width=\linewidth]{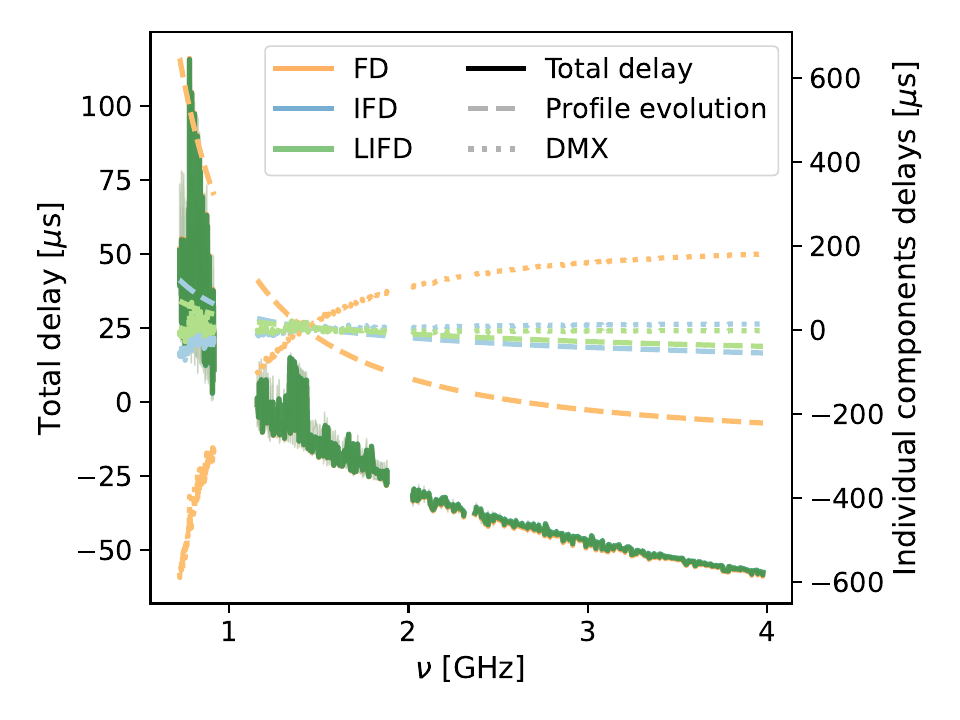}
        \subcaption{J1643$-$1224}
    \end{subfigure}
    \hfill
    \begin{subfigure}{0.325\linewidth}
        \includegraphics[width=\linewidth]{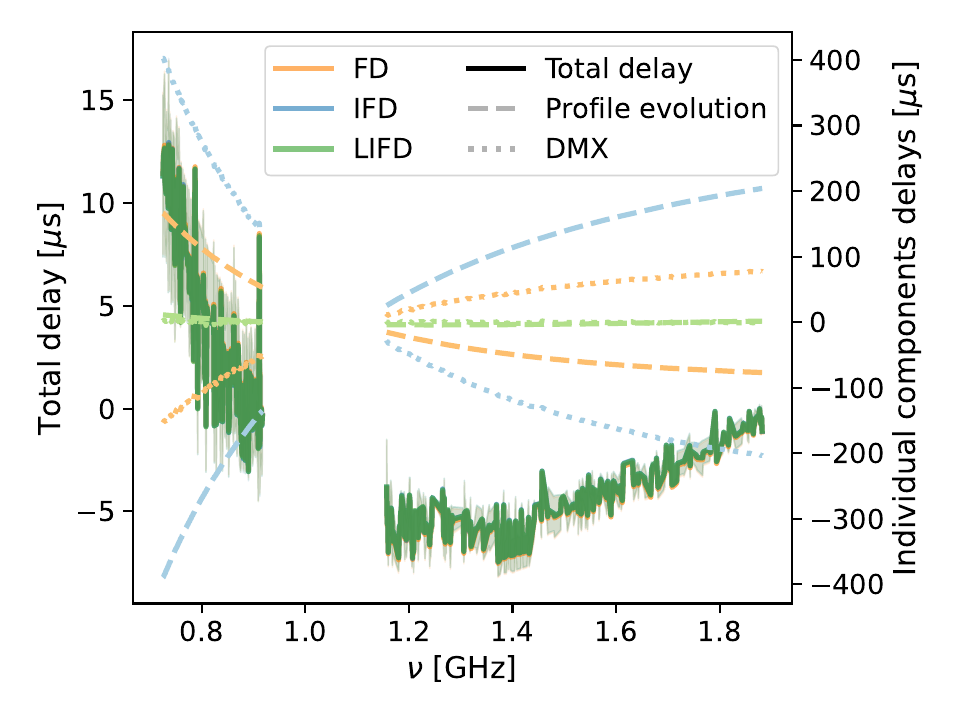}
        \subcaption{J2145$-$0750}
    \end{subfigure}
\end{subfigure}

\caption{For a subset of our sampled MSPs, and using different models of PE, we present: 1) the epoch-average of the PE-induced timing delays at each TOA's frequency (dashed lines), 2) the epoch-average of the DM-induced delays at each TOA's epoch and frequency (dotted lines), and 3) the epoch-average of total chromatic delays (PE + DM; solid lines) and their standard deviation (shaded region). The DMX and PE parameters were obtained simultaneously by fitting a timing model to the pulsar's TOAs, using an FD (orange), IFD (blue), or LIFD (green) model. While all three models predict the same total chromatic delay for each pulsar, FD/IFD fit significantly larger delay amplitudes for the individual effects than LIFD does.}
\label{fig:chromatic_delays_breakdown}
\end{figure*}

\noindent We tested this hypothesis by calculating, for each of our sampled MSPs, 1) the PE-induced timing delay predicted by the fitted FD/IFD/LIFD model, at the observing frequency of each of its TOAs, 2) the DM-induced timing delay predicted by the DM model that was fitted with each PE model, at each TOA's observing frequency and epoch, and 3) the total chromatic delay given by the sum of the previous two terms at each TOA epoch and frequency. Since each observing frequency contains TOAs across different observing epochs, each corresponding to a different DMX correction, we epoch-averaged the delays for TOAs at each frequency. We present the results in Fig.~\ref{fig:chromatic_delays_breakdown}.

We see that the total chromatic delay (DM + PE) stays essentially unchanged by the choice of PE model. However, the contributions of the individual chromatic components change significantly: in the FD formulation, the large delay amplitudes ($\gtrsim 100~\mathrm{\mu}$s) predicted for PE are balanced by large DM delays with opposite sign. LIFD models achieve the same total chromatic delay while fitting smaller PE- and DM-induced delays. IFD models fit individual component delays generally intermediate to FD and LIFD. 

\begin{figure*}[ht]
\centering

\begin{subfigure}{\linewidth}
    \centering
    \begin{subfigure}{0.3\linewidth}
        \includegraphics[width=\linewidth]{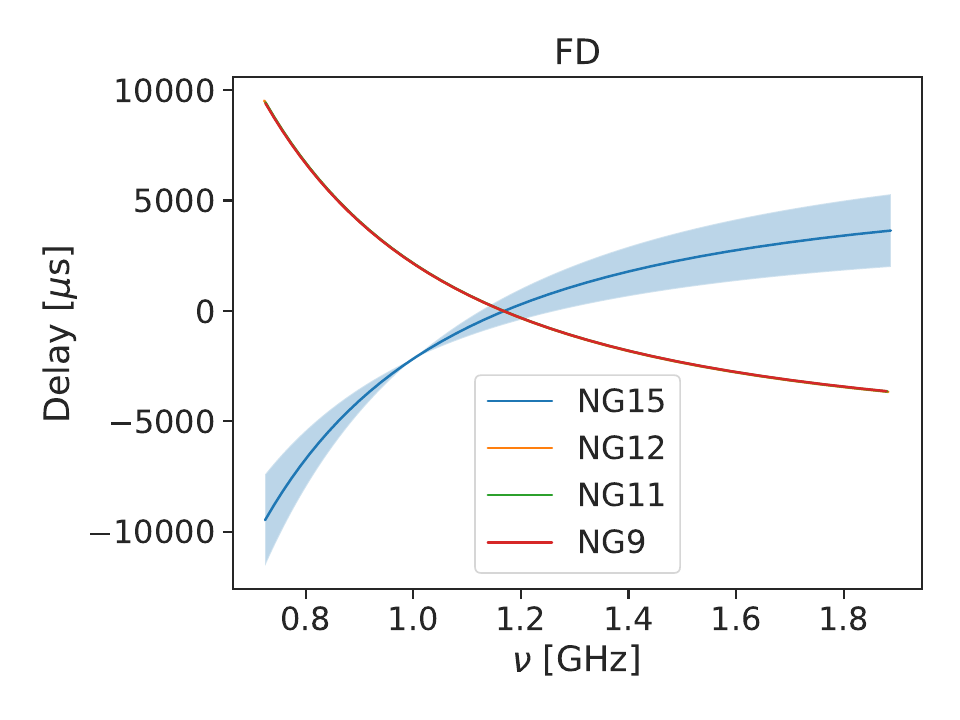}
    \end{subfigure}
    \hfill
    \begin{subfigure}{0.3\linewidth}
        \includegraphics[width=\linewidth]{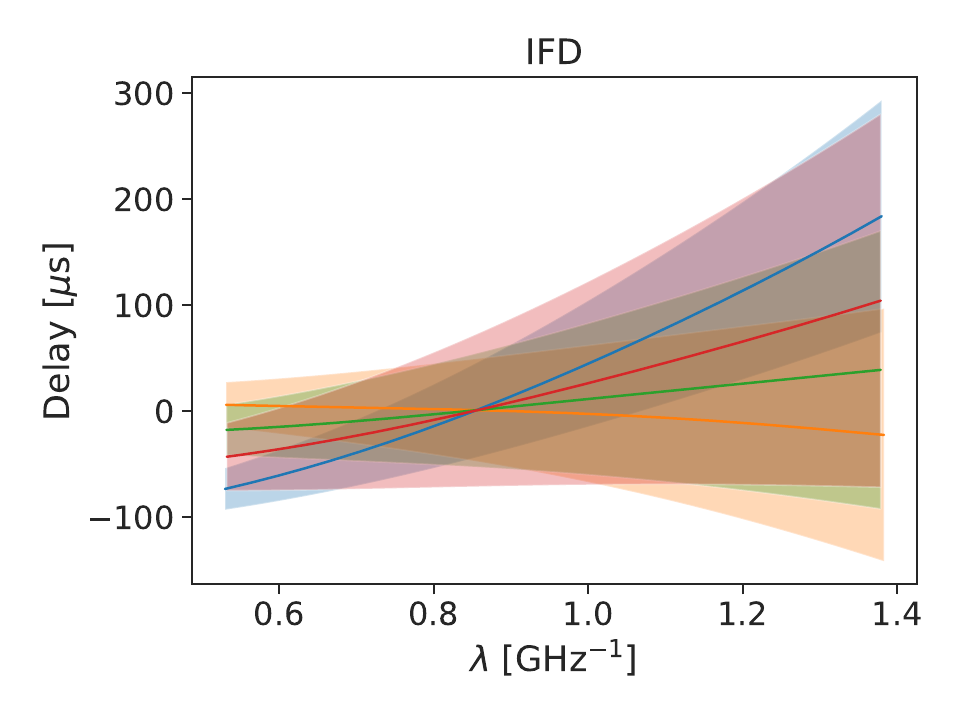}
    \end{subfigure}
    \hfill
    \begin{subfigure}{0.3\linewidth}
        \includegraphics[width=\linewidth]{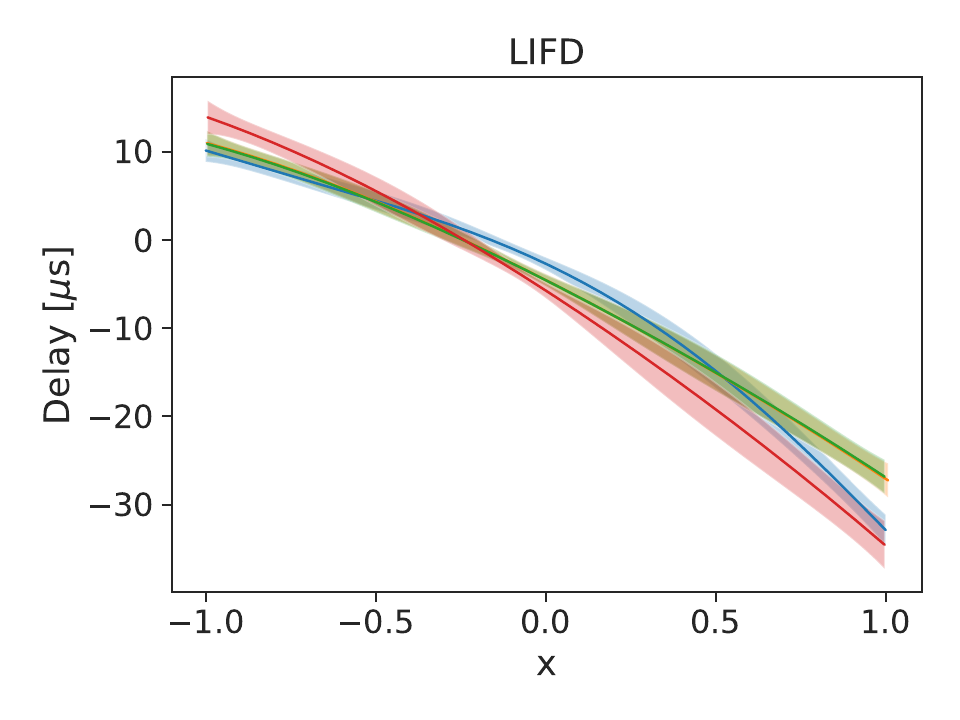}
    \end{subfigure}
    \vspace{-2mm}
    \subcaption{J1024$-$0719}

\end{subfigure}

\begin{subfigure}{\linewidth}
    \centering
    \begin{subfigure}{0.3\linewidth}
        \includegraphics[width=\linewidth]{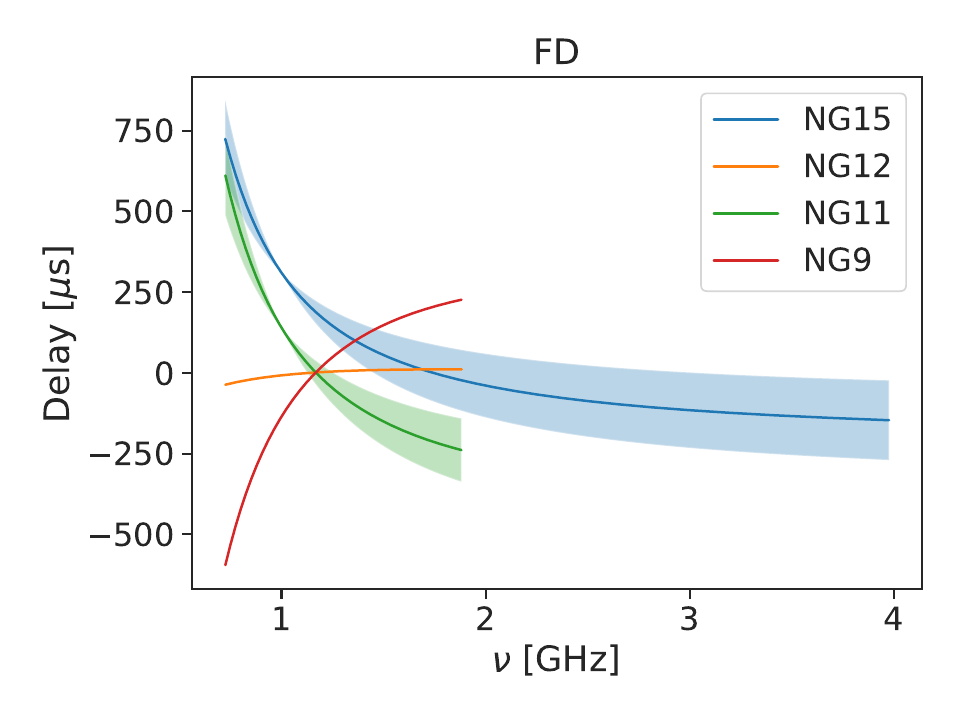}
    \end{subfigure}
    \hfill
    \begin{subfigure}{0.3\linewidth}
        \includegraphics[width=\linewidth]{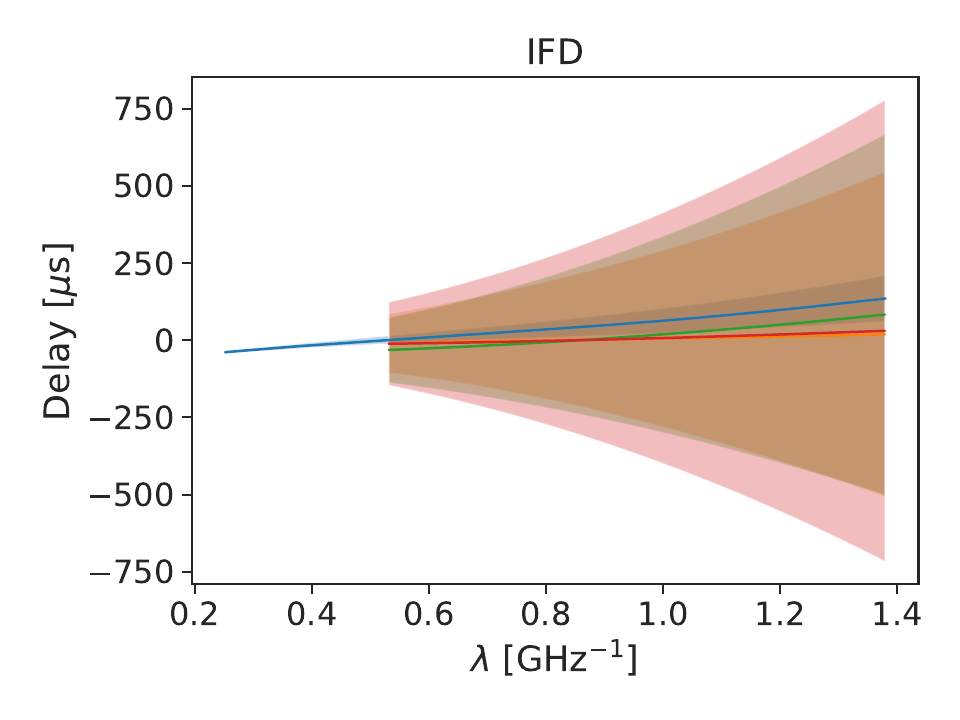}
    \end{subfigure}
    \hfill
    \begin{subfigure}{0.3\linewidth}
        \includegraphics[width=\linewidth]{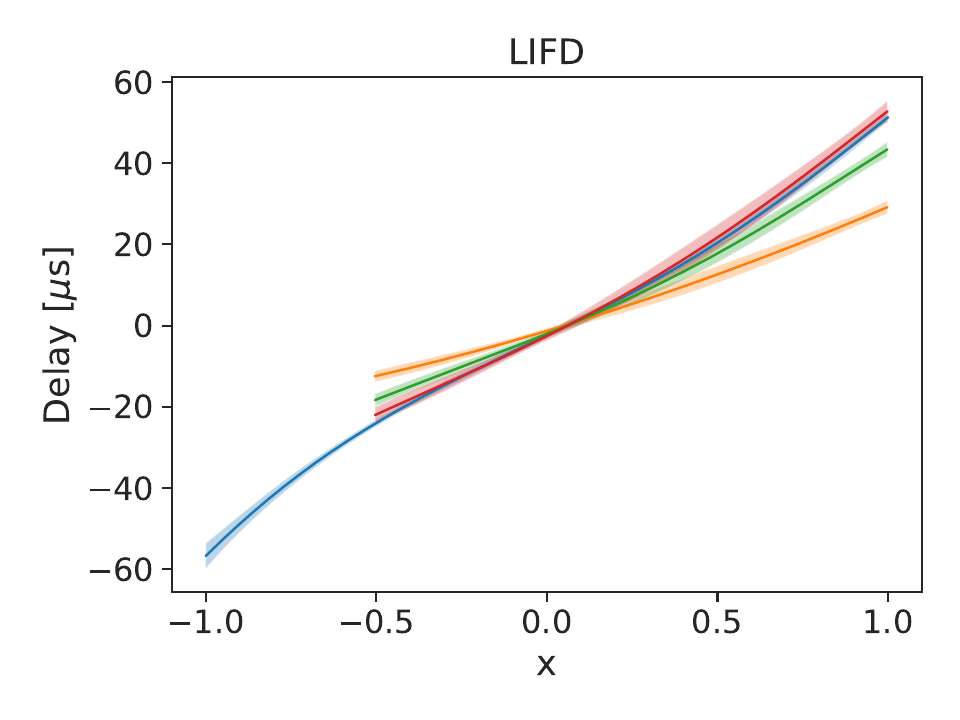}
    \end{subfigure}
    \vspace{-2mm}
    \subcaption{J1643$-$1224}
\end{subfigure}

\begin{subfigure}{\linewidth}
    \centering
    \begin{subfigure}{0.3\linewidth}
        \includegraphics[width=\linewidth]{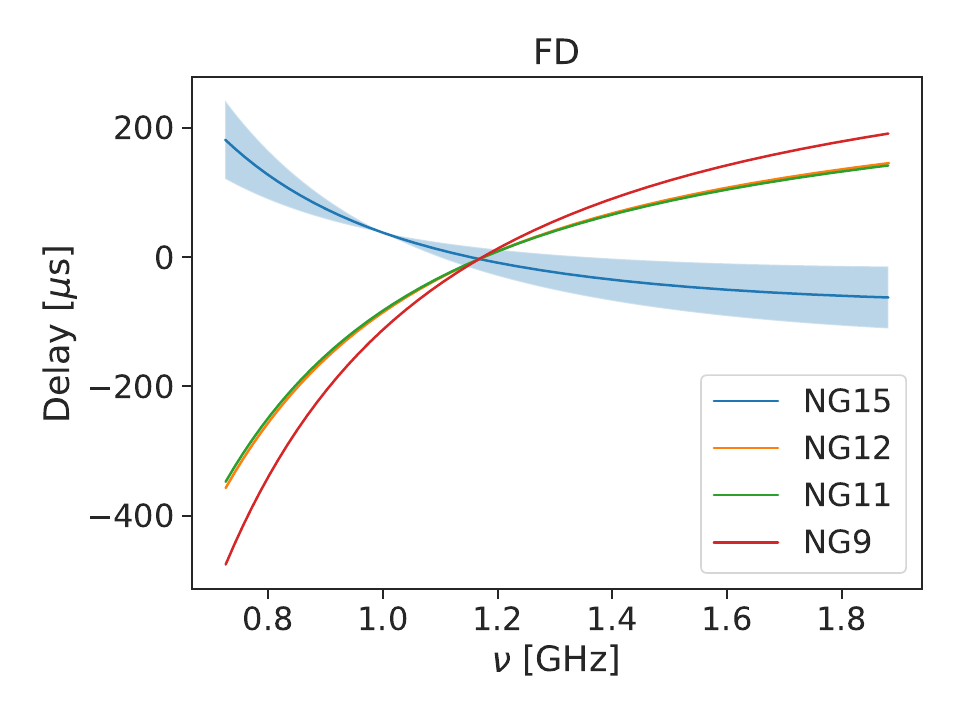}
    \end{subfigure}
    \hfill
    \begin{subfigure}{0.3\linewidth}
        \includegraphics[width=\linewidth]{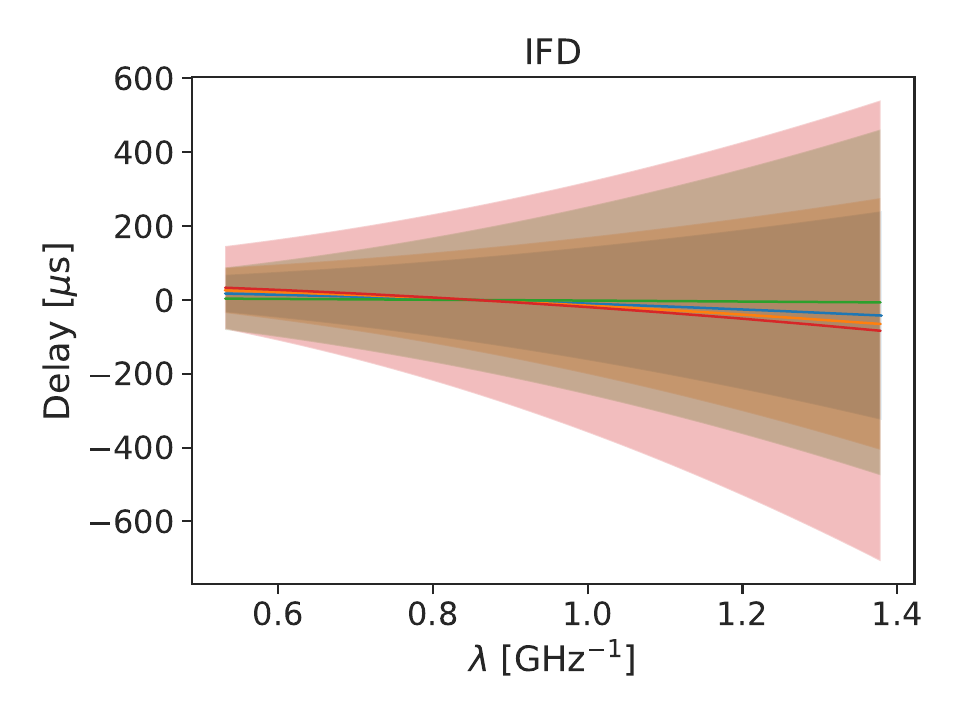}
    \end{subfigure}
    \hfill
    \begin{subfigure}{0.3\linewidth}
        \includegraphics[width=\linewidth]{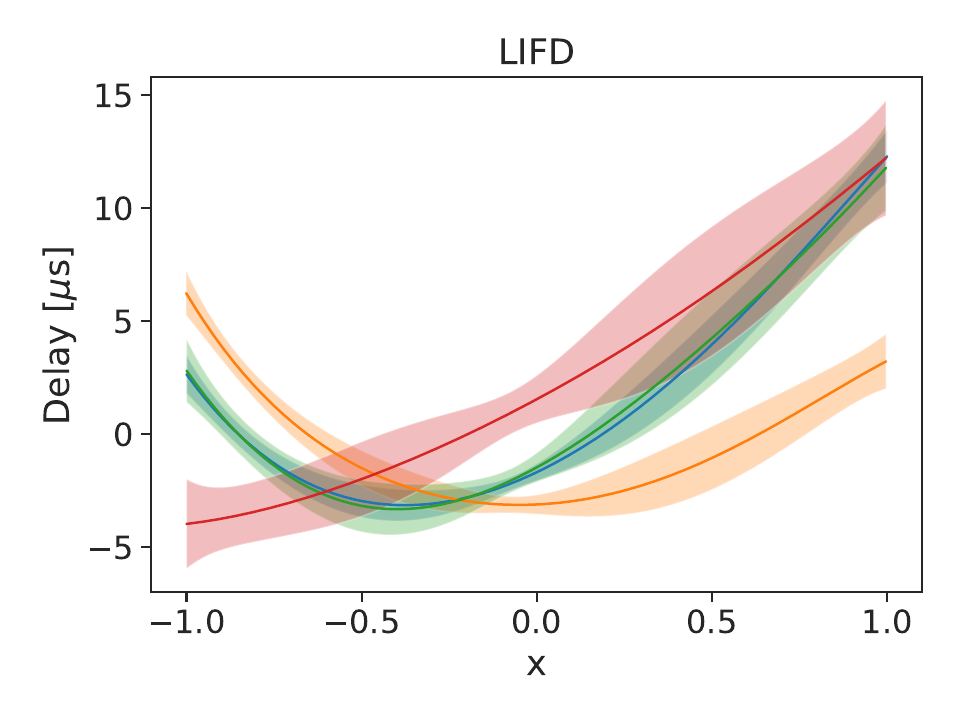}
    \end{subfigure}
    \vspace{-2mm}
    \subcaption{J2145$-$0750}
\end{subfigure}

\begin{subfigure}{\linewidth}
    \centering
    \begin{subfigure}{0.3\linewidth}
        \includegraphics[width=\linewidth]{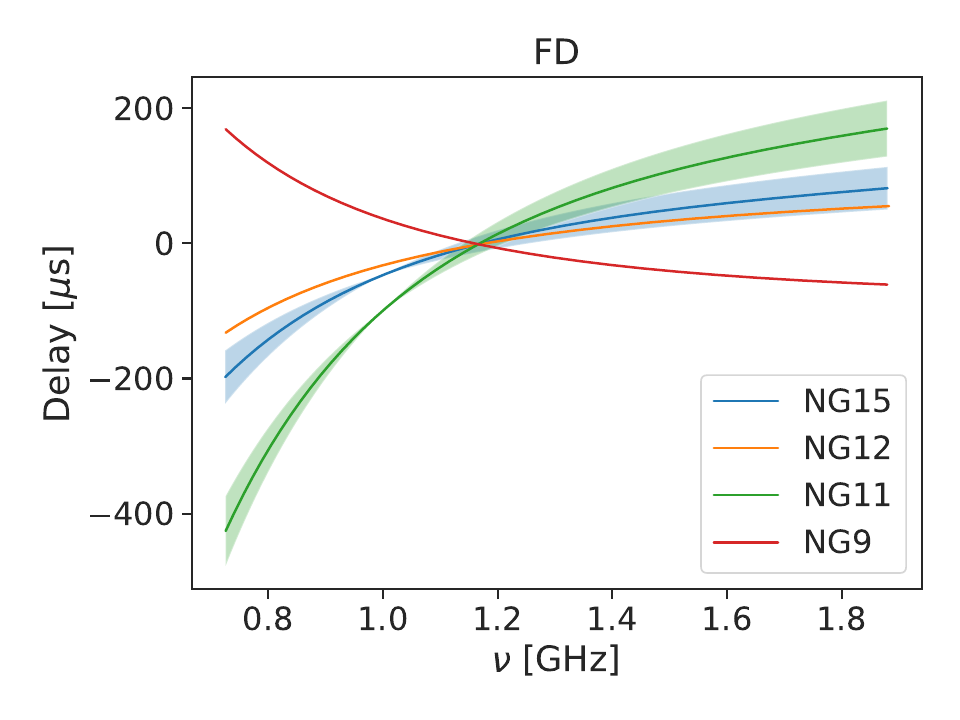}
    \end{subfigure}
    \hfill
    \begin{subfigure}{0.3\linewidth}
        \includegraphics[width=\linewidth]{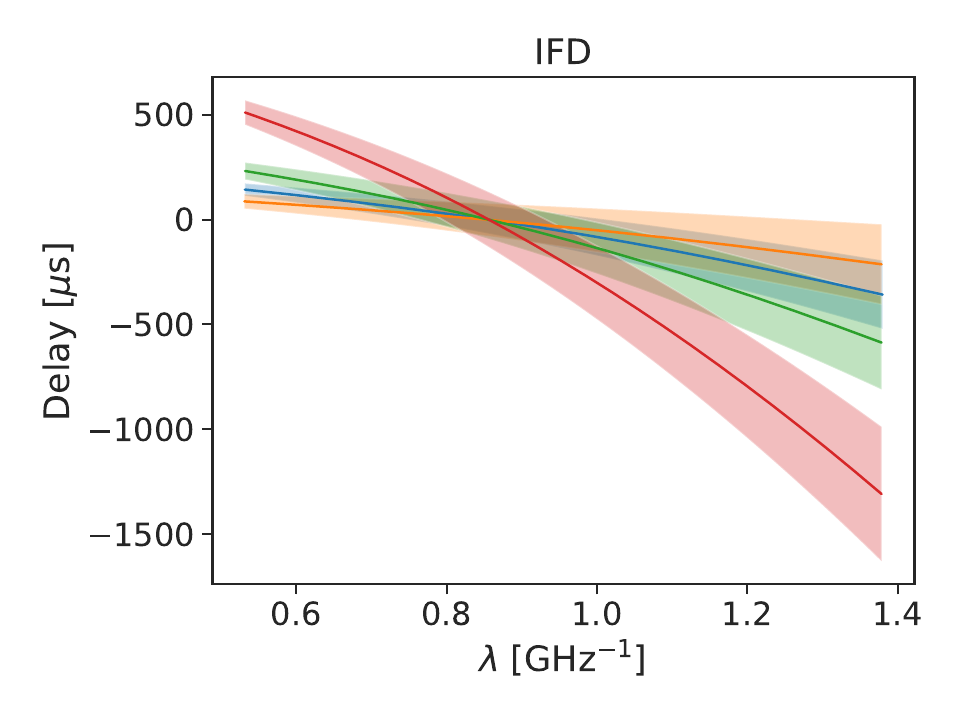}
    \end{subfigure}
    \hfill
    \begin{subfigure}{0.3\linewidth}
        \includegraphics[width=\linewidth]{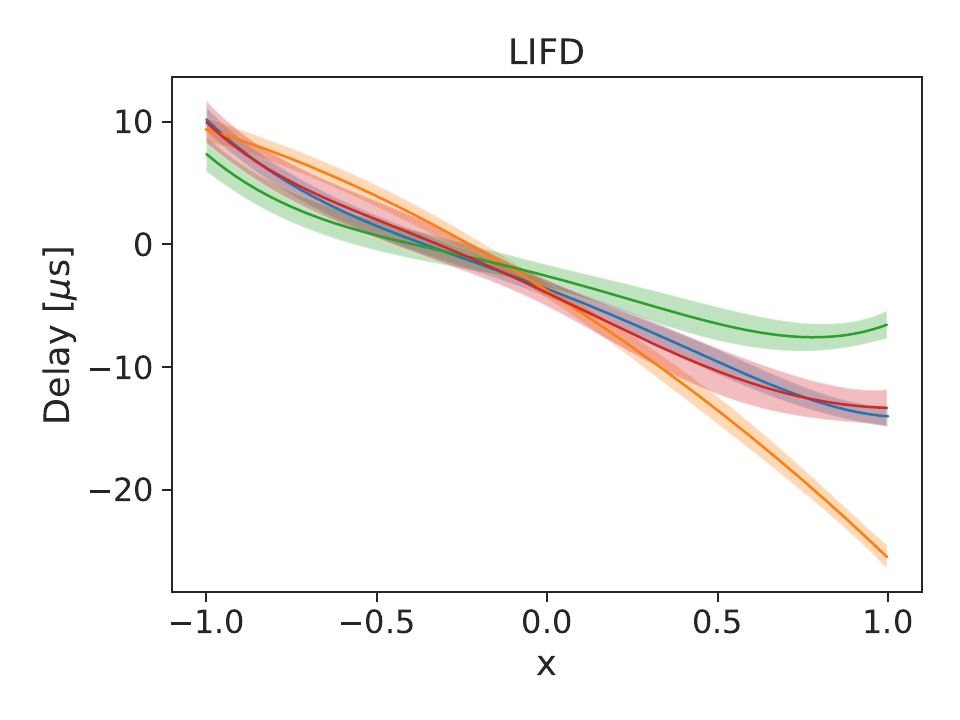}
    \end{subfigure}
    \vspace{-2mm}
    \subcaption{J2302+4442}
\end{subfigure}

\caption{Comparison of FD, IFD, and LIFD delay curves across NANOGrav datasets. In each subplot, we present the modeled timing delay induced by PE when using an FD (Eq.~\ref{eq:old_FD_formulation}), IFD (Eq.~\ref{eq:IFD_polynomial}), or LIFD (Eq.~\ref{eq:legendre_basis_x}) model. For each model, its polynomial coefficients were fitted using the pulsar's TOAs from the NG15 (blue), NG12 (orange), NG11 (green), or NG9 (red) dataset. The 1-$\sigma$ confidence intervals (shaded areas) were obtained using the full parameter covariance matrix after sampling 1000 realizations of the coefficients from a normal distribution and computing the spread of resulting curves. We see that FD's delay curves flip concavity across datasets, IFD produce large errors at larger wavelengths, and they both predict unrealistic PE-induced timing delays, but LIFD largely mitigates these shortcomings.}
\label{fig:curves_across_datasets}
\end{figure*}

\subsection{Stability Across Datasets}

As explained in Sec.~\ref{subsec:FD_mis_estimations}, one of the main indicators of mis-estimations in the FD parameters is that the corresponding PE delay curves show unrealistically large amplitudes and also flip concavity depending on the NANOGrav dataset used to fit them (see Fig.~\ref{fig:FD_curves}). Robust estimations of pulse profile evolution should give delay curves that are not only realistic in amplitude but also relatively stable across datasets for a given pulsar.

To test the stability of the IFD- and LIFD-derived PE delay curves across datasets, we repeated the timing fits from Sec.~\ref{subsec:evolution_across_models} but using the NANOGrav 9-, 11-, 12.5-, and 15-year datasets (from now on referred to as the NG9, NG11, NG12, and NG15 datasets, respectively). We present the results for the four MSPs in Fig.~\ref{fig:curves_across_datasets}.


The FD curves predict unrealistically large delay amplitudes and flip concavity across datasets. For PSRs J1643$-$0719 and J2302+4442, the NG11/15 curves, which use 5 FD parameters, show significantly larger errors and different concavity than the NG9/12 curves, which use 2 FD parameters. A 5-parameter polynomial can fit a more complex PE behavior, so it may capture real chromatic structure that the 2-parameter fits could not model, explaining the concavity flip. However, with 5 parameters, the higher-order terms become partially degenerate, inflating individual parameter uncertainties.
    
The IFD curves have smaller delay amplitudes, do not change dramatically in shape, and are largely consistent within each other's uncertainties. The error bars also diminish towards later datasets and, consequently, larger observing baselines. Their drawback is large error bars, especially at larger wavelengths. This indicates that the IFD parameters are non-orthogonal, leading to large, poorly constrained fitted coefficients.    
    
The LIFD curves have stable concavities and smaller delay amplitudes than the IFD models. Moreover, the errors are significantly better-constrained across frequency than in the IFD model for the same pulsars and datasets, even though both use order-6 polynomials. This can be attributed to the base orthogonality, which means that the fitter can adjust each coefficient separately, keeping individual uncertainties small.


\subsection{Performance on Simulated Data}\label{subsec:simulated_data}

In real pulsar data, the exact contributions from dispersion and PE to the total chromatic delay can never be fully disentangled a posteriori due to their degeneracy in the $\nu^{-2}$-direction of vector space. To reliably quantify the ability of different models to separate the two effects, we created simulated data where the contributions of dispersion and PE can be set a priori.

\begin{figure}[ht]
\centering
    \includegraphics[width=\linewidth]{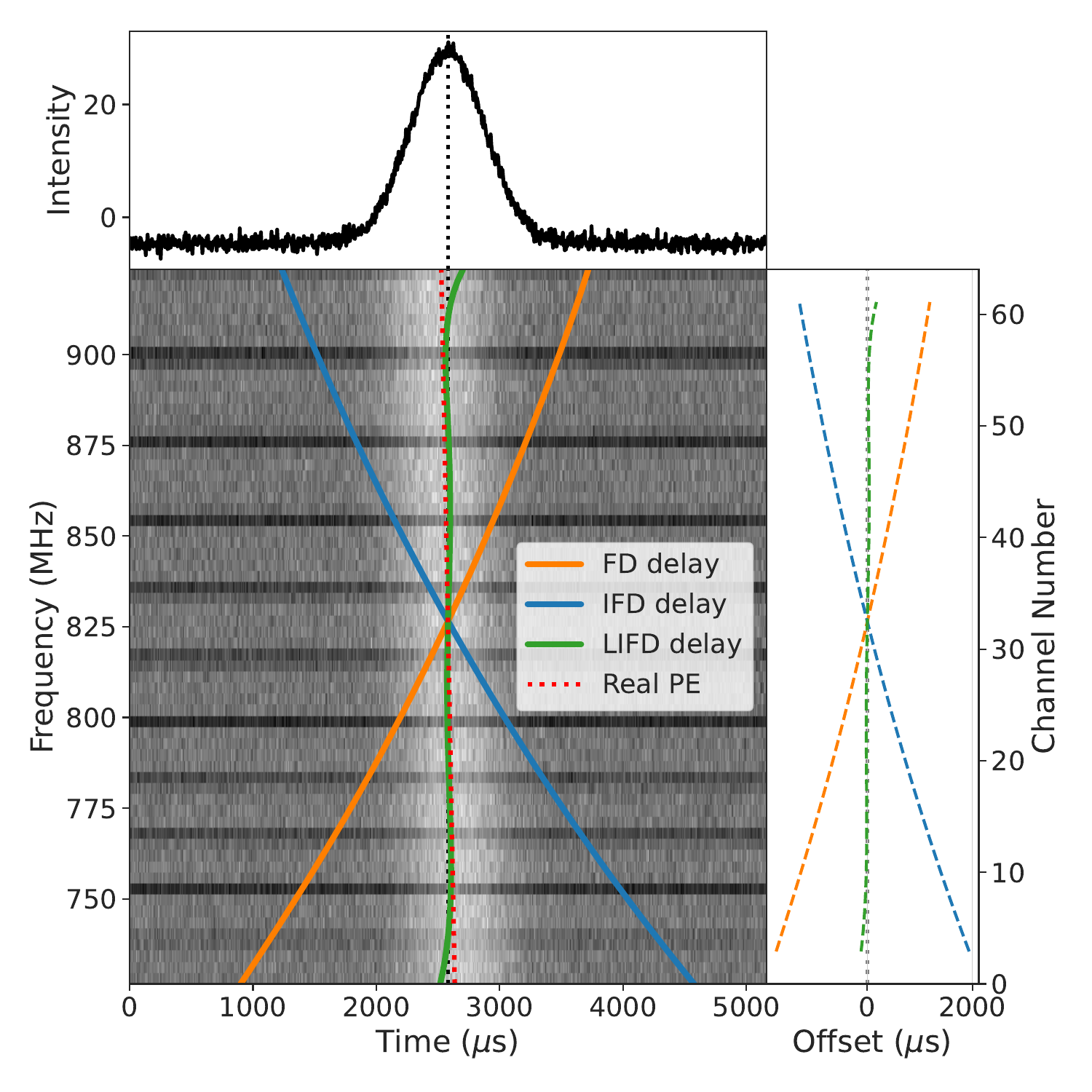}
\caption{Bottom left: intensity of the (de-dispersed) pulse profile as a function of frequency and rotational phase from the highest-S/N simulated observation created by injecting both PE and DM delays. The color curves show the PE delay fitted with an FD, IFD, or LIFD model, and the dotted red line shows the (real) injected PE delay. Top: time- and frequency-integrated profile. Right: offset between the PE delay curve fitted by each model and the injected PE delay curve, as a function of frequency. Of the three models, LIFD results in the delay curve that best follows the injected PE.}\label{fig:synt_waterfall}
\end{figure}

The simulated data consist of time-, frequency-, and phase-resolved observations of a synthetic pulsar with a period of $5.16~\mathrm{ms}$ to mimic that of PSR J1024-0719. We controlled the number, widths, positions, and relative amplitudes of the pulse components to emulate PE effects. In particular, we used a unimodal Gaussian profile with a PE sweep of $110~\mathrm{\mu s}$ (see Fig.~\ref{fig:synt_waterfall}). Moreover, we applied a $\nu^{-2}$-chromatic timing delay to each frequency channel to emulate a slow-varying DM time series in the range $18.412 \pm 0.0005~\mathrm{pc/cm^{3}}$ (see red curve Fig.~\ref{subfig:synthethic_psr_2D_template_comparison}). We also injected white noise with a median of $2.8~\mathrm{\mu s}$. Using these parameters, we simulated 477 observations spanning 10.38 years with weekly cadence; each observation has 32 evenly spaced frequency channels between 728.1 and 921.9 MHz, resulting in 15264 unique TOAs.

\begin{figure*}
\begin{subfigure}{\linewidth}
    \centering
    \begin{subfigure}{0.325\linewidth}
        \includegraphics[width=\linewidth]{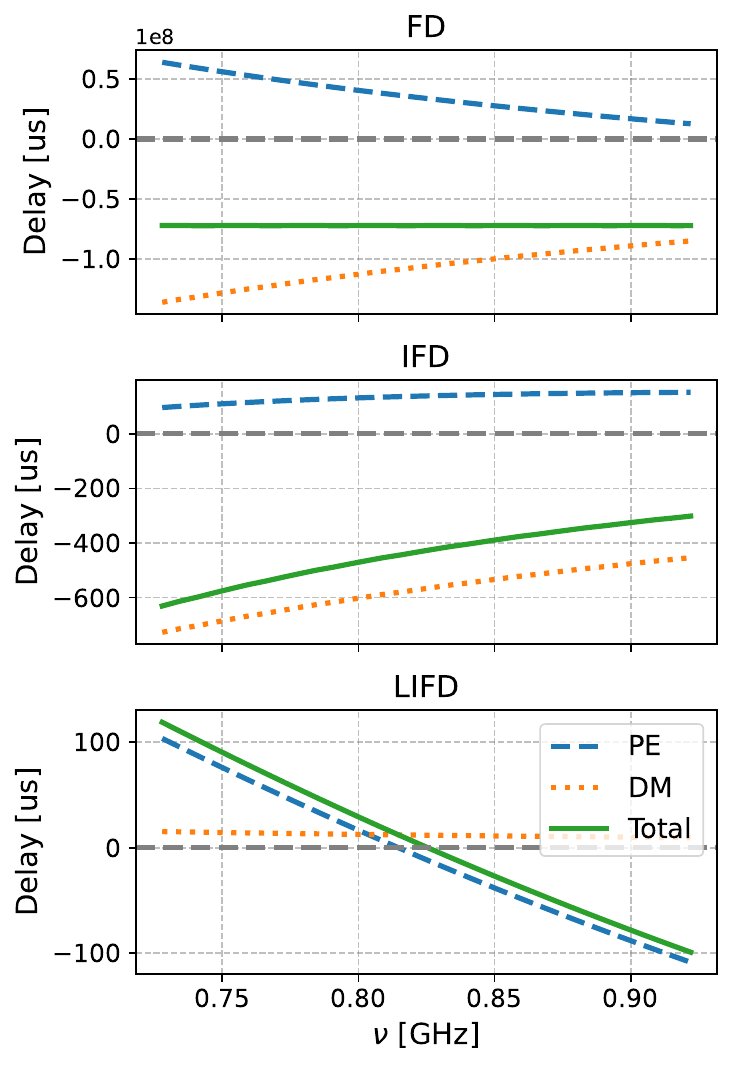}
        \subcaption{Profile evolution and dispersion.}\label{subfig:simulations_chromatic_delays_breakdown_a}
    \end{subfigure}
    \hfill
    \begin{subfigure}{0.325\linewidth}
        \includegraphics[width=\linewidth]{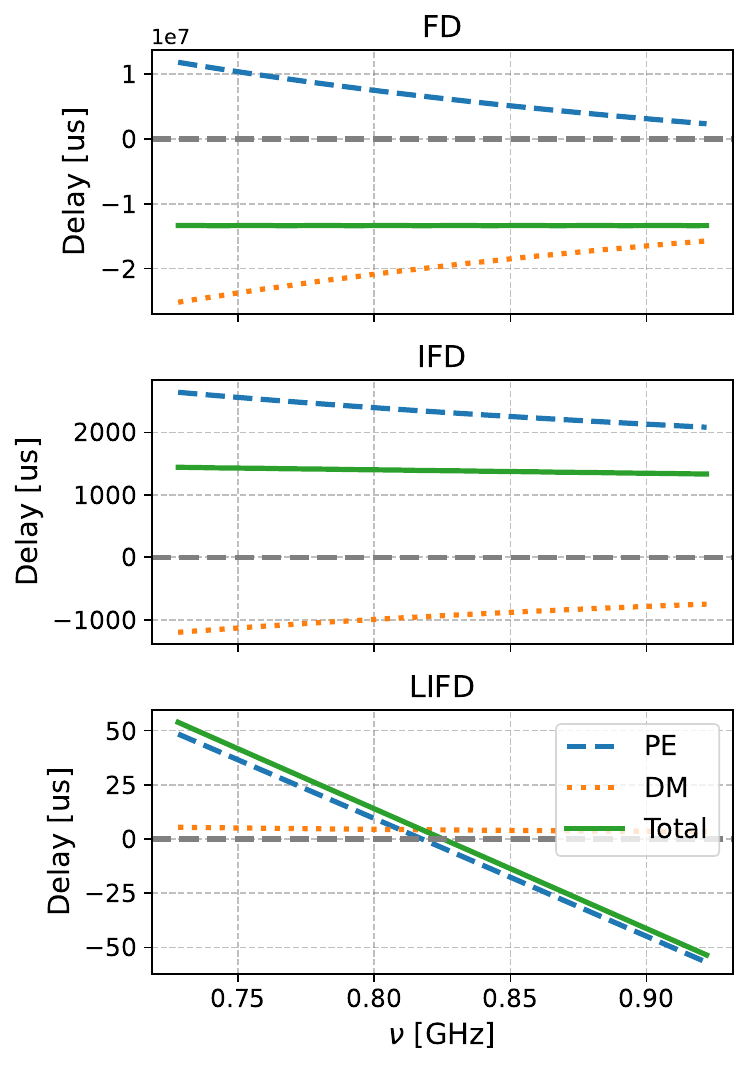}
        \subcaption{Profile evolution and no dispersion.}\label{subfig:simulations_chromatic_delays_breakdown_b}
    \end{subfigure}
    \hfill
    \begin{subfigure}{0.325\linewidth}
        \includegraphics[width=\linewidth]{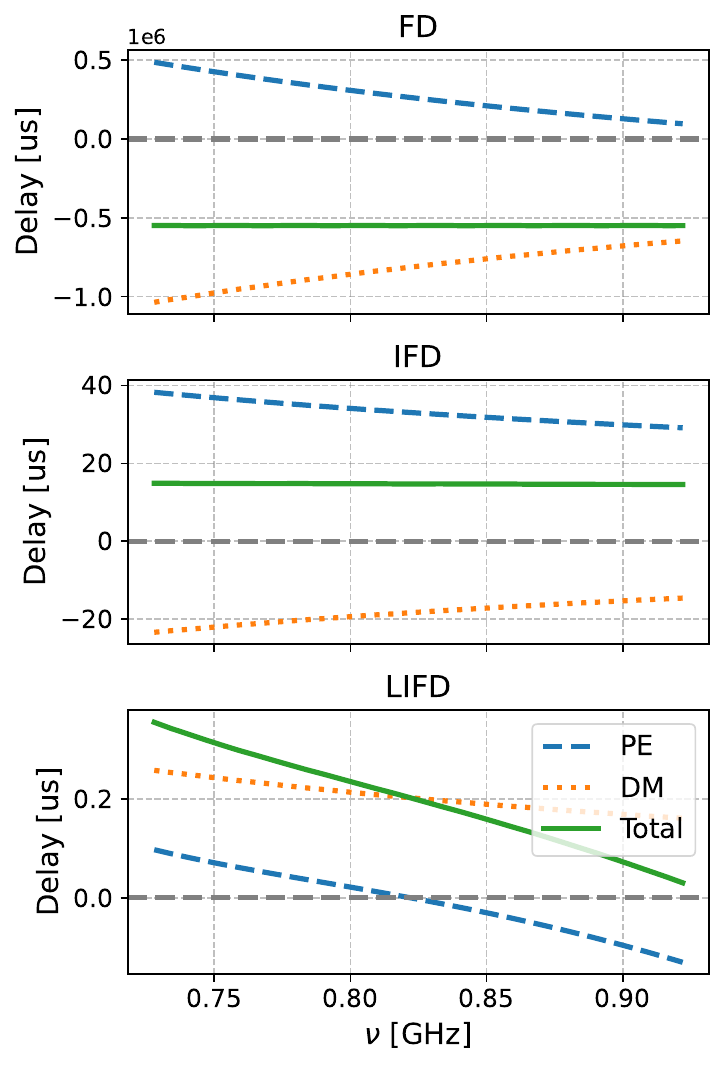}
        \subcaption{Dispersion and no profile evolution.}\label{subfig:simulations_chromatic_delays_breakdown_c}
    \end{subfigure}    
\end{subfigure}

\caption{Timing delay as a function of frequency predicted for PE (blue), dispersion (orange), or both (green) when using a timing model with either an FD, IFD, or LIFD model to fit simulated TOAs that were created by injecting both PE and DM (first column), only PE (second), or only DM timing delays (third). Both FD and IFD models fit PE-induced and DM-induced delays even in datasets free of such effects, and only LIFD recovers PE and DM delays consistent with the injected values.}
\label{fig:simulations_chromatic_delays_breakdown}
\end{figure*}

We simulated three datasets, incorporating different combinations of chromatic delays: 1) PE and dispersion, 2) PE and no dispersion, and 3) dispersion and no PE. For each dataset, we created frequency-resolved TOAs and fit them using a timing model with either an FD, IFD, or LIFD term. We plotted the timing delay predicted by the resulting PE and DM models as a function of observing frequency in Fig.~\ref{fig:simulations_chromatic_delays_breakdown}. We see that:
\begin{itemize}
    \item When we injected both PE and dispersive delays in the simulated TOAs (Fig.~\ref{subfig:simulations_chromatic_delays_breakdown_a}), FD and IFD models fitted PE delay amplitudes significantly larger than the injected PE sweep, compensated by large, oppositely-signed DM delay. Only the LIFD model resulted in realistic delay amplitudes.
    \item For the dataset with no injected dispersion (Fig.~\ref{subfig:simulations_chromatic_delays_breakdown_b}), the DM delay curve should be consistent with zero. However, FD and IFD still produced significant DM-induced timing delays. LIFD resulted in a significantly smaller dispersive delay than FD or IFD models, and in a PE delay whose amplitude, $110~\mathrm{\mu s}$, is in remarkable agreement with our injected PE delay sweep.
    \item When we simulated pulse profiles with no PE (Fig.~\ref{subfig:simulations_chromatic_delays_breakdown_c}), both FD and IFD models still produced a timing fit with a non-negligible PE term. Only the LIFD model fitted a PE delay that, while not strictly zero, has zero mean and small amplitudes.
\end{itemize}

\begin{figure*}[ht]
\centering
\begin{subfigure}{0.49\linewidth}
    \centering
    \includegraphics[width=0.8\linewidth]{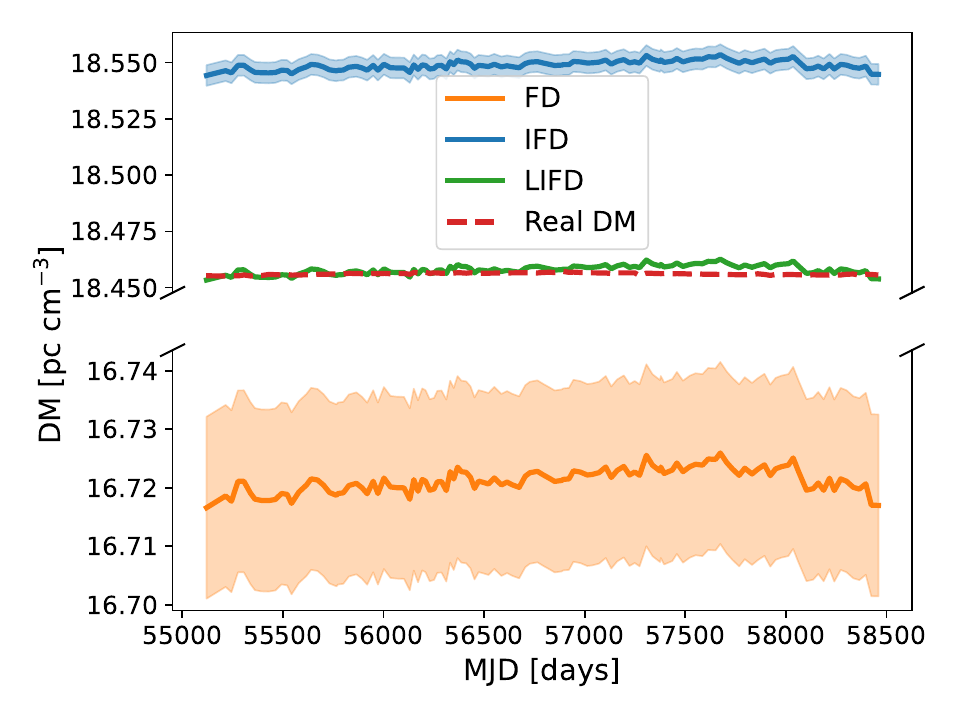}
    \caption{Profile evolution and dispersion}
    \label{subfig:synt_PE_DM_DM_timeseries}
\end{subfigure}
\begin{subfigure}{0.49\linewidth}
    \centering
    \includegraphics[width=0.8\linewidth]{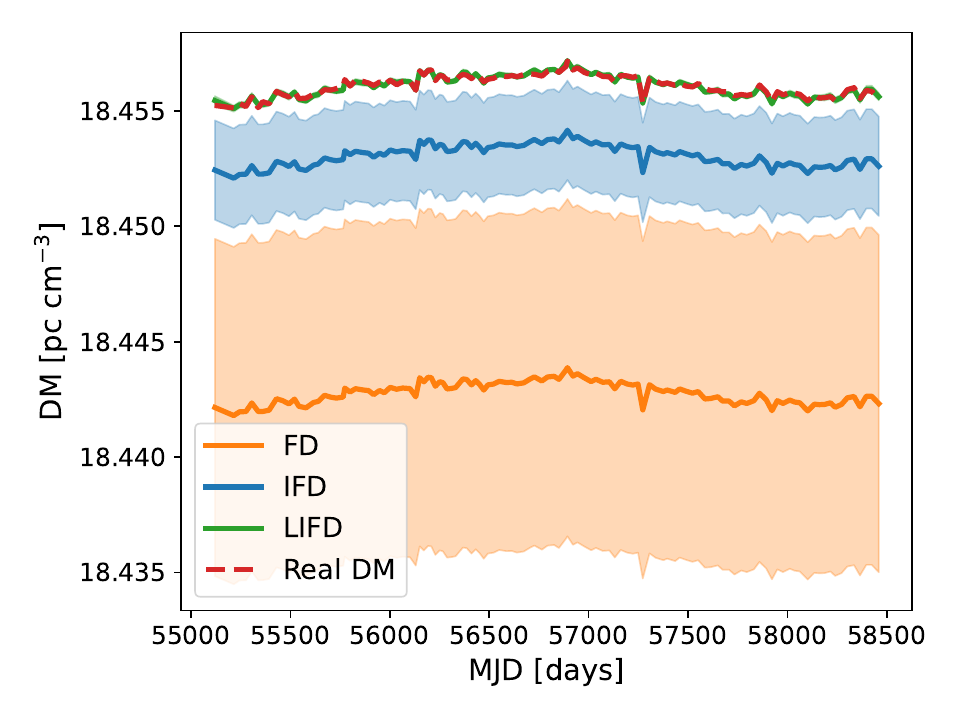}
    \caption{Dispersion and no profile evolution.}
    \label{subfig:synt_noPE_DM_DM_timeseries}
\end{subfigure}
\caption{DM time series recovered using simulated TOAs that were created by injecting both PE and DM timing delays (left) or only DM delays (right). For each dataset, we fitted a DM time series using a timing model that accounts for PE using either an FD (orange), IFD (blue), or LIFD (green) model. The real (injected) DM time series is presented in red. All timing models recover similar temporal DM trends, but only LIFD recovers a DM time series consistent with the injected values.}\label{fig:synt_DM_timeseries}
\end{figure*}

For the dataset with PE and dispersion delays, Fig.~\ref{fig:synt_waterfall} shows the PE predicted by each model, overlaid on the actual frequency-resolved pulse profiles, and the offset of each model's delay curve with respect to the injected PE. We find that the FD model overestimates the injected PE evolution by as much as $1191~\mathrm{\mu s}$. By contrast, the LIFD model resulted in a fitted PE delay curve that, in the worst case, overestimates the injected PE by $174.5~\mathrm{\mu s}$, thus reducing the PE error by $85\%$.

In Fig.~\ref{fig:synt_DM_timeseries}, we present the DM time series recovered from the two datasets with injected dispersion. We see that all FD, IFD, and LIFD models result in similar temporal trends in the fitted $\mathrm{DM}(t)$ (i.e., their peaks and valleys line up). This is reasonable, since these PE models are functions of frequency only and do not include time-dependent corrections; therefore, any time-dependent structure in the injected signal must be absorbed by the $\mathrm{DM}(t)$ model, which is the same in all three cases. This means that for capturing relative DM variations (the component of noise modeling relevant to GWB detection) all PE models are equivalent. However, when it comes to the absolute fitted DM values, we find significant offsets between the DM time series: FD models have a systematic offset from the injected DM time series (mean error of $0.013~\mathrm{pc~cm^{-3}}$), while LIFD models provide a significantly better approximation (mean error of $0.0002~\mathrm{pc~cm^{-3}}$) and smaller errors in the DMX parameters. This offset can be attributed to the degeneracy between dispersion and PE: overestimated PE delays are degenerate with a shift in DM's mean level, resulting in a vertical shift in the recovered DM time series. This effect would affect studies that require absolute DM values, such as comparing electron-density models and using DM for distance estimates.

\subsection{Comparison with Discrete 2D Templates}\label{subsec:2D_templates}

An alternative to accounting for PE in the timing model is to remove this effect a priori during TOA-generation by using frequency-resolved templates. We will thus use this approach as a benchmark test for our PE models. In that respect, wideband timing employs template portraits, where the pulse shape is a continuous function of frequency and phase, and all frequencies are fitted simultaneously to produce a single wideband TOA (see Sec.~\ref{subsec:prof_evolution_theory}). However, since our observations are in the narrowband regime, we used discrete 2D templates, where the pulse shape varies with frequency but only in discrete steps, with one pulse profile shape per frequency sub-band. In this approach, each frequency channel in the frequency-resolved observation is cross-correlated with the corresponding frequency slice of the discrete 2D template, resulting in as many TOAs per observation as frequency channels in it.

Starting from a dataset of time- and frequency-resolved observations, we generate a 2D template by: 
\begin{enumerate}
    \item Selecting a subset of the highest-S/N observations.
    \item Aligning them in phase (e.g., using \texttt{pam}). 
    \item Appending the observations in time (using \texttt{psradd}) and then time-averaging them to obtain a high-S/N frequency-resolved pulse profile.
    \item Frequency-averaging the resulting observation to the desired number of frequency channels, and 
    \item Applying a pulse-smoothing algorithm to the profile in each channel (using \texttt{psrsmooth}).
\end{enumerate}
\noindent Each observation is then cross-correlated against this 2D template using \texttt{pat}. For a complete description of the 2D TOA generation process, see \cite{Melfor2026}.

\begin{figure*}[ht]
\centering
\begin{subfigure}{0.49\linewidth}
    \centering
    \includegraphics[width=0.8\linewidth]{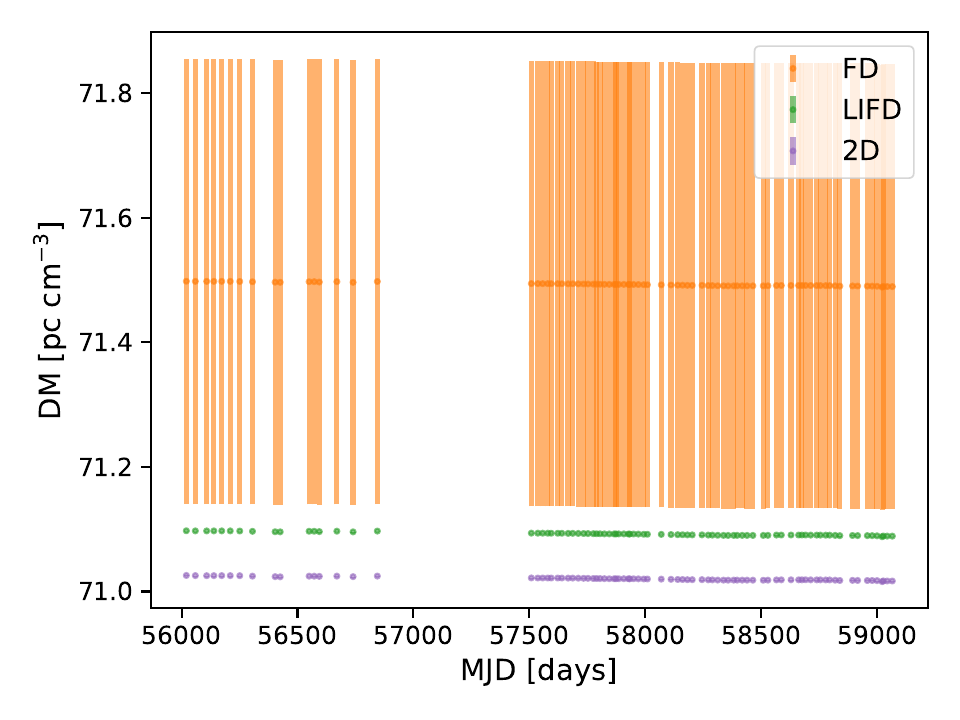}
    \caption{B1937+21}
    \label{subfig:B1937+21_2D_template_comparison}
\end{subfigure}
\begin{subfigure}{0.49\linewidth}
    \centering
    \includegraphics[width=0.8\linewidth]{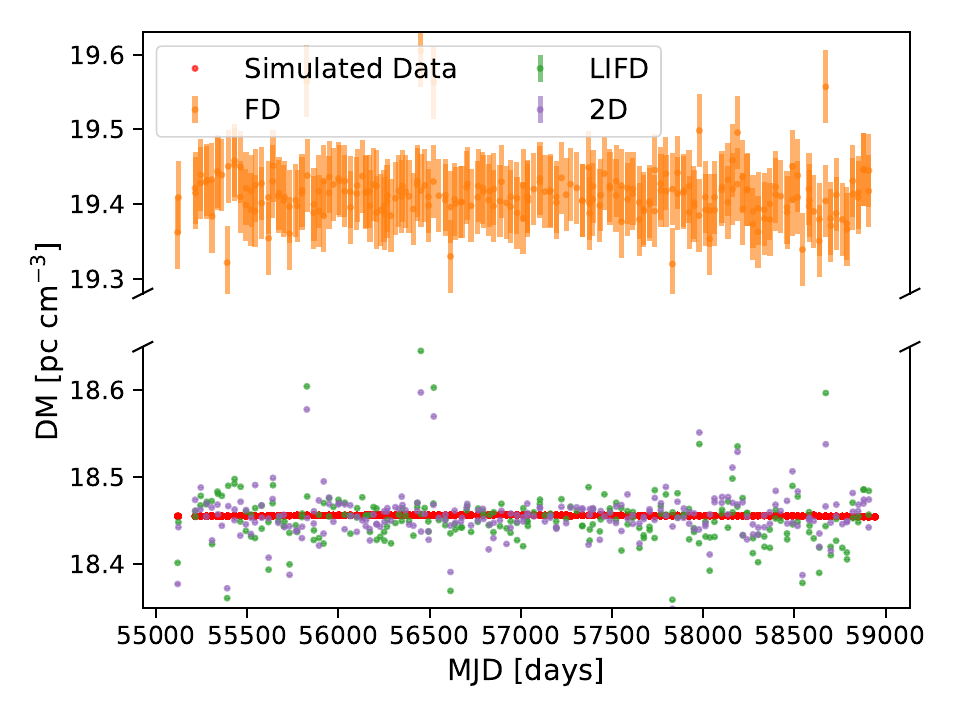}
    \caption{Simulated observations}
    \label{subfig:synthethic_psr_2D_template_comparison}
\end{subfigure}
\caption{DM time series recovered using observations of PSR B1937+21 (left panel) or simulated observations (right panel). For each dataset, we fitted a DM time series using: 1) TOAs generated using a discrete 2D template and a timing model with no corrections for PE (blue dots), or TOAs generated using a 1D template and a timing model with either 2) an FD (green dots) or 3) an LIFD (orange dots). For the simulated data, we also show the real (injected) DM time series (red dots). LIFD models and 2D templates result in recovered DM time series with the smallest error bars and that are consistent with the injected DM.}\label{fig:2D_template_comparison}
\end{figure*}

We then compared the DM time series recovered from performing a timing fit using either: 1) TOAs created using a 1D template and a timing model that includes an FD or LIFD term (a posteriori fitting), or 2) TOAs generated using a 2D template and a timing model that does not include a PE term (since that effect was already accounted for during the TOA generation). We perform this comparison using: a) real NG15 observations of PSR B1937+21 for its large PE, and b) simulated observations that include pulse PE and dispersion. We present the recovered DM time series in Fig.~\ref{fig:2D_template_comparison}.

We observe that, for both real and simulated data, the DM time series recovered using 2D templates and LIFD models have significantly smaller errors than the DM time series obtained using FD models. Most notably, for the simulated observations, where we can control the injected dispersion, both the LIFD and the 2D template methods recover significantly better approximations to the real (injected) DM time series than the FD model.

\section{Conclusions}\label{sec:conclusions}

\subsection{Discussion of Results}

In this work, we tackled the mis-estimations in the FD models of pulse profile evolution with frequency, which lead to predictions of PE-induced timing delays that are not only inconsistent with the observed dispersion-corrected pulse profiles (see Fig.~\ref{fig:FD_evolution}), but also change significantly across NANOGrav datasets (see Fig.~\ref{fig:FD_curves}). We attributed this discrepancy to FD's log-polynomial coefficients being covariant with one another and with parameters describing other chromatic effects, most notably dispersion. These covariances produce an ill-conditioned design matrix for the timing fit and, consequently, statistically correlated and poorly constrained fitted parameters that can shift spectral power across chromatic effects and bias the fitted timing parameters.

To mitigate these covariances, we introduced two novel PE formulations, using monomial (IFD) and Legendre (LIFD) bases, and tested them on the NG15 TOA measurements of 11 MSPs and on simulated data. LIFD consistently predicted PE delay curves that more closely resemble the observed pulse profiles (see Fig.~\ref{fig:profile_evolution_curves}) and are more stable and better constrained across datasets (see Fig.~\ref{fig:curves_across_datasets}) than those obtained using FD models. The fitted IFD models generally predicted delay curves with amplitudes intermediate between FD and LIFD, but larger variances than the LIFD curves. 

We also tested these models on simulated data, where we can control the injected DM and PE. Once again, we found FD and IFD models are susceptible to misattributing spectral power and fitting significant dispersive/PE timing delays when none was injected. LIFD models provided the results most consistent with the injected chromatic delays, and their performance was comparable to that obtained using discrete frequency-resolved templates to account for PE a priori during TOA generation.

These results support the idea that the choice of the basis used to describe PE determines the ability of a timing model to separate and constrain the contributions of the different chromatic effects:

\begin{itemize}
    \item FD's log-polynomial basis elements are not only covariant with one another, but also do not align with the directions of power-law chromatic effects in vector space, including dispersion. Therefore, spectral power can inadvertently move across the dispersion and PE models, leading to FD parameters that overestimate PE and vary significantly with the input data.
    \item IFD models better approximate the geometry of the problem by using basis elements that align with the power-law vectors and, in turn, constrain power leakage from other chromatic effects to specific IFD terms. However, the fitted parameters are covariant with one another, and thus susceptible to numerical instability and uncertainty propagation, resulting in poorly constrained fitted values and larger variances.     
    \item LIFD models encode the PE delay into an orthogonal basis where the fitted polynomial coefficients are independent, and errors in one direction of vector space do not propagate into other basis directions. This approach constrains spectral power leakage while minimizing parameter covariances, leading to more realistic PE.
\end{itemize}

We also found that, over NANOGrav's observing bandwidths, the fitted total chromatic delay is independent of the PE model (see Fig.~\ref{fig:DMX_response}). This suggests that an accurate phenomenological description of the global chromatic behavior in timing observations, as required by high-precision timing experiments \citep{NG15_NB}, can be achieved even if the individual chromatic terms are biased. Because TOAs provide the sum of all chromatic terms at each observed frequency and epoch, and not how that sum is separated between dispersion and PE, this allows for degenerate solutions reproducing the same total chromatic delay. 


Accurate measurements of PE are extremely important for precision pulsar timing and for understanding
noise in the timing data. In particular, GWs are not radio-frequency dependent, so PE-induced TOA biases do not prevent potential GW detections, but they could induce additional red noise processes in the data \citep{2021ApJ...909..219S}. Moreover, advanced noise modeling techniques, such as those discussed in \cite{2020ApJ...905L..34A}, will benefit greatly from disentangling PE from other chromatic effects when precision down to $\sim 100~\mathrm{ns}$ is needed. 

\subsection{Limitations}\label{subsec:limitations}

Among the limitations of these methods, we recall that LIFD models do not completely isolate PE from other chromatic effects. In general, any algebraic basis used to describe PE will contain linear combinations of terms that scale as dispersion-like $\nu^{-2}$ and scattering-like $\nu^{-4}$. These components are therefore intrinsically covariant with propagation effects, and no choice of basis can fully disentangle them. Instead, the basis determines how this degeneracy spreads among the PE parameters and how covariant they are with each other. In the LIFD models, the projection of an excess dispersive-$\nu^{-2}$ signal is confined to the $\mathrm{LIFD}_0$, $\mathrm{LIFD}_1$, and $\mathrm{LIFD}_2$ terms, and all other fitted parameters are decoupled from dispersion (see Sec.~\ref{sec:advantages_Legendre}). This is already an advantage over FD models, where all $\mathrm{FD}_k$ parameters are covariant with dispersion, leading to a more numerically unstable representation of the same underlying degeneracy.

Another limitation is that we only analyzed the degeneracy of PE with dispersion because it is often the most prominent propagation effect in pulsar observations. Other chromatic effects, such as scattering, refraction, and scintillation, act as noise processes and are not represented by deterministic parameters in NANOGrav's timing models. However, they will still introduce covariances at different levels in each of our PE formulations. For example, assuming that scattering delays scale as $\nu^{-4}$, they will be covariant with the $\mathrm{IFD}_4$ parameter and with all $\mathrm{LIFD}_\mathrm{k}$ parameters of order $\mathrm{k} \leq 4$.

Building on this point, we also assumed the weak-scattering regime, under which refraction-induced geometric delays scale as $t_\mathrm{refrac} \propto \nu^{-4}$, and a Gaussian medium, under which scattering delays scale as $\tau_\mathrm{scat} \propto \nu^{-4}$. However, most MSPs showing diffractive scintillation are in the strong scattering regime \citep[e.g.,][]{Cordes_2016}, where $t_\mathrm{refrac} \propto \nu^{-2\beta+4}$, with $\beta$ the spectral index of the electron-density power spectrum \citep[][]{2017MNRAS.464.2075S}. For a Kolmogorov medium ($\beta=11/3$), this becomes $t_\mathrm{refrac} \propto\nu^{-10/3} \approx \nu^{-3.33}$ (shallower than $\nu^{-4}$) and $\tau_\mathrm{scat} \propto \nu^{-4.4}$ (steeper than $\nu^{-4}$). We then expect this leakage, which was confined to the $\mathrm{IFD}_4 \lambda^4$ term, to become degenerate with PE across several IFD terms, hindering the IFD model's ability to isolate chromatic effects. Nonetheless, LIFD models, which use an orthogonal, well-conditioned polynomial basis, should still have lower coefficient covariance and better mitigate the effects of mis-specified exponent leakage than FD or IFD models.

Finally, we treated the LIFD polynomial basis as orthogonal and its coefficients as uncorrelated. Strictly speaking, however, Legendre polynomials are only orthogonal under the inner product defined by Eq.~\ref{eq:orthogonal} with a flat weighting function $w(x)=1$. In pulsar timing, TOAs at different frequencies are included in the fit with weight proportional to their inverse variance, i.e., $w(\lambda) \propto 1/\sigma^2(\lambda)$. For typical TOA uncertainties, where $\sigma(\lambda)$ varies only weakly with frequency, this weighting remains close to uniform, and the LIFD basis is approximately orthogonal in practice. Some covariance between LIFD parameters is still expected, but it remains significantly smaller than in non-orthogonal bases.

\subsection{Future Work}

With the advent of wideband radio receivers, such as the ultra-wideband receiver (UWBR) at the Green Bank Telescope \citep{2026AAS...24741606R} and the ultra-wideband low (UWL) receiver at the Parkes radio telescope \citep{2020PASA...37...12H}, accounting for PE across broad frequency ranges has become increasingly important \citep[e.g.,][]{2012A&A...543A..66H}. For these receivers, pulse portraits have emerged as the preferred PE-modeling scheme (see Sec.~\ref{subsec:prof_evolution_theory}). However, wideband timing residuals still show evidence of unmodeled frequency dependence, which is typically corrected by adding a-posteriori PE parameters to the timing model \citep[e.g.,][]{Curylo_2023, 2021ApJS..252....5A}. This residual frequency dependence suggests that PE models capable of minimizing parameter covariances, such as LIFD, may still play a relevant role in high-precision timing experiments in the era of wideband pulsar timing.


In our analysis, we assumed that pulse profiles show no secular evolution, so any PE-induced timing delay is independent of the observing epoch. While this is generally a good practical approximation \citep{Lam_2016}, some pulsars exhibit temporal variations in pulse shape, sometimes with long-term, quasi-periodic, and correlated behavior \citep[e.g.,][]{2000Natur.406..484S, 10.1093/mnras/staf886}. These variations may be due to ISM effects \citep{Brook_2018}, magnetospheric state changes \citep{2010Sci...329..408L}, geodetic precession \citep{Perera_2010}, or long-term evolution in component separation \citep{doi:10.1126/science.1243254}, among others. 
This suggests that the profile at a given frequency can also depend on time. Although these temporal changes are more pronounced in young pulsars than in MSPs, they can still introduce timing systematics in profile evolution models \citep[e.g.,][]{Pennucci_2014}. Therefore, future work could extend the current analysis to include epoch-dependent piecewise corrections to the IFD and LIFD parameters.

The ability to effectively isolate PE will be key to testing and constraining theoretical models of pulsar emission. An open question \citep{10.1093/mnras/190.4.945} is whether pulsar emission is inherently narrowband (with different frequencies generated at distinct altitudes, giving rise to RFM) or broadband (with a single emission region producing radiation across a wide frequency range). The frequency dependence of pulse microstructure may differ markedly between the two cases \citep{1978ApJ...222.1006C}, suggesting that constraining how profile morphology evolves with frequency (in particular, whether components shift in position or only in width) can provide a direct empirical test of these scenarios.

These differences in pulse morphology should be more easily observed at lower frequencies, where PE is more pronounced. However, that is also where ISM propagation effects, such as dispersion and scattering, are most severe. Scattering broadens and shifts the profile, while residual dispersive smearing contributes additional broadening, thus mimicking PE effects expected from RFM. Therefore, these effects must be separated before PE can be used to test theoretical models.

In particular, low-frequency pulsar observations from the LOFAR array \citep{2013A&A...556A...2V} have already been proven useful for analyzing pulse profiles \citep{2016A&A...585A.128K} and constraining ISM-induced chromatic effects \citep{2025A&A...704A.109I}. Future studies could leverage this data, combining LIFD models and techniques such as cyclic spectroscopy \citep{Dolch_2021, Turner_2026}, or alternative methods of quantifying time-variable scattering 
\citep{2020MNRAS.499.1468M} to break the covariance between the different chromatic effects.

\section*{Acknowledgments}

S.V.S.F. undertook the analysis, developed the code pipeline, and prepared the ﬁgures, tables, and the majority of the text. M.T.L. developed the mathematical framework for this work, selected the analyzed dataset, assisted with the preparation of the manuscript, provided advice on interpreting the results, and supervised the project development. O.W. created the 2D templates and TOAs, and recovered the DM time series presented in Sec.~\ref{subsec:2D_templates}. R.V. created the simulated data sets used in Secs.~\ref{subsec:simulated_data} and \ref{subsec:2D_templates}. All other authors were involved in curating the NANOGrav 15-year dataset; specific contributions are summarized in \cite{NG15}.

We thank NANOGrav’s Noise Budget and Timing Working
Groups for their valuable input and feedback on the manuscript. We thank the anonymous referee for comments that helped to improve the manuscript.

S.V.S.F. is supported by the National Science Foundation Graduate Research Fellowship Program under Grant No. 2139292, and acknowledges partial support from the NASA New York Space Grant. S.V.S.F. and R.V. acknowledge financial support from the European Research Council (ERC) starting grant ``GIGA'' under grant agreement number
101116134. T.D.\ and M.T.L.\ received support from an NSF Astronomy and Astrophysics Grant (AAG) award number 2511107 during this work. P.R.B.\ is supported by the Science and Technology Facilities Council, grant number ST/W000946/1. Pulsar research at UBC is supported by an NSERC Discovery Grant and by CIFAR. K.C.\ is supported by a UBC Four Year Fellowship. M.E.D.\ acknowledges support from the Naval Research Laboratory by NASA under contract S-15633Y. E.C.F.\ is supported by NASA under award number 80GSFC24M0006. D.C.G.\ is supported by NSF Astronomy and Astrophysics Grant (AAG) award \#2406919. D.R.L.\ and M.A.M.\ are supported by NSF \#1458952. M.A.M.\ is supported by NSF \#2009425. The Dunlap Institute is funded by an endowment established by the David Dunlap family and the University of Toronto. T.T.P.\ acknowledges support from the Extragalactic Astrophysics Research Group at E\"{o}tv\"{o}s Lor\'{a}nd University, funded by the E\"{o}tv\"{o}s Lor\'{a}nd Research Network (ELKH), which was used during the development of this research. H.A.R.\ is supported by NSF Partnerships for Research and Education in Physics (PREP) award No.\ 2216793. S.M.R.\ and I.H.S.\ are CIFAR Fellows. Portions of this work performed at NRL were supported by ONR 6.1 basic research funding.

We acknowledge support of the NANOGrav Physics Frontiers Center through NSF award numbers 2020265 and 2607948. The authors acknowledge Research Computing at the Rochester Institute of Technology for providing computational resources and support that have contributed to the research results reported in this publication. The Green Bank Observatory and National Radio Astronomy Observatory are facilities of the NSF operated under a cooperative agreement by Associated Universities, Inc.



This work made use of Astropy:\footnote{http://www.astropy.org} a community-developed core Python package and an ecosystem of tools and resources for astronomy \citep{astropy:2013, astropy:2018, astropy:2022}. 


%

\vspace{5mm}
\facilities{Green Bank Observatory (GBO), Arecibo Observatory (AO), Very Large Array (VLA).}


\software{PINT \citep{Luo_2021}, Astropy \citep{astropy:2013, astropy:2018, astropy:2022}.}



\appendix

\section{Reproducing our work}
\label{sec:appendix}

Our software scripts, input data files, and the order in which they must be executed are available from the GitHub repository for this project, \url{https://github.com/sophiasosafiscella/LIFD_PINT}). Our Python code requires the \texttt{astropy} and \texttt{PINT} packages. The input \texttt{.par} and \texttt{.tim} pulsar timing data files, and the 1D pulsar templates, are available from \url{https://zenodo.org/records/16051178}.

As part of this work, we developed numerical implementations of the IFD and LIFD models as \texttt{DelayComponent} classes within the Python-based pulsar timing software \texttt{PINT}. These implementations are available in the project’s GitHub repository and can be readily incorporated into any \texttt{PINT}-based timing pipeline as follows:

\begin{enumerate}
    \item Starting from a \texttt{TimingModel} object (from \texttt{pint.models.timing\_model}), the existing FD component can be removed using the built-in method \texttt{remove\_component("FD")}. This ensures that no frequency-dependent delay terms from the standard FD model remain in the timing model.
   \item Mapping to the new independent variable ($\lambda$ for IFD, $x$ for LIFD) requires the TOA's frequency range. Therefore, we attach the TOAs to the model by assigning the \texttt{toas} attribute of the timing model.
   \item We add the \texttt{IFD} or \texttt{LIFD} class to the timing model via \texttt{add\_component(IFD(order))} or \texttt{add\_component(LIFD(order))}, respectively. Here, \texttt{order} specifies the polynomial order of the chosen model.
    \item Once added, the corresponding IFD or LIFD coefficients are automatically registered as free parameters that can then be fit simultaneously with the rest of the model using any of \texttt{PINT}'s \texttt{fitter} classes.
\end{enumerate}

\noindent A simple implementation would be:

\begin{verbatim}
import pint
from LIFD_class import LIFD

timing_model = pint.models.get_model(parfile)
toas = pint.toa.get_TOAs(timfile)

timing_model.toas = toas
timing_model.remove_component("FD")
timing_model.add_component(LIFD(order))

f = pint.fitter.WLSFitter(toas, timing_model)
f.fit_toas()
f.print_summary()
\end{verbatim}

\noindent For a complete, working example of this procedure, please refer to the \texttt{main.py} script included in the project’s GitHub repository, which demonstrates the integration of these components within a full timing analysis pipeline.


\bibliographystyle{mnras}
\bibliography{main}



\end{document}

%% file: authorsLIFD.tex
\author[0000-0002-5176-2924]{Sofia V. Sosa Fiscella}
\affiliation{ASTRON, Netherlands Institute for Radio Astronomy, Oude Hoogeveensedijk 4, 7991 PD Dwingeloo, The Netherlands}
\email{sophia.sosa@nanograv.org}
\author[0000-0003-0721-651X]{Michael T. Lam}
\affiliation{SETI Institute, 339 N Bernardo Ave Suite 200, Mountain View, CA 94043, USA}
\email{michael.lam@nanograv.org}
\author[0009-0006-2876-5613]{Olivia West}
\affiliation{Centre for Extragalactic Astronomy, Department of Physics, University of Durham, Durham, DH1 3LE, UK}
\email{liv.h.west@gmail.com}
\author{Ruggero Valdata}
\affiliation{ASTRON, Netherlands Institute for Radio Astronomy, Oude Hoogeveensedijk 4, 7991 PD Dwingeloo, The Netherlands}
\affiliation{Anton Pannekoek Institute for Astronomy, University of Amsterdam, Science Park 904, 1098 XH Amsterdam, The Netherlands}
\email{valdata@astron.nl}
\author[0000-0001-5134-3925]{Gabriella Agazie}
\affiliation{Center for Gravitation, Cosmology and Astrophysics, Department of Physics and Astronomy, University of Wisconsin-Milwaukee,\\ P.O. Box 413, Milwaukee, WI 53201, USA}
\email{gabriella.agazie@nanograv.org}
\author[0000-0002-8935-9882]{Akash Anumarlapudi}
\affiliation{Department of Physics and Astronomy, University of North Carolina, Chapel Hill, NC 27599, USA}
\email{akasha@unc.edu}
\author[0000-0003-0638-3340]{Anne M. Archibald}
\affiliation{Newcastle University, NE1 7RU, UK}
\email{anne.archibald@nanograv.org}
\author[0009-0008-6187-8753]{Zaven Arzoumanian}
\affiliation{X-Ray Astrophysics Laboratory, NASA Goddard Space Flight Center, Code 662, Greenbelt, MD 20771, USA}
\email{zaven.arzoumanian@nanograv.org}
\author[0000-0003-2745-753X]{Paul T. Baker}
\affiliation{Department of Physics and Astronomy, Widener University, One University Place, Chester, PA 19013, USA}
\email{paul.baker@nanograv.org}
\author[0000-0003-3053-6538]{Paul R. Brook}
\affiliation{Institute for Gravitational Wave Astronomy and School of Physics and Astronomy, University of Birmingham, Edgbaston, Birmingham B15 2TT, UK}
\email{paul.brook@nanograv.org}
\author[0000-0002-6039-692X]{H. Thankful Cromartie}
\affiliation{Department of Physics and Astronomy, Vanderbilt University, 2301 Vanderbilt Place, Nashville, TN 37235, USA}
\email{thankful.cromartie@nanograv.org}
\author[0000-0002-1529-5169]{Kathryn Crowter}
\affiliation{Department of Physics and Astronomy, University of British Columbia, 6224 Agricultural Road, Vancouver, BC V6T 1Z1, Canada}
\email{kathryn.crowter@nanograv.org}
\author[0000-0002-2185-1790]{Megan E. DeCesar}
\altaffiliation{Resident at the Naval Research Laboratory}
\affiliation{Department of Physics and Astronomy, George Mason University, Fairfax, VA 22030}
\email{megan.decesar@nanograv.org}
\author[0000-0002-6664-965X]{Paul B. Demorest}
\affiliation{National Radio Astronomy Observatory, 1003 Lopezville Rd., Socorro, NM 87801, USA}
\email{paul.demorest@nanograv.org}
\author[0000-0001-8885-6388]{Timothy Dolch}
\affiliation{Department of Physics and Astronomy, University of New Mexico, Albuquerque, NM 87131, USA}
\affiliation{Department of Physics, Hillsdale College, 33 E. College Street, Hillsdale, MI 49242, USA}
\affiliation{Eureka Scientific, 2452 Delmer Street, Suite 100, Oakland, CA 94602-3017, USA}
\affiliation{SETI Institute, 339 N Bernardo Ave Suite 200, Mountain View, CA 94043, USA}
\email{timothy.dolch@nanograv.org}
\author[0000-0001-7828-7708]{Elizabeth C. Ferrara}
\affiliation{Department of Astronomy, University of Maryland, College Park, MD 20742, USA}
\affiliation{Center for Research and Exploration in Space Science and Technology, NASA/GSFC, Greenbelt, MD 20771}
\affiliation{NASA Goddard Space Flight Center, Greenbelt, MD 20771, USA}
\email{elizabeth.ferrara@nanograv.org}
\author[0000-0001-5645-5336]{William Fiore}
\affiliation{Department of Physics and Astronomy, University of British Columbia, 6224 Agricultural Road, Vancouver, BC V6T 1Z1, Canada}
\email{william.fiore@nanograv.org}
\author[0000-0001-8384-5049]{Emmanuel Fonseca}
\affiliation{Department of Physics and Astronomy, West Virginia University, P.O. Box 6315, Morgantown, WV 26506, USA}
\affiliation{Center for Gravitational Waves and Cosmology, West Virginia University, Chestnut Ridge Research Building, Morgantown, WV 26505, USA}
\email{emmanuel.fonseca@nanograv.org}
\author[0000-0001-7624-4616]{Gabriel E. Freedman}
\affiliation{NASA Goddard Space Flight Center, Greenbelt, MD 20771, USA}
\email{gabriel.freedman@nanograv.org}
\author[0000-0001-6166-9646]{Nate Garver-Daniels}
\affiliation{Department of Physics and Astronomy, West Virginia University, P.O. Box 6315, Morgantown, WV 26506, USA}
\affiliation{Center for Gravitational Waves and Cosmology, West Virginia University, Chestnut Ridge Research Building, Morgantown, WV 26505, USA}
\email{nathaniel.garver-daniels@nanograv.org}
\author[0000-0001-8158-683X]{Peter A. Gentile}
\affiliation{Department of Physics and Astronomy, West Virginia University, P.O. Box 6315, Morgantown, WV 26506, USA}
\affiliation{Center for Gravitational Waves and Cosmology, West Virginia University, Chestnut Ridge Research Building, Morgantown, WV 26505, USA}
\email{peter.gentile@nanograv.org}
\author[0000-0003-4090-9780]{Joseph Glaser}
\affiliation{Department of Physics and Astronomy, West Virginia University, P.O. Box 6315, Morgantown, WV 26506, USA}
\affiliation{Center for Gravitational Waves and Cosmology, West Virginia University, Chestnut Ridge Research Building, Morgantown, WV 26505, USA}
\email{joseph.glaser@nanograv.org}
\author[0000-0003-1884-348X]{Deborah C. Good}
\affiliation{Department of Physics and Astronomy, University of Montana, 32 Campus Drive, Missoula, MT 59812}
\email{deborah.good@nanograv.org}
\author[0000-0003-2742-3321]{Jeffrey S. Hazboun}
\affiliation{Department of Physics, Oregon State University, Corvallis, OR 97331, USA}
\email{jeffrey.hazboun@nanograv.org}
\author[0000-0003-1082-2342]{Ross J. Jennings}
\altaffiliation{NANOGrav Physics Frontiers Center Postdoctoral Fellow}
\affiliation{Department of Physics and Astronomy, West Virginia University, P.O. Box 6315, Morgantown, WV 26506, USA}
\affiliation{Center for Gravitational Waves and Cosmology, West Virginia University, Chestnut Ridge Research Building, Morgantown, WV 26505, USA}
\email{ross.jennings@nanograv.org}
\author[0000-0001-6607-3710]{Megan L. Jones}
\affiliation{Center for Gravitation, Cosmology and Astrophysics, Department of Physics and Astronomy, University of Wisconsin-Milwaukee,\\ P.O. Box 413, Milwaukee, WI 53201, USA}
\email{megan.jones@nanograv.org}
\author[0000-0001-6295-2881]{David L. Kaplan}
\affiliation{Center for Gravitation, Cosmology and Astrophysics, Department of Physics and Astronomy, University of Wisconsin-Milwaukee,\\ P.O. Box 413, Milwaukee, WI 53201, USA}
\email{kaplan@uwm.edu}
\author[0000-0002-0893-4073]{Matthew Kerr}
\affiliation{Space Science Division, Naval Research Laboratory, Washington, DC 20375-5352, USA}
\email{matthew.kerr@nanograv.org}
\author[0000-0003-1301-966X]{Duncan R. Lorimer}
\affiliation{Department of Physics and Astronomy, West Virginia University, P.O. Box 6315, Morgantown, WV 26506, USA}
\affiliation{Center for Gravitational Waves and Cosmology, West Virginia University, Chestnut Ridge Research Building, Morgantown, WV 26505, USA}
\email{duncan.lorimer@nanograv.org}
\author[0000-0001-5373-5914]{Jing Luo}
\altaffiliation{Deceased}
\affiliation{Department of Astronomy \& Astrophysics, University of Toronto, 50 Saint George Street, Toronto, ON M5S 3H4, Canada}
\email{jing.luo@nanograv.org}
\author[0000-0001-5229-7430]{Ryan S. Lynch}
\affiliation{Green Bank Observatory, P.O. Box 2, Green Bank, WV 24944, USA}
\email{ryan.lynch@nanograv.org}
\author[0000-0001-5481-7559]{Alexander McEwen}
\affiliation{Center for Gravitation, Cosmology and Astrophysics, Department of Physics and Astronomy, University of Wisconsin-Milwaukee,\\ P.O. Box 413, Milwaukee, WI 53201, USA}
\email{alexander.mcewen@nanograv.org}
\author[0000-0001-7697-7422]{Maura A. McLaughlin}
\affiliation{Department of Physics and Astronomy, West Virginia University, P.O. Box 6315, Morgantown, WV 26506, USA}
\affiliation{Center for Gravitational Waves and Cosmology, West Virginia University, Chestnut Ridge Research Building, Morgantown, WV 26505, USA}
\email{maura.mclaughlin@nanograv.org}
\author[0000-0002-4642-1260]{Natasha McMann}
\affiliation{Department of Physics and Astronomy, Vanderbilt University, 2301 Vanderbilt Place, Nashville, TN 37235, USA}
\email{natasha.mcmann@nanograv.org}
\author[0000-0001-8845-1225]{Bradley W. Meyers}
\affiliation{Australian SKA Regional Centre (AusSRC), Curtin University, Bentley, WA 6102, Australia}
\affiliation{International Centre for Radio Astronomy Research (ICRAR), Curtin University, Bentley, WA 6102, Australia}
\email{bradley.meyers@nanograv.org}
\author[0000-0002-3616-5160]{Cherry Ng}
\affiliation{Dunlap Institute for Astronomy and Astrophysics, University of Toronto, 50 St. George St., Toronto, ON M5S 3H4, Canada}
\email{cherry.ng@nanograv.org}
\author[0000-0002-6709-2566]{David J. Nice}
\affiliation{Department of Physics, Lafayette College, Easton, PA 18042, USA}
\email{niced@lafayette.edu}
\author[0000-0001-5465-2889]{Timothy T. Pennucci}
\affiliation{Institute of Physics and Astronomy, E\"{o}tv\"{o}s Lor\'{a}nd University, P\'{a}zm\'{a}ny P. s. 1/A, 1117 Budapest, Hungary}
\email{timothy.pennucci@nanograv.org}
\author[0000-0002-8509-5947]{Benetge B. P. Perera}
\affiliation{Arecibo Observatory, HC3 Box 53995, Arecibo, PR 00612, USA}
\email{benetge.perera@nanograv.org}
\author[0000-0002-8826-1285]{Nihan S. Pol}
\affiliation{Department of Physics, Texas Tech University, Box 41051, Lubbock, TX 79409, USA}
\email{nihan.pol@nanograv.org}
\author[0000-0002-2074-4360]{Henri A. Radovan}
\affiliation{Department of Physics, University of Puerto Rico, Mayag\"{u}ez, PR 00681, USA}
\email{henri.radovan@nanograv.org}
\author[0000-0001-5799-9714]{Scott M. Ransom}
\affiliation{National Radio Astronomy Observatory, 520 Edgemont Road, Charlottesville, VA 22903, USA}
\email{sransom@nrao.edu}
\author[0000-0002-5297-5278]{Paul S. Ray}
\affiliation{Space Science Division, Naval Research Laboratory, Washington, DC 20375-5352, USA}
\email{paul.ray@nanograv.org}
\author[0000-0003-4391-936X]{Ann Schmiedekamp}
\affiliation{Department of Physics, Penn State Abington, Abington, PA 19001, USA}
\email{ann.schmiedekamp@nanograv.org}
\author[0000-0002-1283-2184]{Carl Schmiedekamp}
\affiliation{Department of Physics, Penn State Abington, Abington, PA 19001, USA}
\email{carl.schmiedekamp@nanograv.org}
\author[0000-0002-7283-1124]{Brent J. Shapiro-Albert}
\affiliation{Department of Physics and Astronomy, West Virginia University, P.O. Box 6315, Morgantown, WV 26506, USA}
\affiliation{Center for Gravitational Waves and Cosmology, West Virginia University, Chestnut Ridge Research Building, Morgantown, WV 26505, USA}
\affiliation{Giant Army, 915A 17th Ave, Seattle WA 98122}
\email{brent.shapiro-albert@nanograv.org}
\author[0000-0003-3583-3742]{Taylor Starkman}
\affiliation{Department of Physics and Astronomy, University of British Columbia, 6224 Agricultural Road, Vancouver, BC V6T 1Z1, Canada}
\email{trstark@student.ubc.ca}
\author[0000-0001-9784-8670]{Ingrid H. Stairs}
\affiliation{Department of Physics and Astronomy, University of British Columbia, 6224 Agricultural Road, Vancouver, BC V6T 1Z1, Canada}
\email{stairs@astro.ubc.ca}
\author[0000-0002-7261-594X]{Kevin Stovall}
\affiliation{National Radio Astronomy Observatory, 1003 Lopezville Rd., Socorro, NM 87801, USA}
\email{kevin.stovall@nanograv.org}
\author[0000-0002-2820-0931]{Abhimanyu Susobhanan}
\affiliation{Max-Planck-Institut f{\"u}r Gravitationsphysik (Albert-Einstein-Institut), Callinstra{\ss}e 38, D-30167 Hannover, Germany\\Leibniz Universit{\"a}t Hannover, D-30167 Hannover, Germany}
\email{abhimanyu.susobhanan@nanograv.org}
\author[0000-0002-1075-3837]{Joseph K. Swiggum}
\altaffiliation{NANOGrav Physics Frontiers Center Postdoctoral Fellow}
\affiliation{Department of Physics, Lafayette College, Easton, PA 18042, USA}
\email{joseph.swiggum@nanograv.org}
\author[0000-0001-9678-0299]{Haley M. Wahl}
\affiliation{Department of Physics and Astronomy, West Virginia University, P.O. Box 6315, Morgantown, WV 26506, USA}
\affiliation{Center for Gravitational Waves and Cosmology, West Virginia University, Chestnut Ridge Research Building, Morgantown, WV 26505, USA}
\email{haley.wahl@nanograv.org}

%% file: main.bib
@article{NG15,
doi = {10.3847/2041-8213/acda9a},
url = {https://dx.doi.org/10.3847/2041-8213/acda9a},
year = {2023},
month = {jun},
publisher = {The American Astronomical Society},
volume = {951},
number = {1},
pages = {L9},
author = {Agazie, Gabriella and Alam, Md Faisal and Anumarlapudi, Akash and Archibald, Anne M. and Arzoumanian, Zaven and Baker, Paul T. and Blecha, Laura and Bonidie, Victoria and Brazier, Adam and Brook, Paul R. and Burke-Spolaor, Sarah and Bécsy, Bence and Chapman, Christopher and Charisi, Maria and Chatterjee, Shami and Cohen, Tyler and Cordes, James M. and Cornish, Neil J. and Crawford, Fronefield and Cromartie, H. Thankful and Crowter, Kathryn and DeCesar, Megan E. and Demorest, Paul B. and Dolch, Timothy and Drachler, Brendan and Ferrara, Elizabeth C. and Fiore, William and Fonseca, Emmanuel and Freedman, Gabriel E. and Garver-Daniels, Nate and Gentile, Peter A. and Glaser, Joseph and Good, Deborah C. and Gültekin, Kayhan and Hazboun, Jeffrey S. and Jennings, Ross J. and Jessup, Cody and Johnson, Aaron D. and Jones, Megan L. and Kaiser, Andrew R. and Kaplan, David L. and Kelley, Luke Zoltan and Kerr, Matthew and Key, Joey S. and Kuske, Anastasia and Laal, Nima and Lam, Michael T. and Lamb, William G. and W. Lazio, T. Joseph and Lewandowska, Natalia and Lin, Ye and Liu, Tingting and Lorimer, Duncan R. and Luo, Jing and Lynch, Ryan S. and Ma, Chung-Pei and Madison, Dustin R. and Maraccini, Kaleb and McEwen, Alexander and McKee, James W. and McLaughlin, Maura A. and McMann, Natasha and Meyers, Bradley W. and Mingarelli, Chiara M. F. and Mitridate, Andrea and Ng, Cherry and Nice, David J. and Ocker, Stella Koch and Olum, Ken D. and Panciu, Elisa and Pennucci, Timothy T. and Perera, Benetge B. P. and Pol, Nihan S. and Radovan, Henri A. and Ransom, Scott M. and Ray, Paul S. and Romano, Joseph D. and Salo, Laura and Sardesai, Shashwat C. and Schmiedekamp, Carl and Schmiedekamp, Ann and Schmitz, Kai and Shapiro-Albert, Brent J. and Siemens, Xavier and Simon, Joseph and Siwek, Magdalena S. and Stairs, Ingrid H. and Stinebring, Daniel R. and Stovall, Kevin and Susobhanan, Abhimanyu and Swiggum, Joseph K. and Taylor, Stephen R. and Turner, Jacob E. and Unal, Caner and Vallisneri, Michele and Vigeland, Sarah J. and Wahl, Haley M. and Wang, Qiaohong and Witt, Caitlin A. and Young, Olivia and The NANOGrav Collaboration},
title = {The NANOGrav 15 yr Data Set: Observations and Timing of 68 Millisecond Pulsars},
journal = {The Astrophysical Journal Letters}
}

@article{antoniadis2023second,
  title={The second data release from the European Pulsar Timing Array-III. Search for gravitational wave signals},
  author={Antoniadis, J and Arumugam, P and Arumugam, S and Babak, S and Bagchi, M and Nielsen, A-S Bak and Bassa, CG and Bathula, A and Berthereau, A and Bonetti, M and others},
  journal={Astronomy \& Astrophysics},
  volume={678},
  pages={A50},
  year={2023},
  publisher={EDP Sciences}
}

@article{reardon2023search,
  title={Search for an isotropic gravitational-wave background with the Parkes Pulsar Timing Array},
  author={Reardon, Daniel J and Zic, Andrew and Shannon, Ryan M and Hobbs, George B and Bailes, Matthew and Di Marco, Valentina and Kapur, Agastya and Rogers, Axl F and Thrane, Eric and Askew, Jacob and others},
  journal={The Astrophysical Journal Letters},
  volume={951},
  number={1},
  pages={L6},
  year={2023},
  publisher={IOP Publishing}
}

@ARTICLE{2021ApJS..252....5A,
       author = {{Alam}, Md F. and {Arzoumanian}, Zaven and {Baker}, Paul T. and {Blumer}, Harsha and {Bohler}, Keith E. and {Brazier}, Adam and {Brook}, Paul R. and {Burke-Spolaor}, Sarah and {Caballero}, Keeisi and {Camuccio}, Richard S. and {Chamberlain}, Rachel L. and {Chatterjee}, Shami and {Cordes}, James M. and {Cornish}, Neil J. and {Crawford}, Fronefield and {Cromartie}, H. Thankful and {Decesar}, Megan E. and {Demorest}, Paul B. and {Dolch}, Timothy and {Ellis}, Justin A. and {Ferdman}, Robert D. and {Ferrara}, Elizabeth C. and {Fiore}, William and {Fonseca}, Emmanuel and {Garcia}, Yhamil and {Garver-Daniels}, Nathan and {Gentile}, Peter A. and {Good}, Deborah C. and {Gusdorff}, Jordan A. and {Halmrast}, Daniel and {Hazboun}, Jeffrey S. and {Islo}, Kristina and {Jennings}, Ross J. and {Jessup}, Cody and {Jones}, Megan L. and {Kaiser}, Andrew R. and {Kaplan}, David L. and {Kelley}, Luke Zoltan and {Key}, Joey Shapiro and {Lam}, Michael T. and {Lazio}, T. Joseph W. and {Lorimer}, Duncan R. and {Luo}, Jing and {Lynch}, Ryan S. and {Madison}, Dustin R. and {Maraccini}, Kaleb and {McLaughlin}, Maura A. and {Mingarelli}, Chiara M.~F. and {Ng}, Cherry and {Nguyen}, Benjamin M.~X. and {Nice}, David J. and {Pennucci}, Timothy T. and {Pol}, Nihan S. and {Ramette}, Joshua and {Ransom}, Scott M. and {Ray}, Paul S. and {Shapiro-Albert}, Brent J. and {Siemens}, Xavier and {Simon}, Joseph and {Spiewak}, Ren{\'e}e and {Stairs}, Ingrid H. and {Stinebring}, Daniel R. and {Stovall}, Kevin and {Swiggum}, Joseph K. and {Taylor}, Stephen R. and {Tripepi}, Michael and {Vallisneri}, Michele and {Vigeland}, Sarah J. and {Witt}, Caitlin A. and {Zhu}, Weiwei and {Nanograv Collaboration}},
        title = "{The NANOGrav 12.5 yr Data Set: Wideband Timing of 47 Millisecond Pulsars}",
      journal = {\apjs},
         year = 2021,
        month = jan,
       volume = {252},
       number = {1},
          eid = {5},
        pages = {5},
          doi = {10.3847/1538-4365/abc6a1},
archivePrefix = {arXiv},
       eprint = {2005.06495},
 primaryClass = {astro-ph.HE},
       adsurl = {https://ui.adsabs.harvard.edu/abs/2021ApJS..252....5A}
}

@ARTICLE{2023RAA....23g5024X,
       author = {{Xu}, Heng and {Chen}, Siyuan and {Guo}, Yanjun and {Jiang}, Jinchen and {Wang}, Bojun and {Xu}, Jiangwei and {Xue}, Zihan and {Caballero}, R. Nicolas and {Yuan}, Jianping and {Xu}, Yonghua and {Wang}, Jingbo and {Hao}, Longfei and {Luo}, Jingtao and {Lee}, Kejia and {Han}, Jinlin and {Jiang}, Peng and {Shen}, Zhiqiang and {Wang}, Min and {Wang}, Na and {Xu}, Renxin and {Wu}, Xiangping and {Manchester}, Richard and {Qian}, Lei and {Guan}, Xin and {Huang}, Menglin and {Sun}, Chun and {Zhu}, Yan},
        title = "{Searching for the Nano-Hertz Stochastic Gravitational Wave Background with the Chinese Pulsar Timing Array Data Release I}",
      journal = {Research in Astronomy and Astrophysics},
         year = 2023,
        month = jul,
       volume = {23},
       number = {7},
          eid = {075024},
        pages = {075024},
          doi = {10.1088/1674-4527/acdfa5},
archivePrefix = {arXiv},
       eprint = {2306.16216},
 primaryClass = {astro-ph.HE},
       adsurl = {https://ui.adsabs.harvard.edu/abs/2023RAA....23g5024X}
}

@ARTICLE{2016A&A...585A.128K,
       author = {{Kondratiev}, V.~I. and {Verbiest}, J.~P.~W. and {Hessels}, J.~W.~T. and {Bilous}, A.~V. and {Stappers}, B.~W. and {Kramer}, M. and {Keane}, E.~F. and {Noutsos}, A. and {Os{\l}owski}, S. and {Breton}, R.~P. and {Hassall}, T.~E. and {Alexov}, A. and {Cooper}, S. and {Falcke}, H. and {Grie{\ss}meier}, J.-M. and {Karastergiou}, A. and {Kuniyoshi}, M. and {Pilia}, M. and {Sobey}, C. and {ter Veen}, S. and {van Leeuwen}, J. and {Weltevrede}, P. and {Bell}, M.~E. and {Broderick}, J.~W. and {Corbel}, S. and {Eisl{\"o}ffel}, J. and {Markoff}, S. and {Rowlinson}, A. and {Swinbank}, J.~D. and {Wijers}, R.~A.~M.~J. and {Wijnands}, R. and {Zarka}, P.},
        title = "{A LOFAR census of millisecond pulsars}",
      journal = {\aap},
         year = 2016,
        month = jan,
       volume = {585},
          eid = {A128},
        pages = {A128},
          doi = {10.1051/0004-6361/201527178},
archivePrefix = {arXiv},
       eprint = {1508.02948},
 primaryClass = {astro-ph.HE},
       adsurl = {https://ui.adsabs.harvard.edu/abs/2016A&A...585A.128K}
}

@ARTICLE{2013A&A...556A...2V,
       author = {{van Haarlem}, M.~P. and {Wise}, M.~W. and {Gunst}, A.~W. and {Heald}, G. and {McKean}, J.~P. and {Hessels}, J.~W.~T. and {de Bruyn}, A.~G. and {Nijboer}, R. and {Swinbank}, J. and {Fallows}, R. and {Brentjens}, M. and {Nelles}, A. and {Beck}, R. and {Falcke}, H. and {Fender}, R. and {H{\"o}randel}, J. and {Koopmans}, L.~V.~E. and {Mann}, G. and {Miley}, G. and {R{\"o}ttgering}, H. and {Stappers}, B.~W. and {Wijers}, R.~A.~M.~J. and {Zaroubi}, S. and {van den Akker}, M. and {Alexov}, A. and {Anderson}, J. and {Anderson}, K. and {van Ardenne}, A. and {Arts}, M. and {Asgekar}, A. and {Avruch}, I.~M. and {Batejat}, F. and {B{\"a}hren}, L. and {Bell}, M.~E. and {Bell}, M.~R. and {van Bemmel}, I. and {Bennema}, P. and {Bentum}, M.~J. and {Bernardi}, G. and {Best}, P. and {B{\^\i}rzan}, L. and {Bonafede}, A. and {Boonstra}, A.-J. and {Braun}, R. and {Bregman}, J. and {Breitling}, F. and {van de Brink}, R.~H. and {Broderick}, J. and {Broekema}, P.~C. and {Brouw}, W.~N. and {Br{\"u}ggen}, M. and {Butcher}, H.~R. and {van Cappellen}, W. and {Ciardi}, B. and {Coenen}, T. and {Conway}, J. and {Coolen}, A. and {Corstanje}, A. and {Damstra}, S. and {Davies}, O. and {Deller}, A.~T. and {Dettmar}, R.-J. and {van Diepen}, G. and {Dijkstra}, K. and {Donker}, P. and {Doorduin}, A. and {Dromer}, J. and {Drost}, M. and {van Duin}, A. and {Eisl{\"o}ffel}, J. and {van Enst}, J. and {Ferrari}, C. and {Frieswijk}, W. and {Gankema}, H. and {Garrett}, M.~A. and {de Gasperin}, F. and {Gerbers}, M. and {de Geus}, E. and {Grie{\ss}meier}, J.-M. and {Grit}, T. and {Gruppen}, P. and {Hamaker}, J.~P. and {Hassall}, T. and {Hoeft}, M. and {Holties}, H.~A. and {Horneffer}, A. and {van der Horst}, A. and {van Houwelingen}, A. and {Huijgen}, A. and {Iacobelli}, M. and {Intema}, H. and {Jackson}, N. and {Jelic}, V. and {de Jong}, A. and {Juette}, E. and {Kant}, D. and {Karastergiou}, A. and {Koers}, A. and {Kollen}, H. and {Kondratiev}, V.~I. and {Kooistra}, E. and {Koopman}, Y. and {Koster}, A. and {Kuniyoshi}, M. and {Kramer}, M. and {Kuper}, G. and {Lambropoulos}, P. and {Law}, C. and {van Leeuwen}, J. and {Lemaitre}, J. and {Loose}, M. and {Maat}, P. and {Macario}, G. and {Markoff}, S. and {Masters}, J. and {McFadden}, R.~A. and {McKay-Bukowski}, D. and {Meijering}, H. and {Meulman}, H. and {Mevius}, M. and {Middelberg}, E. and {Millenaar}, R. and {Miller-Jones}, J.~C.~A. and {Mohan}, R.~N. and {Mol}, J.~D. and {Morawietz}, J. and {Morganti}, R. and {Mulcahy}, D.~D. and {Mulder}, E. and {Munk}, H. and {Nieuwenhuis}, L. and {van Nieuwpoort}, R. and {Noordam}, J.~E. and {Norden}, M. and {Noutsos}, A. and {Offringa}, A.~R. and {Olofsson}, H. and {Omar}, A. and {Orr{\'u}}, E. and {Overeem}, R. and {Paas}, H. and {Pandey-Pommier}, M. and {Pandey}, V.~N. and {Pizzo}, R. and {Polatidis}, A. and {Rafferty}, D. and {Rawlings}, S. and {Reich}, W. and {de Reijer}, J.-P. and {Reitsma}, J. and {Renting}, G.~A. and {Riemers}, P. and {Rol}, E. and {Romein}, J.~W. and {Roosjen}, J. and {Ruiter}, M. and {Scaife}, A. and {van der Schaaf}, K. and {Scheers}, B. and {Schellart}, P. and {Schoenmakers}, A. and {Schoonderbeek}, G. and {Serylak}, M. and {Shulevski}, A. and {Sluman}, J. and {Smirnov}, O. and {Sobey}, C. and {Spreeuw}, H. and {Steinmetz}, M. and {Sterks}, C.~G.~M. and {Stiepel}, H.-J. and {Stuurwold}, K. and {Tagger}, M. and {Tang}, Y. and {Tasse}, C. and {Thomas}, I. and {Thoudam}, S. and {Toribio}, M.~C. and {van der Tol}, B. and {Usov}, O. and {van Veelen}, M. and {van der Veen}, A.-J. and {ter Veen}, S. and {Verbiest}, J.~P.~W. and {Vermeulen}, R. and {Vermaas}, N. and {Vocks}, C. and {Vogt}, C. and {de Vos}, M. and {van der Wal}, E. and {van Weeren}, R. and {Weggemans}, H. and {Weltevrede}, P. and {White}, S. and {Wijnholds}, S.~J. and {Wilhelmsson}, T. and {Wucknitz}, O. and {Yatawatta}, S. and {Zarka}, P. and {Zensus}, A.},
        title = "{LOFAR: The LOw-Frequency ARray}",
      journal = {\aap},
         year = 2013,
        month = aug,
       volume = {556},
          eid = {A2},
        pages = {A2},
          doi = {10.1051/0004-6361/201220873},
archivePrefix = {arXiv},
       eprint = {1305.3550},
 primaryClass = {astro-ph.IM},
       adsurl = {https://ui.adsabs.harvard.edu/abs/2013A&A...556A...2V}
}

@INPROCEEDINGS{2026AAS...24741606R,
       author = {{Ransom}, Scott and {Lynch}, Ryan and {White}, Steven and {Simon}, Bob},
        title = "{An Ultrawideband Receiver for the Green Bank Telescope}",
    booktitle = {American Astronomical Society Meeting Abstracts},
         year = 2026,
       series = {American Astronomical Society Meeting Abstracts},
       volume = {247},
        month = feb,
          eid = {416.06},
        pages = {416.06},
       adsurl = {https://ui.adsabs.harvard.edu/abs/2026AAS...24741606R}
}

@ARTICLE{1982ApJ...258..776B,
       author = {{Bartel}, N. and {Morris}, D. and {Sieber}, W. and {Hankins}, T.~H.},
        title = "{The mode-switching phenomenon in pulsars.}",
      journal = {\apj},
         year = 1982,
        month = jul,
       volume = {258},
        pages = {776-789},
          doi = {10.1086/160125},
       adsurl = {https://ui.adsabs.harvard.edu/abs/1982ApJ...258..776B}
}

@ARTICLE{2022MNRAS.510.5908M,
       author = {{Miles}, M.~T. and {Shannon}, R.~M. and {Bailes}, M. and {Reardon}, D.~J. and {Buchner}, S. and {Middleton}, H. and {Spiewak}, R.},
        title = "{Mode changing in J1909 - 3744: the most precisely timed pulsar}",
      journal = {\mnras},
         year = 2022,
        month = mar,
       volume = {510},
       number = {4},
        pages = {5908-5915},
          doi = {10.1093/mnras/stab3549},
archivePrefix = {arXiv},
       eprint = {2112.00897},
 primaryClass = {astro-ph.HE},
       adsurl = {https://ui.adsabs.harvard.edu/abs/2022MNRAS.510.5908M}
}

@article{Dolch_2021,
doi = {10.3847/1538-4357/abf48b},
url = {https://doi.org/10.3847/1538-4357/abf48b},
year = {2021},
month = {may},
publisher = {The American Astronomical Society},
volume = {913},
number = {2},
pages = {98},
author = {Dolch, Timothy and Stinebring, Dan R. and Jones, Glenn and Zhu, Hengrui and Lynch, Ryan S. and Cohen, Tyler and Demorest, Paul B. and Lam, Michael T. and Levin, Lina and McLaughlin, Maura A. and Palliyaguru, Nipuni T.},
title = {Deconvolving Pulsar Signals with Cyclic Spectroscopy: A Systematic Evaluation},
journal = {The Astrophysical Journal}
}

@ARTICLE{2021ApJ...909..219S,
       author = {{Shapiro-Albert}, B.~J. and {Hazboun}, J.~S. and {McLaughlin}, M.~A. and {Lam}, M.~T.},
        title = "{A Study in Frequency-dependent Effects on Precision Pulsar Timing Parameters with the Pulsar Signal Simulator}",
      journal = {\apj},
         year = 2021,
        month = mar,
       volume = {909},
       number = {2},
          eid = {219},
        pages = {219},
          doi = {10.3847/1538-4357/abdc29},
archivePrefix = {arXiv},
       eprint = {2010.07301},
 primaryClass = {astro-ph.HE},
       adsurl = {https://ui.adsabs.harvard.edu/abs/2021ApJ...909..219S}
}

@article{Turner_2026,
doi = {10.3847/1538-4357/ae578c},
url = {https://doi.org/10.3847/1538-4357/ae578c},
year = {2026},
month = {apr},
publisher = {The American Astronomical Society},
volume = {1001},
number = {2},
pages = {231},
author = {Turner, Jacob E.},
title = {Evaluating the Prospects of Cyclic Deconvolution across 312 Pulsars},
journal = {The Astrophysical Journal}
}

@article{Lam_2018,
doi = {10.3847/1538-4357/aac48d},
url = {https://doi.org/10.3847/1538-4357/aac48d},
year = {2018},
month = {jun},
publisher = {The American Astronomical Society},
volume = {861},
number = {1},
pages = {12},
author = {Lam, M. T. and McLaughlin, M. A. and Cordes, J. M. and Chatterjee, S. and Lazio, T. J. W.},
title = {Optimal Frequency Ranges for Submicrosecond Precision Pulsar Timing},
journal = {The Astrophysical Journal}
}

@ARTICLE{2025A&A...704A.109I,
       author = {{Iraci}, F. and {Chalumeau}, A. and {Tiburzi}, C. and {Verbiest}, J.~P.~W. and {Possenti}, A. and {Susarla}, S.~C. and {Krishnakumar}, M.~A. and {Shaifullah}, G.~M. and {Antoniadis}, J. and {Bagchi}, M. and {Bassa}, C. and {Caballero}, R.~N. and {Cecconi}, B. and {Chen}, S. and {Chowdhury}, S. and {Ciardi}, B. and {Cognard}, I. and {Corbel}, S. and {Desai}, S. and {Deb}, D. and {Girard}, J. and {Golden}, A. and {Grie{\ss}meier}, J.-M. and {Guillemot}, L. and {Hoeft}, M. and {Hu}, H. and {Jankowski}, F. and {Janssen}, G. and {Joshi}, B.~C. and {Kala}, S. and {Keane}, E. and {Nobleson}, K. and {Konovalenko}, A. and {Kravtsov}, I. and {Kramer}, M. and {Liu}, K. and {Parthasarathy}, A. and {Rana}, P. and {Schwarz}, D. and {Singha}, J. and {Srivastava}, A. and {Takahashi}, K. and {Tarafdar}, P. and {Theureau}, G. and {Ulyanov}, O. and {Vocks}, C. and {Wang}, J. and {Zakharenko}, V. and {Zarka}, P.},
        title = "{Combining the second data release of the European Pulsar Timing Array with low-frequency pulsar data}",
      journal = {\aap},
         year = 2025,
        month = dec,
       volume = {704},
          eid = {A109},
        pages = {A109},
          doi = {10.1051/0004-6361/202555516},
archivePrefix = {arXiv},
       eprint = {2510.04639},
 primaryClass = {astro-ph.HE},
       adsurl = {https://ui.adsabs.harvard.edu/abs/2025A&A...704A.109I}
}

@ARTICLE{2012A&A...543A..66H,
       author = {{Hassall}, T.~E. and {Stappers}, B.~W. and {Hessels}, J.~W.~T. and {Kramer}, M. and {Alexov}, A. and {Anderson}, K. and {Coenen}, T. and {Karastergiou}, A. and {Keane}, E.~F. and {Kondratiev}, V.~I. and {Lazaridis}, K. and {van Leeuwen}, J. and {Noutsos}, A. and {Serylak}, M. and {Sobey}, C. and {Verbiest}, J.~P.~W. and {Weltevrede}, P. and {Zagkouris}, K. and {Fender}, R. and {Wijers}, R.~A.~M.~J. and {B{\"a}hren}, L. and {Bell}, M.~E. and {Broderick}, J.~W. and {Corbel}, S. and {Daw}, E.~J. and {Dhillon}, V.~S. and {Eisl{\"o}ffel}, J. and {Falcke}, H. and {Grie{\ss}meier}, J.-M. and {Jonker}, P. and {Law}, C. and {Markoff}, S. and {Miller-Jones}, J.~C.~A. and {Osten}, R. and {Rol}, E. and {Scaife}, A.~M.~M. and {Scheers}, B. and {Schellart}, P. and {Spreeuw}, H. and {Swinbank}, J. and {ter Veen}, S. and {Wise}, M.~W. and {Wijnands}, R. and {Wucknitz}, O. and {Zarka}, P. and {Asgekar}, A. and {Bell}, M.~R. and {Bentum}, M.~J. and {Bernardi}, G. and {Best}, P. and {Bonafede}, A. and {Boonstra}, A.~J. and {Brentjens}, M. and {Brouw}, W.~N. and {Br{\"u}ggen}, M. and {Butcher}, H.~R. and {Ciardi}, B. and {Garrett}, M.~A. and {Gerbers}, M. and {Gunst}, A.~W. and {van Haarlem}, M.~P. and {Heald}, G. and {Hoeft}, M. and {Holties}, H. and {de Jong}, A. and {Koopmans}, L.~V.~E. and {Kuniyoshi}, M. and {Kuper}, G. and {Loose}, G.~M. and {Maat}, P. and {Masters}, J. and {McKean}, J.~P. and {Meulman}, H. and {Mevius}, M. and {Munk}, H. and {Noordam}, J.~E. and {Orr{\'u}}, E. and {Paas}, H. and {Pandey-Pommier}, M. and {Pandey}, V.~N. and {Pizzo}, R. and {Polatidis}, A. and {Reich}, W. and {R{\"o}ttgering}, H. and {Sluman}, J. and {Steinmetz}, M. and {Sterks}, C.~G.~M. and {Tagger}, M. and {Tang}, Y. and {Tasse}, C. and {Vermeulen}, R. and {van Weeren}, R.~J. and {Wijnholds}, S.~J. and {Yatawatta}, S.},
        title = "{Wide-band simultaneous observations of pulsars: disentangling dispersion measure and profile variations}",
      journal = {\aap},
         year = 2012,
        month = jul,
       volume = {543},
          eid = {A66},
        pages = {A66},
          doi = {10.1051/0004-6361/201218970},
archivePrefix = {arXiv},
       eprint = {1204.3864},
 primaryClass = {astro-ph.HE},
       adsurl = {https://ui.adsabs.harvard.edu/abs/2012A&A...543A..66H}
}

@article{10.1098/rsta.1992.0088,
    author = {Taylor, J. H.},
    title = {Pulsar timing and relativistic gravity},
    journal = {Philosophical Transactions of the Royal Society of London, Series A: Physical and Engineering Sciences},
    volume = {341},
    number = {1660},
    pages = {117-134},
    year = {1992},
    month = {10},
    issn = {0962-8428},
    doi = {10.1098/rsta.1992.0088},
    url = {https://doi.org/10.1098/rsta.1992.0088},
    eprint = {https://royalsocietypublishing.org/rsta/article-pdf/341/1660/117/281626/rsta.1992.0088.pdf},
}

@article{Melfor2026,
  author  = {Melfor, S. J. I. and Parthasarathy, A. and Johnston, S. and
             Chalumeau, A. and Dai, S. and Kerr, M. and Lower, M. E. and
             Manchester, R. N. and Oswald, L. S. and Roch, S. and
             Sobey, C. A. and Weltevrede, P.},
  title   = {Probing turbulence in the ionized interstellar medium using young pulsars. I. Dispersion measure variations},
  journal = {Astronomy \& Astrophysics},
  year    = {2026},
  doi     = {10.1051/0004-6361/202660370},
  note    = {Forthcoming article}
}

@ARTICLE{1970Natur.225..612K,
       author = {{Komesaroff}, M.~M.},
        title = "{Possible Mechanism for the Pulsar Radio Emission}",
      journal = {\nat},
         year = 1970,
        month = feb,
       volume = {225},
       number = {5233},
        pages = {612-614},
          doi = {10.1038/225612a0},
       adsurl = {https://ui.adsabs.harvard.edu/abs/1970Natur.225..612K}
}

@ARTICLE{1978ApJ...222.1006C,
       author = {{Cordes}, J.~M.},
        title = "{Observational limits on the location of pulsar emission regions.}",
      journal = {\apj},
         year = 1978,
        month = jun,
       volume = {222},
        pages = {1006-1011},
          doi = {10.1086/156218},
       adsurl = {https://ui.adsabs.harvard.edu/abs/1978ApJ...222.1006C}
}

@ARTICLE{1999ApJ...526..957K,
       author = {{Kramer}, Michael and {Lange}, Christoph and {Lorimer}, Duncan R. and {Backer}, Donald C. and {Xilouris}, Kiriaki M. and {Jessner}, Axel and {Wielebinski}, Richard},
        title = "{The Characteristics of Millisecond Pulsar Emission. III. From Low to High Frequencies}",
      journal = {\apj},
         year = 1999,
        month = dec,
       volume = {526},
       number = {2},
        pages = {957-975},
          doi = {10.1086/308042},
archivePrefix = {arXiv},
       eprint = {astro-ph/9906442},
 primaryClass = {astro-ph},
       adsurl = {https://ui.adsabs.harvard.edu/abs/1999ApJ...526..957K}
}

@article{Kramer_1998,
doi = {10.1086/305790},
url = {https://doi.org/10.1086/305790},
year = {1998},
month = {jul},
publisher = {},
volume = {501},
number = {1},
pages = {270},
author = {Kramer, Michael and Xilouris, Kiriaki M. and Lorimer, Duncan R. and Doroshenko, Oleg and Jessner, Axel and Wielebinski, Richard and Wolszczan, Alexander and Camilo, Fernando},
title = {The Characteristics of Millisecond Pulsar Emission. I. Spectra, Pulse Shapes, and the Beaming Fraction},
journal = {The Astrophysical Journal}
}

@ARTICLE{Lam2016,
       author = {{Lam}, M.~T. and {Cordes}, J.~M. and {Chatterjee}, S. and {Jones}, M.~L. and {McLaughlin}, M.~A. and {Armstrong}, J.~W.},
        title = "{Systematic and Stochastic Variations in Pulsar Dispersion Measures}",
      journal = {\apj},
         year = 2016,
        month = apr,
       volume = {821},
       number = {1},
          eid = {66},
        pages = {66},
          doi = {10.3847/0004-637X/821/1/66},
archivePrefix = {arXiv},
       eprint = {1512.02203},
 primaryClass = {astro-ph.HE},
       adsurl = {https://ui.adsabs.harvard.edu/abs/2016ApJ...821...66L}
}

@article{Cordes_1998,
doi = {10.1086/306358},
url = {https://dx.doi.org/10.1086/306358},
year = {1998},
month = {nov},
publisher = {},
volume = {507},
number = {2},
pages = {846},
author = {J. M. Cordes and B. J. Rickett},
title = {Diffractive Interstellar Scintillation Timescales and Velocities},
journal = {The Astrophysical Journal}
}

@ARTICLE{1991ApJ...382L..27P,
       author = {{Phillips}, J.~A. and {Wolszczan}, A.},
        title = "{Time Variability of Pulsar Dispersion Measures}",
      journal = {\apjl},
         year = 1991,
        month = nov,
       volume = {382},
        pages = {L27},
          doi = {10.1086/186206},
       adsurl = {https://ui.adsabs.harvard.edu/abs/1991ApJ...382L..27P}
}

@article{Zhu_2015,
doi = {10.1088/0004-637X/809/1/41},
url = {https://doi.org/10.1088/0004-637X/809/1/41},
year = {2015},
month = {aug},
publisher = {The American Astronomical Society},
volume = {809},
number = {1},
pages = {41},
author = {Zhu, W. W. and Stairs, I. H. and Demorest, P. B. and Nice, D. J. and Ellis, J. A. and Ransom, S. M. and Arzoumanian, Z. and Crowter, K. and Dolch, T. and Ferdman, R. D. and Fonseca, E. and Gonzalez, M. E. and Jones, G. and Jones, M. L. and Lam, M. T. and Levin, L. and McLaughlin, M. A. and Pennucci, T. and Stovall, K. and Swiggum, J.},
title = {TESTING THEORIES OF GRAVITATION USING 21-YEAR TIMING OF PULSAR BINARY J1713+0747},
journal = {The Astrophysical Journal}
}

@ARTICLE{2020MNRAS.499.1468M,
       author = {{Main}, R.~A. and {Sanidas}, S.~A. and {Antoniadis}, J. and {Bassa}, C. and {Chen}, S. and {Cognard}, I. and {Gaikwad}, M. and {Hu}, H. and {Janssen}, G.~H. and {Karuppusamy}, R. and {Kramer}, M. and {Lee}, K.~J. and {Liu}, K. and {Mall}, G. and {McKee}, J.~W. and {Mickaliger}, M.~B. and {Perrodin}, D. and {Stappers}, B.~W. and {Tiburzi}, C. and {Wucknitz}, O. and {Wang}, L. and {Zhu}, W.~W.},
        title = "{Measuring interstellar delays of PSR J0613-0200 over 7 yr, using the Large European Array for Pulsars}",
      journal = {\mnras},
         year = 2020,
        month = nov,
       volume = {499},
       number = {1},
        pages = {1468-1479},
          doi = {10.1093/mnras/staa2955},
archivePrefix = {arXiv},
       eprint = {2009.10707},
 primaryClass = {astro-ph.HE},
       adsurl = {https://ui.adsabs.harvard.edu/abs/2020MNRAS.499.1468M}
}

@ARTICLE{2020ApJ...905L..34A,
       author = {{Arzoumanian}, Zaven and {Baker}, Paul T. and {Blumer}, Harsha and {B{\'e}csy}, Bence and {Brazier}, Adam and {Brook}, Paul R. and {Burke-Spolaor}, Sarah and {Chatterjee}, Shami and {Chen}, Siyuan and {Cordes}, James M. and {Cornish}, Neil J. and {Crawford}, Fronefield and {Cromartie}, H. Thankful and {Decesar}, Megan E. and {Demorest}, Paul B. and {Dolch}, Timothy and {Ellis}, Justin A. and {Ferrara}, Elizabeth C. and {Fiore}, William and {Fonseca}, Emmanuel and {Garver-Daniels}, Nathan and {Gentile}, Peter A. and {Good}, Deborah C. and {Hazboun}, Jeffrey S. and {Holgado}, A. Miguel and {Islo}, Kristina and {Jennings}, Ross J. and {Jones}, Megan L. and {Kaiser}, Andrew R. and {Kaplan}, David L. and {Kelley}, Luke Zoltan and {Key}, Joey Shapiro and {Laal}, Nima and {Lam}, Michael T. and {Lazio}, T. Joseph W. and {Lorimer}, Duncan R. and {Luo}, Jing and {Lynch}, Ryan S. and {Madison}, Dustin R. and {McLaughlin}, Maura A. and {Mingarelli}, Chiara M.~F. and {Ng}, Cherry and {Nice}, David J. and {Pennucci}, Timothy T. and {Pol}, Nihan S. and {Ransom}, Scott M. and {Ray}, Paul S. and {Shapiro-Albert}, Brent J. and {Siemens}, Xavier and {Simon}, Joseph and {Spiewak}, Ren{\'e}e and {Stairs}, Ingrid H. and {Stinebring}, Daniel R. and {Stovall}, Kevin and {Sun}, Jerry P. and {Swiggum}, Joseph K. and {Taylor}, Stephen R. and {Turner}, Jacob E. and {Vallisneri}, Michele and {Vigeland}, Sarah J. and {Witt}, Caitlin A. and {Nanograv Collaboration}},
        title = "{The NANOGrav 12.5 yr Data Set: Search for an Isotropic Stochastic Gravitational-wave Background}",
      journal = {\apjl},
         year = 2020,
        month = dec,
       volume = {905},
       number = {2},
          eid = {L34},
        pages = {L34},
          doi = {10.3847/2041-8213/abd401},
archivePrefix = {arXiv},
       eprint = {2009.04496},
 primaryClass = {astro-ph.HE},
       adsurl = {https://ui.adsabs.harvard.edu/abs/2020ApJ...905L..34A}
}

@article{10.1093/mnras/stt989,
    author = {Lewandowski, Wojciech and Dembska, Marta and Kijak, Jaroslaw and Kowalińska, Magdalena},
    title = "{Pulse broadening analysis for several new pulsars and anomalous scattering}",
    journal = {Monthly Notices of the Royal Astronomical Society},
    volume = {434},
    number = {1},
    pages = {69-83},
    year = {2013},
    month = {07},
    issn = {0035-8711},
    doi = {10.1093/mnras/stt989},
    url = {https://doi.org/10.1093/mnras/stt989},
    eprint = {https://academic.oup.com/mnras/article-pdf/434/1/69/18497796/stt989.pdf},
}

@article{Bansal_2019,
doi = {10.3847/1538-4357/ab0d8f},
url = {https://dx.doi.org/10.3847/1538-4357/ab0d8f},
year = {2019},
month = {apr},
publisher = {The American Astronomical Society},
volume = {875},
number = {2},
pages = {146},
author = {K. Bansal and G. B. Taylor and Kevin Stovall and Jayce Dowell},
title = {Scattering Study of Pulsars below 100 MHz Using LWA1},
journal = {The Astrophysical Journal}
}

@article{10.1093/mnras/220.1.19,
    author = {Romani, Roger W. and Narayan, Ramesh and Blandford, Roger},
    title = "{Refractive effects in pulsar scintillation}",
    journal = {Monthly Notices of the Royal Astronomical Society},
    volume = {220},
    number = {1},
    pages = {19-49},
    year = {1986},
    month = {05},
    issn = {0035-8711},
    doi = {10.1093/mnras/220.1.19},
    url = {https://doi.org/10.1093/mnras/220.1.19},
    eprint = {https://academic.oup.com/mnras/article-pdf/220/1/19/2894151/mnras220-0019.pdf},
}

@ARTICLE{1971ApJ...164..249L,
       author = {{Lang}, Kenneth R.},
        title = "{Interstellar Scintillation of Pulsar Radiation}",
      journal = {\apj},
         year = 1971,
        month = mar,
       volume = {164},
        pages = {249},
          doi = {10.1086/150836},
       adsurl = {https://ui.adsabs.harvard.edu/abs/1971ApJ...164..249L}
}

@ARTICLE{2017MNRAS.464.2075S,
       author = {{Shannon}, R.~M. and {Cordes}, J.~M.},
        title = "{Modelling and mitigating refractive propagation effects in precision pulsar timing observations}",
      journal = {\mnras},
         year = 2017,
        month = jan,
       volume = {464},
       number = {2},
        pages = {2075-2089},
          doi = {10.1093/mnras/stw2449},
archivePrefix = {arXiv},
       eprint = {1609.07573},
 primaryClass = {astro-ph.IM},
       adsurl = {https://ui.adsabs.harvard.edu/abs/2017MNRAS.464.2075S}
}

@ARTICLE{2022A&A...664A.116L,
       author = {{Liu}, Yulan and {Verbiest}, Joris P.~W. and {Main}, Robert A. and {Wu}, Ziwei and {Ambalappat}, Krishnakumar Moochickal and {Champion}, David J. and {Cognard}, Isma{\"e}l and {Guillemot}, Lucas and {Gaikwad}, Madhuri and {Janssen}, Gemma H. and {Kramer}, Michael and {Keith}, Michael J. and {Karuppusamy}, Ramesh and {K{\"u}nkel}, Lars and {Liu}, Kuo and {McKee}, James W. and {Mickaliger}, Mitchell B. and {Stappers}, Ben W. and {Shaifullah}, Golam. M. and {Theureau}, Gilles},
        title = "{Long-term scintillation studies of EPTA pulsars. I. Observations and basic results}",
      journal = {Astronomy & Astrophysics},
         year = 2022,
        month = aug,
       volume = {664},
          eid = {A116},
        pages = {A116},
          doi = {10.1051/0004-6361/202142552},
archivePrefix = {arXiv},
       eprint = {2203.16950},
 primaryClass = {astro-ph.HE},
       adsurl = {https://ui.adsabs.harvard.edu/abs/2022A&A...664A.116L}
}

@article{Brown2025,
author = {Brown, Grayce and Sheikh, Sofia and Cruz, Luigi and Farah, Wael and Gajjar, Vishal and Gilbertson, Christian and Grimaldo, Brandon and Lam, Michael and Marquez, Sofia and McLaughlin, Maura and Pollak, Alexander and Siemion, Andrew and Singh, Gurmehar},
year = {2025},
month = {12},
pages = {120},
title = {Long-term Monitoring of Scintillation in the Pulsar J0332+5434},
volume = {995},
journal = {The Astrophysical Journal},
doi = {10.3847/1538-4357/ae0fff}
}

@article{Lam_2016,
doi = {10.3847/0004-637X/819/2/155},
url = {https://doi.org/10.3847/0004-637X/819/2/155},
year = {2016},
month = {mar},
publisher = {The American Astronomical Society},
volume = {819},
number = {2},
pages = {155},
author = {Lam, M. T. and Cordes, J. M. and Chatterjee, S. and Arzoumanian, Z. and Crowter, K. and Demorest, P. B. and Dolch, T. and Ellis, J. A. and Ferdman, R. D. and Fonseca, E. F. and Gonzalez, M. E. and Jones, G. and Jones, M. L. and Levin, L. and Madison, D. R. and McLaughlin, M. A. and Nice, D. J. and Pennucci, T. T. and Ransom, S. M. and Siemens, X. and Stairs, I. H. and Stovall, K. and Swiggum, J. K. and Zhu, W. W.},
title = {THE NANOGRAV NINE-YEAR DATA SET: NOISE BUDGET FOR PULSAR ARRIVAL TIMES ON INTRADAY TIMESCALES},
journal = {The Astrophysical Journal}
}

@article{Lam_2016b,
doi = {10.3847/0004-637X/821/1/66},
url = {https://doi.org/10.3847/0004-637X/821/1/66},
year = {2016},
month = {apr},
publisher = {The American Astronomical Society},
volume = {821},
number = {1},
pages = {66},
author = {Lam, M. T. and Cordes, J. M. and Chatterjee, S. and Jones, M. L. and McLaughlin, M. A. and Armstrong, J. W.},
title = {SYSTEMATIC AND STOCHASTIC VARIATIONS IN PULSAR DISPERSION MEASURES},
journal = {The Astrophysical Journal}
}

@ARTICLE{1990ApJ...364..123F,
       author = {{Foster}, R.~S. and {Cordes}, J.~M.},
        title = "{Interstellar Propagation Effects and the Precision of Pulsar Timing}",
      journal = {\apj},
         year = 1990,
        month = nov,
       volume = {364},
        pages = {123},
          doi = {10.1086/169393},
       adsurl = {https://ui.adsabs.harvard.edu/abs/1990ApJ...364..123F}
}

@ARTICLE{1986ApJ...310..737C,
       author = {{Cordes}, J.~M. and {Pidwerbetsky}, A. and {Lovelace}, R.~V.~E.},
        title = "{Refractive and Diffractive Scattering in the Interstellar Medium}",
      journal = {\apj},
         year = 1986,
        month = nov,
       volume = {310},
        pages = {737},
          doi = {10.1086/164728},
       adsurl = {https://ui.adsabs.harvard.edu/abs/1986ApJ...310..737C}
}

@article{Cordes_2016,
doi = {10.3847/0004-637X/817/1/16},
url = {https://doi.org/10.3847/0004-637X/817/1/16},
year = {2016},
month = {jan},
publisher = {The American Astronomical Society},
volume = {817},
number = {1},
pages = {16},
author = {Cordes, J. M. and Shannon, R. M. and Stinebring, D. R.},
title = {FREQUENCY-DEPENDENT DISPERSION MEASURES AND IMPLICATIONS FOR PULSAR TIMING},
journal = {The Astrophysical Journal}
}

@article{Lam_2017,
doi = {10.3847/1538-4357/834/1/35},
url = {https://doi.org/10.3847/1538-4357/834/1/35},
year = {2016},
month = {dec},
publisher = {The American Astronomical Society},
volume = {834},
number = {1},
pages = {35},
author = {Lam, M. T. and Cordes, J. M. and Chatterjee, S. and Arzoumanian, Z. and Crowter, K. and Demorest, P. B. and Dolch, T. and Ellis, J. A and Ferdman, R. D. and Fonseca, E. and Gonzalez, M. E. and Jones, G. and Jones, M. L. and Levin, L. and Madison, D. R. and McLaughlin, M. A. and Nice, D. J. and Pennucci, T. T. and Ransom, S. M. and Shannon, R. M. and Siemens, X. and Stairs, I. H. and Stovall, K. and Swiggum, J. K. and Zhu, W. W.},
title = {THE NANOGRAV NINE-YEAR DATA SET: EXCESS NOISE IN MILLISECOND PULSAR ARRIVAL TIMES},
journal = {The Astrophysical Journal}
}

@ARTICLE{1985ApJ...288..221C,
       author = {{Cordes}, J.~M. and {Weisberg}, J.~M. and {Boriakoff}, V.},
        title = "{Small-scale electron density turbulence in the interstellar medium.}",
      journal = {\apj},
         year = 1985,
        month = jan,
       volume = {288},
        pages = {221-247},
          doi = {10.1086/162784},
       adsurl = {https://ui.adsabs.harvard.edu/abs/1985ApJ...288..221C}
}

@ARTICLE{NG12,
       author = {{Alam}, Md F. and {Arzoumanian}, Zaven and {Baker}, Paul T. and {Blumer}, Harsha and {Bohler}, Keith E. and {Brazier}, Adam and {Brook}, Paul R. and {Burke-Spolaor}, Sarah and {Caballero}, Keeisi and {Camuccio}, Richard S. and {Chamberlain}, Rachel L. and {Chatterjee}, Shami and {Cordes}, James M. and {Cornish}, Neil J. and {Crawford}, Fronefield and {Cromartie}, H. Thankful and {Decesar}, Megan E. and {Demorest}, Paul B. and {Dolch}, Timothy and {Ellis}, Justin A. and {Ferdman}, Robert D. and {Ferrara}, Elizabeth C. and {Fiore}, William and {Fonseca}, Emmanuel and {Garcia}, Yhamil and {Garver-Daniels}, Nathan and {Gentile}, Peter A. and {Good}, Deborah C. and {Gusdorff}, Jordan A. and {Halmrast}, Daniel and {Hazboun}, Jeffrey S. and {Islo}, Kristina and {Jennings}, Ross J. and {Jessup}, Cody and {Jones}, Megan L. and {Kaiser}, Andrew R. and {Kaplan}, David L. and {Kelley}, Luke Zoltan and {Key}, Joey Shapiro and {Lam}, Michael T. and {Lazio}, T. Joseph W. and {Lorimer}, Duncan R. and {Luo}, Jing and {Lynch}, Ryan S. and {Madison}, Dustin R. and {Maraccini}, Kaleb and {McLaughlin}, Maura A. and {Mingarelli}, Chiara M.~F. and {Ng}, Cherry and {Nguyen}, Benjamin M.~X. and {Nice}, David J. and {Pennucci}, Timothy T. and {Pol}, Nihan S. and {Ramette}, Joshua and {Ransom}, Scott M. and {Ray}, Paul S. and {Shapiro-Albert}, Brent J. and {Siemens}, Xavier and {Simon}, Joseph and {Spiewak}, Ren{\'e}e and {Stairs}, Ingrid H. and {Stinebring}, Daniel R. and {Stovall}, Kevin and {Swiggum}, Joseph K. and {Taylor}, Stephen R. and {Tripepi}, Michael and {Vallisneri}, Michele and {Vigeland}, Sarah J. and {Witt}, Caitlin A. and {Zhu}, Weiwei and {Nanograv Collaboration}},
        title = "{The NANOGrav 12.5 yr Data Set: Observations and Narrowband Timing of 47 Millisecond Pulsars}",
      journal = {\apjs},
         year = 2021,
        month = jan,
       volume = {252},
       number = {1},
          eid = {4},
        pages = {4},
          doi = {10.3847/1538-4365/abc6a0},
archivePrefix = {arXiv},
       eprint = {2005.06490},
 primaryClass = {astro-ph.HE},
       adsurl = {https://ui.adsabs.harvard.edu/abs/2021ApJS..252....4A}
}

@article{10.1093/mnras/sts486,
    author = {Keith, M. J. and Coles, W. and Shannon, R. M. and Hobbs, G. B. and Manchester, R. N. and Bailes, M. and Bhat, N. D. R. and Burke-Spolaor, S. and Champion, D. J. and Chaudhary, A. and Hotan, A. W. and Khoo, J. and Kocz, J. and Osłowski, S. and Ravi, V. and Reynolds, J. E. and Sarkissian, J. and van Straten, W. and Yardley, D. R. B.},
    title = {Measurement and correction of variations in interstellar dispersion in high-precision pulsar timing},
    journal = {Monthly Notices of the Royal Astronomical Society},
    volume = {429},
    number = {3},
    pages = {2161-2174},
    year = {2012},
    month = {12},
    issn = {0035-8711},
    doi = {10.1093/mnras/sts486},
    url = {https://doi.org/10.1093/mnras/sts486},
    eprint = {https://academic.oup.com/mnras/article-pdf/429/3/2161/3222740/sts486.pdf},
}

@ARTICLE{NG11,
       author = {{Arzoumanian}, Zaven and {Brazier}, Adam and {Burke-Spolaor}, Sarah and {Chamberlin}, Sydney and {Chatterjee}, Shami and {Christy}, Brian and {Cordes}, James M. and {Cornish}, Neil J. and {Crawford}, Fronefield and {Thankful Cromartie}, H. and {Crowter}, Kathryn and {DeCesar}, Megan E. and {Demorest}, Paul B. and {Dolch}, Timothy and {Ellis}, Justin A. and {Ferdman}, Robert D. and {Ferrara}, Elizabeth C. and {Fonseca}, Emmanuel and {Garver-Daniels}, Nathan and {Gentile}, Peter A. and {Halmrast}, Daniel and {Huerta}, E.~A. and {Jenet}, Fredrick A. and {Jessup}, Cody and {Jones}, Glenn and {Jones}, Megan L. and {Kaplan}, David L. and {Lam}, Michael T. and {Lazio}, T. Joseph W. and {Levin}, Lina and {Lommen}, Andrea and {Lorimer}, Duncan R. and {Luo}, Jing and {Lynch}, Ryan S. and {Madison}, Dustin and {Matthews}, Allison M. and {McLaughlin}, Maura A. and {McWilliams}, Sean T. and {Mingarelli}, Chiara and {Ng}, Cherry and {Nice}, David J. and {Pennucci}, Timothy T. and {Ransom}, Scott M. and {Ray}, Paul S. and {Siemens}, Xavier and {Simon}, Joseph and {Spiewak}, Ren{\'e}e and {Stairs}, Ingrid H. and {Stinebring}, Daniel R. and {Stovall}, Kevin and {Swiggum}, Joseph K. and {Taylor}, Stephen R. and {Vallisneri}, Michele and {van Haasteren}, Rutger and {Vigeland}, Sarah J. and {Zhu}, Weiwei and {NANOGrav Collaboration}},
        title = "{The NANOGrav 11-year Data Set: High-precision Timing of 45 Millisecond Pulsars}",
      journal = {\apjs},
         year = 2018,
        month = apr,
       volume = {235},
       number = {2},
          eid = {37},
        pages = {37},
          doi = {10.3847/1538-4365/aab5b0},
archivePrefix = {arXiv},
       eprint = {1801.01837},
 primaryClass = {astro-ph.HE},
       adsurl = {https://ui.adsabs.harvard.edu/abs/2018ApJS..235...37A}
}

@ARTICLE{NG9,
       author = {{NANOGrav Collaboration} and {Arzoumanian}, Zaven and {Brazier}, Adam and {Burke-Spolaor}, Sarah and {Chamberlin}, Sydney and {Chatterjee}, Shami and {Christy}, Brian and {Cordes}, James M. and {Cornish}, Neil and {Crowter}, Kathryn and {Demorest}, Paul B. and {Dolch}, Timothy and {Ellis}, Justin A. and {Ferdman}, Robert D. and {Fonseca}, Emmanuel and {Garver-Daniels}, Nathan and {Gonzalez}, Marjorie E. and {Jenet}, Fredrick A. and {Jones}, Glenn and {Jones}, Megan L. and {Kaspi}, Victoria M. and {Koop}, Michael and {Lam}, Michael T. and {Lazio}, T. Joseph W. and {Levin}, Lina and {Lommen}, Andrea N. and {Lorimer}, Duncan R. and {Luo}, Jing and {Lynch}, Ryan S. and {Madison}, Dustin and {McLaughlin}, Maura A. and {McWilliams}, Sean T. and {Nice}, David J. and {Palliyaguru}, Nipuni and {Pennucci}, Timothy T. and {Ransom}, Scott M. and {Siemens}, Xavier and {Stairs}, Ingrid H. and {Stinebring}, Daniel R. and {Stovall}, Kevin and {Swiggum}, Joseph K. and {Vallisneri}, Michele and {van Haasteren}, Rutger and {Wang}, Yan and {Zhu}, Weiwei},
        title = "{The NANOGrav Nine-year Data Set: Observations, Arrival Time Measurements, and Analysis of 37 Millisecond Pulsars}",
      journal = {\apj},
         year = 2015,
        month = nov,
       volume = {813},
       number = {1},
          eid = {65},
        pages = {65},
          doi = {10.1088/0004-637X/813/1/65},
archivePrefix = {arXiv},
       eprint = {1505.07540},
 primaryClass = {astro-ph.IM},
       adsurl = {https://ui.adsabs.harvard.edu/abs/2015ApJ...813...65N}
}

@ARTICLE{2016MNRAS.458.3341D,
       author = {{Desvignes}, G. and {Caballero}, R.~N. and {Lentati}, L. and {Verbiest}, J.~P.~W. and {Champion}, D.~J. and {Stappers}, B.~W. and {Janssen}, G.~H. and {Lazarus}, P. and {Os{\l}owski}, S. and {Babak}, S. and {Bassa}, C.~G. and {Brem}, P. and {Burgay}, M. and {Cognard}, I. and {Gair}, J.~R. and {Graikou}, E. and {Guillemot}, L. and {Hessels}, J.~W.~T. and {Jessner}, A. and {Jordan}, C. and {Karuppusamy}, R. and {Kramer}, M. and {Lassus}, A. and {Lazaridis}, K. and {Lee}, K.~J. and {Liu}, K. and {Lyne}, A.~G. and {McKee}, J. and {Mingarelli}, C.~M.~F. and {Perrodin}, D. and {Petiteau}, A. and {Possenti}, A. and {Purver}, M.~B. and {Rosado}, P.~A. and {Sanidas}, S. and {Sesana}, A. and {Shaifullah}, G. and {Smits}, R. and {Taylor}, S.~R. and {Theureau}, G. and {Tiburzi}, C. and {van Haasteren}, R. and {Vecchio}, A.},
        title = "{High-precision timing of 42 millisecond pulsars with the European Pulsar Timing Array}",
      journal = {\mnras},
         year = 2016,
        month = may,
       volume = {458},
       number = {3},
        pages = {3341-3380},
          doi = {10.1093/mnras/stw483},
archivePrefix = {arXiv},
       eprint = {1602.08511},
 primaryClass = {astro-ph.HE},
       adsurl = {https://ui.adsabs.harvard.edu/abs/2016MNRAS.458.3341D}
}

@ARTICLE{2013ApJ...762...94D,
       author = {{Demorest}, P.~B. and {Ferdman}, R.~D. and {Gonzalez}, M.~E. and {Nice}, D. and {Ransom}, S. and {Stairs}, I.~H. and {Arzoumanian}, Z. and {Brazier}, A. and {Burke-Spolaor}, S. and {Chamberlin}, S.~J. and {Cordes}, J.~M. and {Ellis}, J. and {Finn}, L.~S. and {Freire}, P. and {Giampanis}, S. and {Jenet}, F. and {Kaspi}, V.~M. and {Lazio}, J. and {Lommen}, A.~N. and {McLaughlin}, M. and {Palliyaguru}, N. and {Perrodin}, D. and {Shannon}, R.~M. and {Siemens}, X. and {Stinebring}, D. and {Swiggum}, J. and {Zhu}, W.~W.},
        title = "{Limits on the Stochastic Gravitational Wave Background from the North American Nanohertz Observatory for Gravitational Waves}",
      journal = {\apj},
         year = 2013,
        month = jan,
       volume = {762},
       number = {2},
          eid = {94},
        pages = {94},
          doi = {10.1088/0004-637X/762/2/94},
archivePrefix = {arXiv},
       eprint = {1201.6641},
 primaryClass = {astro-ph.CO},
       adsurl = {https://ui.adsabs.harvard.edu/abs/2013ApJ...762...94D}
}

@ARTICLE{2013PASA...30...17M,
       author = {{Manchester}, R.~N. and {Hobbs}, G. and {Bailes}, M. and {Coles}, W.~A. and {van Straten}, W. and {Keith}, M.~J. and {Shannon}, R.~M. and {Bhat}, N.~D.~R. and {Brown}, A. and {Burke-Spolaor}, S.~G. and {Champion}, D.~J. and {Chaudhary}, A. and {Edwards}, R.~T. and {Hampson}, G. and {Hotan}, A.~W. and {Jameson}, A. and {Jenet}, F.~A. and {Kesteven}, M.~J. and {Khoo}, J. and {Kocz}, J. and {Maciesiak}, K. and {Oslowski}, S. and {Ravi}, V. and {Reynolds}, J.~R. and {Sarkissian}, J.~M. and {Verbiest}, J.~P.~W. and {Wen}, Z.~L. and {Wilson}, W.~E. and {Yardley}, D. and {Yan}, W.~M. and {You}, X.~P.},
        title = "{The Parkes Pulsar Timing Array Project}",
      journal = {\pasa},
         year = 2013,
        month = jan,
       volume = {30},
          eid = {e017},
        pages = {e017},
          doi = {10.1017/pasa.2012.017},
archivePrefix = {arXiv},
       eprint = {1210.6130},
 primaryClass = {astro-ph.IM},
       adsurl = {https://ui.adsabs.harvard.edu/abs/2013PASA...30...17M}
}

@ARTICLE{2020PASA...37...20K,
       author = {{Kerr}, Matthew and {Reardon}, Daniel J. and {Hobbs}, George and {Shannon}, Ryan M. and {Manchester}, Richard N. and {Dai}, Shi and {Russell}, Christopher J. and {Zhang}, Songbo and {van Straten}, Willem and {Os{\l}owski}, Stefan and {Parthasarathy}, Aditya and {Spiewak}, Renee and {Bailes}, Matthew and {Bhat}, N.~D. Ramesh and {Cameron}, Andrew D. and {Coles}, William A. and {Dempsey}, James and {Deng}, Xinping and {Goncharov}, Boris and {Kaczmarek}, Jane F. and {Keith}, Michael J. and {Lasky}, Paul D. and {Lower}, Marcus E. and {Preisig}, Brett and {Sarkissian}, John Mihran and {Toomey}, Lawrence and {Wang}, Hongguang and {Wang}, Jingbo and {Zhang}, Lei and {Zhu}, Xingjiang},
        title = "{The Parkes Pulsar Timing Array project: second data release}",
      journal = {\pasa},
         year = 2020,
        month = jun,
       volume = {37},
          eid = {e020},
        pages = {e020},
          doi = {10.1017/pasa.2020.11},
archivePrefix = {arXiv},
       eprint = {2003.09780},
 primaryClass = {astro-ph.IM},
       adsurl = {https://ui.adsabs.harvard.edu/abs/2020PASA...37...20K}
}

@ARTICLE{2020PASA...37...12H,
       author = {{Hobbs}, George and {Manchester}, Richard N. and {Dunning}, Alex and {Jameson}, Andrew and {Roberts}, Paul and {George}, Daniel and {Green}, J.~A. and {Tuthill}, John and {Toomey}, Lawrence and {Kaczmarek}, Jane F. and {Mader}, Stacy and {Marquarding}, Malte and {Ahmed}, Azeem and {Amy}, Shaun W. and {Bailes}, Matthew and {Beresford}, Ron and {Bhat}, N.~D.~R. and {Bock}, Douglas C.-J. and {Bourne}, Michael and {Bowen}, Mark and {Brothers}, Michael and {Cameron}, Andrew D. and {Carretti}, Ettore and {Carter}, Nick and {Castillo}, Santy and {Chekkala}, Raji and {Cheng}, Wan and {Chung}, Yoon and {Craig}, Daniel A. and {Dai}, Shi and {Dawson}, Joanne and {Dempsey}, James and {Doherty}, Paul and {Dong}, Bin and {Edwards}, Philip and {Ergesh}, Tuohutinuer and {Gao}, Xuyang and {Han}, JinLin and {Hayman}, Douglas and {Indermuehle}, Balthasar and {Jeganathan}, Kanapathippillai and {Johnston}, Simon and {Kanoniuk}, Henry and {Kesteven}, Michael and {Kramer}, Michael and {Leach}, Mark and {Mcintyre}, Vince and {Moss}, Vanessa and {Os{\l}owski}, Stefan and {Phillips}, Chris and {Pope}, Nathan and {Preisig}, Brett and {Price}, Daniel and {Reeves}, Ken and {Reilly}, Les and {Reynolds}, John and {Robishaw}, Tim and {Roush}, Peter and {Ruckley}, Tim and {Sadler}, Elaine and {Sarkissian}, John and {Severs}, Sean and {Shannon}, Ryan and {Smart}, Ken and {Smith}, Malcolm and {Smith}, Stephanie and {Sobey}, Charlotte and {Staveley-Smith}, Lister and {Tzioumis}, Anastasios and {van Straten}, Willem and {Wang}, Nina and {Wen}, Linqing and {Whiting}, Matthew},
        title = "{An ultra-wide bandwidth (704 to 4 032 MHz) receiver for the Parkes radio telescope}",
      journal = {\pasa},
         year = 2020,
        month = apr,
       volume = {37},
          eid = {e012},
        pages = {e012},
          doi = {10.1017/pasa.2020.2},
archivePrefix = {arXiv},
       eprint = {1911.00656},
 primaryClass = {astro-ph.IM},
       adsurl = {https://ui.adsabs.harvard.edu/abs/2020PASA...37...12H}
}

@article{Curylo_2023,
doi = {10.3847/1538-4357/aca535},
url = {https://doi.org/10.3847/1538-4357/aca535},
year = {2023},
month = {feb},
publisher = {The American Astronomical Society},
volume = {944},
number = {2},
pages = {128},
author = {Curyło, Małgorzata and Pennucci, Timothy T. and Bailes, Matthew and Bhat, N. D. Ramesh and Cameron, Andrew D. and Dai, Shi and Hobbs, George and Kapur, Agastya and Manchester, Richard N. and Mandow, Rami and Miles, Matthew T. and Russell, Christopher J. and Reardon, Daniel J. and Shannon, Ryan M. and Spiewak, Renée and van Straten, Willem and Zhu, Xing-Jiang and Zic, Andrew},
title = {Wide-band Timing of the Parkes Pulsar Timing Array UWL Data},
journal = {The Astrophysical Journal}
}

@article{NG15_NB,
doi = {10.3847/2041-8213/acda88},
url = {https://doi.org/10.3847/2041-8213/acda88},
year = {2023},
month = {jun},
publisher = {The American Astronomical Society},
volume = {951},
number = {1},
pages = {L10},
author = {Agazie, Gabriella and Anumarlapudi, Akash and Archibald, Anne M. and Arzoumanian, Zaven and Baker, Paul T. and Bécsy, Bence and Blecha, Laura and Brazier, Adam and Brook, Paul R. and Burke-Spolaor, Sarah and Charisi, Maria and Chatterjee, Shami and Cohen, Tyler and Cordes, James M. and Cornish, Neil J. and Crawford, Fronefield and Cromartie, H. Thankful and Crowter, Kathryn and DeCesar, Megan E. and Demorest, Paul B. and Dolch, Timothy and Drachler, Brendan and Ferrara, Elizabeth C. and Fiore, William and Fonseca, Emmanuel and Freedman, Gabriel E. and Garver-Daniels, Nate and Gentile, Peter A. and Glaser, Joseph and Good, Deborah C. and Guertin, Lydia and Gültekin, Kayhan and Hazboun, Jeffrey S. and Jennings, Ross J. and Johnson, Aaron D. and Jones, Megan L. and Kaiser, Andrew R. and Kaplan, David L. and Kelley, Luke Zoltan and Kerr, Matthew and Key, Joey S. and Laal, Nima and Lam, Michael T. and Lamb, William G. and W. Lazio, T. Joseph and Lewandowska, Natalia and Liu, Tingting and Lorimer, Duncan R. and Luo, Jing and Lynch, Ryan S. and Ma, Chung-Pei and Madison, Dustin R. and McEwen, Alexander and McKee, James W. and McLaughlin, Maura A. and McMann, Natasha and Meyers, Bradley W. and Mingarelli, Chiara M. F. and Mitridate, Andrea and Ng, Cherry and Nice, David J. and Ocker, Stella Koch and Olum, Ken D. and Pennucci, Timothy T. and Perera, Benetge B. P. and Pol, Nihan S. and Radovan, Henri A. and Ransom, Scott M. and Ray, Paul S. and Romano, Joseph D. and Sardesai, Shashwat C. and Schmiedekamp, Ann and Schmiedekamp, Carl and Schmitz, Kai and Shapiro-Albert, Brent J. and Siemens, Xavier and Simon, Joseph and Siwek, Magdalena S. and Stairs, Ingrid H. and Stinebring, Daniel R. and Stovall, Kevin and Susobhanan, Abhimanyu and Swiggum, Joseph K. and Taylor, Stephen R. and Turner, Jacob E. and Unal, Caner and Vallisneri, Michele and Vigeland, Sarah J. and Wahl, Haley M. and Witt, Caitlin A. and Young, Olivia and The NANOGrav Collaboration},
title = {The NANOGrav 15 yr Data Set: Detector Characterization and Noise Budget},
journal = {The Astrophysical Journal Letters}
}

@article{Brook_2018,
doi = {10.3847/1538-4357/aae9e3},
url = {https://doi.org/10.3847/1538-4357/aae9e3},
year = {2018},
month = {nov},
publisher = {The American Astronomical Society},
volume = {868},
number = {2},
pages = {122},
author = {Brook, P. R. and Karastergiou, A. and McLaughlin, M. A. and Lam, M. T. and Arzoumanian, Z. and Chatterjee, S. and Cordes, J. M. and Crowter, K. and DeCesar, M. and Demorest, P. B. and Dolch, T. and Ellis, J. A. and Ferdman, R. D. and Ferrara, E. and Fonseca, E. and Gentile, P. A. and Jones, G. and Jones, M. L. and Lazio, T. J. W. and Levin, L. and Lorimer, D. R. and Lynch, R. S. and Ng, C. and Nice, D. J. and Pennucci, T. T. and Ransom, S. M. and Ray, P. S. and Spiewak, R. and Stairs, I. H. and Stinebring, D. R. and Stovall, K. and Swiggum, J. K. and Zhu, W. W.},
title = {The NANOGrav 11-year Data Set: Pulse Profile Variability},
journal = {The Astrophysical Journal}
}

@article{NG15_GWB,
doi = {10.3847/2041-8213/acdac6},
url = {https://doi.org/10.3847/2041-8213/acdac6},
year = {2023},
month = {jun},
publisher = {The American Astronomical Society},
volume = {951},
number = {1},
pages = {L8},
author = {Agazie, Gabriella and Anumarlapudi, Akash and Archibald, Anne M. and Arzoumanian, Zaven and Baker, Paul T. and Bécsy, Bence and Blecha, Laura and Brazier, Adam and Brook, Paul R. and Burke-Spolaor, Sarah and Burnette, Rand and Case, Robin and Charisi, Maria and Chatterjee, Shami and Chatziioannou, Katerina and Cheeseboro, Belinda D. and Chen, Siyuan and Cohen, Tyler and Cordes, James M. and Cornish, Neil J. and Crawford, Fronefield and Cromartie, H. Thankful and Crowter, Kathryn and Cutler, Curt J. and DeCesar, Megan E. and DeGan, Dallas and Demorest, Paul B. and Deng, Heling and Dolch, Timothy and Drachler, Brendan and Ellis, Justin A. and Ferrara, Elizabeth C. and Fiore, William and Fonseca, Emmanuel and Freedman, Gabriel E. and Garver-Daniels, Nate and Gentile, Peter A. and Gersbach, Kyle A. and Glaser, Joseph and Good, Deborah C. and Gültekin, Kayhan and Hazboun, Jeffrey S. and Hourihane, Sophie and Islo, Kristina and Jennings, Ross J. and Johnson, Aaron D. and Jones, Megan L. and Kaiser, Andrew R. and Kaplan, David L. and Kelley, Luke Zoltan and Kerr, Matthew and Key, Joey S. and Klein, Tonia C. and Laal, Nima and Lam, Michael T. and Lamb, William G. and W. Lazio, T. Joseph and Lewandowska, Natalia and Littenberg, Tyson B. and Liu, Tingting and Lommen, Andrea and Lorimer, Duncan R. and Luo, Jing and Lynch, Ryan S. and Ma, Chung-Pei and Madison, Dustin R. and Mattson, Margaret A. and McEwen, Alexander and McKee, James W. and McLaughlin, Maura A. and McMann, Natasha and Meyers, Bradley W. and Meyers, Patrick M. and Mingarelli, Chiara M. F. and Mitridate, Andrea and Natarajan, Priyamvada and Ng, Cherry and Nice, David J. and Ocker, Stella Koch and Olum, Ken D. and Pennucci, Timothy T. and Perera, Benetge B. P. and Petrov, Polina and Pol, Nihan S. and Radovan, Henri A. and Ransom, Scott M. and Ray, Paul S. and Romano, Joseph D. and Sardesai, Shashwat C. and Schmiedekamp, Ann and Schmiedekamp, Carl and Schmitz, Kai and Schult, Levi and Shapiro-Albert, Brent J. and Siemens, Xavier and Simon, Joseph and Siwek, Magdalena S. and Stairs, Ingrid H. and Stinebring, Daniel R. and Stovall, Kevin and Sun, Jerry P. and Susobhanan, Abhimanyu and Swiggum, Joseph K. and Taylor, Jacob and Taylor, Stephen R. and Turner, Jacob E. and Unal, Caner and Vallisneri, Michele and van Haasteren, Rutger and Vigeland, Sarah J. and Wahl, Haley M. and Wang, Qiaohong and Witt, Caitlin A. and Young, Olivia and The NANOGrav Collaboration},
title = {The NANOGrav 15 yr Data Set: Evidence for a Gravitational-wave Background},
journal = {The Astrophysical Journal Letters}
}

@article{Pennucci_2014,
doi = {10.1088/0004-637X/790/2/93},
url = {https://doi.org/10.1088/0004-637X/790/2/93},
year = {2014},
month = {jul},
publisher = {The American Astronomical Society},
volume = {790},
number = {2},
pages = {93},
author = {Pennucci, Timothy T. and Demorest, Paul B. and Ransom, Scott M.},
title = {ELEMENTARY WIDEBAND TIMING OF RADIO PULSARS},
journal = {The Astrophysical Journal}
}

@ARTICLE{1968Natur.220..676C,
       author = {{Craft}, H.~D. and {Comella}, J.~M.},
        title = "{Frequency Dependent Pulse Widths for CP 1133}",
      journal = {\nat},
         year = 1968,
        month = nov,
       volume = {220},
       number = {5168},
        pages = {676-678},
          doi = {10.1038/220676a0},
       adsurl = {https://ui.adsabs.harvard.edu/abs/1968Natur.220..676C}
}

@article{10.1093/mnras/stu2262,
    author = {Dyks, J. and Rudak, B.},
    title = {The origin of the frequency-dependent behaviour of pulsar radio profiles},
    journal = {Monthly Notices of the Royal Astronomical Society},
    volume = {446},
    number = {3},
    pages = {2505-2522},
    year = {2014},
    month = {11},
    issn = {0035-8711},
    doi = {10.1093/mnras/stu2262},
    url = {https://doi.org/10.1093/mnras/stu2262},
    eprint = {https://academic.oup.com/mnras/article-pdf/446/3/2505/10447877/stu2262.pdf},
}

@article{10.1093/mnras/stad070,
    author = {Oswald, L S and Johnston, S and Karastergiou, A and Dai, S and Kerr, M and Lower, M E and Manchester, R N and Shannon, R M and Sobey, C and Weltevrede, P},
    title = {Pulsar polarization: a broad-band population view with the Parkes Ultra-Wideband receiver},
    journal = {Monthly Notices of the Royal Astronomical Society},
    volume = {520},
    number = {4},
    pages = {4961-4980},
    year = {2023},
    month = {01},
    issn = {0035-8711},
    doi = {10.1093/mnras/stad070},
    url = {https://doi.org/10.1093/mnras/stad070},
    eprint = {https://academic.oup.com/mnras/article-pdf/520/4/4961/49319316/stad070.pdf},
}

@ARTICLE{2025A&A...704A.214J,
       author = {{Jaroenjittichai}, P. and {Johnston}, S. and {Dai}, S. and {Kerr}, M. and {Lower}, M.~E. and {Manchester}, R.~N. and {Oswald}, L.~S. and {Shannon}, R.~M. and {Sobey}, C. and {Weltevrede}, P.},
        title = "{Frequency evolution of pulsar emission: Further evidence for the fan beam model}",
      journal = {\aap},
         year = 2025,
        month = dec,
       volume = {704},
          eid = {A214},
        pages = {A214},
          doi = {10.1051/0004-6361/202555535},
archivePrefix = {arXiv},
       eprint = {2509.11945},
 primaryClass = {astro-ph.HE},
       adsurl = {https://ui.adsabs.harvard.edu/abs/2025A&A...704A.214J}
}

@ARTICLE{2017MNRAS.468.4389S,
       author = {{Shang}, Lun-Hua and {Lu}, Ji-Guang and {Du}, Yuan-Jie and {Hao}, Long-Fei and {Li}, Di and {Lee}, Ke-Jia and {Li}, Bin and {Li}, Li-Xin and {Qiao}, Guo-Jun and {Shen}, Zhi-Qiang and {Wang}, De-Hua and {Wang}, Min and {Wu}, Xin-Ji and {Wu}, Ya-Jun and {Xu}, Ren-Xin and {Yue}, You-Ling and {Yan}, Zhen and {Zhi}, Qi-Jun and {Zhao}, Rong-Bing and {Zhao}, Ru-Shuang},
        title = "{Investigating the multifrequency pulse profiles of PSRs B0329+54 and B1642-03 in an inverse Compton scattering model}",
      journal = {\mnras},
         year = 2017,
        month = jul,
       volume = {468},
       number = {4},
        pages = {4389-4398},
          doi = {10.1093/mnras/stx815},
archivePrefix = {arXiv},
       eprint = {1703.03582},
 primaryClass = {astro-ph.HE},
       adsurl = {https://ui.adsabs.harvard.edu/abs/2017MNRAS.468.4389S}
}

@article{10.1111/j.1365-2966.2011.19578.x,
    author = {Osłowski, S. and van Straten, W. and Hobbs, G. B. and Bailes, M. and Demorest, P.},
    title = {High signal-to-noise ratio observations and the ultimate limits of precision pulsar timing},
    journal = {Monthly Notices of the Royal Astronomical Society},
    volume = {418},
    number = {2},
    pages = {1258-1271},
    year = {2011},
    month = {11},
    issn = {0035-8711},
    doi = {10.1111/j.1365-2966.2011.19578.x},
    url = {https://doi.org/10.1111/j.1365-2966.2011.19578.x},
    eprint = {https://academic.oup.com/mnras/article-pdf/418/2/1258/3717760/mnras0418-1258.pdf},
}

@article{10.1111/j.1365-2966.2011.19452.x,
    author = {Liu, K. and Verbiest, J. P. W. and Kramer, M. and Stappers, B. W. and van Straten, W. and Cordes, J. M.},
    title = {Prospects for high-precision pulsar timing},
    journal = {Monthly Notices of the Royal Astronomical Society},
    volume = {417},
    number = {4},
    pages = {2916-2926},
    year = {2011},
    month = {11},
    issn = {0035-8711},
    doi = {10.1111/j.1365-2966.2011.19452.x},
    url = {https://doi.org/10.1111/j.1365-2966.2011.19452.x},
    eprint = {https://academic.oup.com/mnras/article-pdf/417/4/2916/3839017/mnras0417-2916.pdf},
}

@article{10.1093/mnras/stx580,
    author = {Lentati, L. and Kerr, M. and Dai, S. and Shannon, R. M. and Hobbs, G. and Osłowski, S.},
    title = {Robust estimation of scattering in pulsar timing analysis},
    journal = {Monthly Notices of the Royal Astronomical Society},
    volume = {468},
    number = {2},
    pages = {1474-1485},
    year = {2017},
    month = {03},
    issn = {0035-8711},
    doi = {10.1093/mnras/stx580},
    url = {https://doi.org/10.1093/mnras/stx580},
    eprint = {https://academic.oup.com/mnras/article-pdf/468/2/1474/11127052/stx580.pdf},
}

@article{Pennucci_2019,
doi = {10.3847/1538-4357/aaf6ef},
url = {https://doi.org/10.3847/1538-4357/aaf6ef},
year = {2019},
month = {jan},
publisher = {The American Astronomical Society},
volume = {871},
number = {1},
pages = {34},
author = {Pennucci, Timothy T.},
title = {Frequency-dependent Template Profiles for High-precision Pulsar Timing},
journal = {The Astrophysical Journal}
}

@article{Xu_2021,
doi = {10.3847/1538-4357/ac0b40},
url = {https://doi.org/10.3847/1538-4357/ac0b40},
year = {2021},
month = {aug},
publisher = {The American Astronomical Society},
volume = {917},
number = {2},
pages = {108},
author = {Xu, X. and Shang, L. H. and Zhi, Q. J. and Qiao, G. J. and Dang, S. J. and Bai, J. T. and Zhao, R. S. and Lu, J. G. and Dong, A. J. and Lin, Q. W. and Zhang, D. D. and Yang, H.},
title = {A Systematic Study of the Frequency Evolution Behavior of Pulsar Pulse Profiles},
journal = {The Astrophysical Journal}
}

@article{10.1093/mnras/stu1213,
    author = {Shannon, R. M. and Osłowski, S. and Dai, S. and Bailes, M. and Hobbs, G. and Manchester, R. N. and van Straten, W. and Raithel, C. A. and Ravi, V. and Toomey, L. and Bhat, N. D. R. and Burke-Spolaor, S. and Coles, W. A. and Keith, M. J. and Kerr, M. and Levin, Y. and Sarkissian, J. M. and Wang, J.-B. and Wen, L. and Zhu, X.-J.},
    title = {Limitations in timing precision due to single-pulse shape variability in millisecond pulsars},
    journal = {Monthly Notices of the Royal Astronomical Society},
    volume = {443},
    number = {2},
    pages = {1463-1481},
    year = {2014},
    month = {07},
    issn = {0035-8711},
    doi = {10.1093/mnras/stu1213},
    url = {https://doi.org/10.1093/mnras/stu1213},
    eprint = {https://academic.oup.com/mnras/article-pdf/443/2/1463/3654347/stu1213.pdf},
}

@article{Shannon_2010,
doi = {10.1088/0004-637X/725/2/1607},
url = {https://dx.doi.org/10.1088/0004-637X/725/2/1607},
year = {2010},
month = {nov},
publisher = {The American Astronomical Society},
volume = {725},
number = {2},
pages = {1607},
author = {Shannon, Ryan M. and Cordes, James M.},
title = {ASSESSING THE ROLE OF SPIN NOISE IN THE PRECISION TIMING OF MILLISECOND PULSARS},
journal = {The Astrophysical Journal}
}

@ARTICLE{2019ApJ...872..150M,
       author = {{Madison}, D.~R. and {Cordes}, J.~M. and {Arzoumanian}, Z. and {Chatterjee}, S. and {Crowter}, K. and {DeCesar}, M.~E. and {Demorest}, P.~B. and {Dolch}, T. and {Ellis}, J.~A. and {Ferdman}, R.~D. and {Ferrara}, E.~C. and {Fonseca}, E. and {Gentile}, P.~A. and {Jones}, G. and {Jones}, M.~L. and {Lam}, M.~T. and {Levin}, L. and {Lorimer}, D.~R. and {Lynch}, R.~S. and {McLaughlin}, M.~A. and {Mingarelli}, C.~M.~F. and {Ng}, C. and {Nice}, D.~J. and {Pennucci}, T.~T. and {Ransom}, S.~M. and {Ray}, P.~S. and {Spiewak}, R. and {Stairs}, I.~H. and {Stovall}, K. and {Swiggum}, J.~K. and {Zhu}, W.~W.},
        title = "{The NANOGrav 11 yr Data Set: Solar Wind Sounding through Pulsar Timing}",
      journal = {\apj},
         year = 2019,
        month = feb,
       volume = {872},
       number = {2},
          eid = {150},
        pages = {150},
          doi = {10.3847/1538-4357/ab01fd},
archivePrefix = {arXiv},
       eprint = {1808.07078},
 primaryClass = {astro-ph.SR},
       adsurl = {https://ui.adsabs.harvard.edu/abs/2019ApJ...872..150M}
}

@article{Jones_2017,
doi = {10.3847/1538-4357/aa73df},
url = {https://doi.org/10.3847/1538-4357/aa73df},
year = {2017},
month = {jun},
publisher = {The American Astronomical Society},
volume = {841},
number = {2},
pages = {125},
author = {Jones, M. L. and McLaughlin, M. A. and Lam, M. T. and Cordes, J. M. and Levin, L. and Chatterjee, S. and Arzoumanian, Z. and Crowter, K. and Demorest, P. B. and Dolch, T. and Ellis, J. A and Ferdman, R. D. and Fonseca, E. and Gonzalez, M. E. and Jones, G. and Lazio, T. J. W. and Nice, D. J. and Pennucci, T. T. and Ransom, S. M. and Stinebring, D. R. and Stairs, I. H. and Stovall, K. and Swiggum, J. K. and Zhu, W. W.},
title = {The NANOGrav Nine-year Data Set: Measurement and Analysis of Variations in Dispersion Measures},
journal = {The Astrophysical Journal}
}

@article{Sosa_Fiscella_2024,
doi = {10.3847/1538-4357/ad2858},
url = {https://dx.doi.org/10.3847/1538-4357/ad2858},
year = {2024},
month = {apr},
publisher = {The American Astronomical Society},
volume = {966},
number = {1},
pages = {95},
author = {Sosa Fiscella, Sophia Valentina and Lam, Michael T. and Arzoumanian, Zaven and Blumer, Harsha and Brook, Paul R. and Cromartie, H. Thankful and DeCesar, Megan E. and Demorest, Paul B. and Dolch, Timothy and Ellis, Justin A. and Ferdman, Robert D. and Ferrara, Elizabeth C. and Fonseca, Emmanuel and Garver-Daniels, Nate and Gentile, Peter A. and Good, Deborah C. and Jones, Megan L. and Lorimer, Duncan R. and Luo, Jing and Lynch, Ryan S. and McLaughlin, Maura A. and Ng, Cherry and Nice, David J. and Pennucci, Timothy T. and Pol, Nihan S. and Ransom, Scott M. and Spiewak, Renée and Stairs, Ingrid H. and Stovall, Kevin and Swiggum, Joseph K. and Vigeland, Sarah J.},
title = {The NANOGrav 12.5-Year Data Set: Dispersion Measure Misestimations with Varying Bandwidths},
journal = {The Astrophysical Journal}
}

@ARTICLE{2025ApJ...986..191G,
       author = {{Geiger}, Abra and {Cordes}, James M. and {Lam}, Michael T. and {Ocker}, Stella Koch and {Chatterjee}, Shami and {Arzoumanian}, Zaven and {Battaglia}, Ava L. and {Blumer}, Harsha and {Brook}, Paul R. and {Combs}, Olivia A. and {Cromartie}, H. Thankful and {DeCesar}, Megan E. and {Demorest}, Paul B. and {Dolch}, Timothy and {Ellis}, Justin A. and {Ferdman}, Robert D. and {Ferrara}, Elizabeth C. and {Fonseca}, Emmanuel and {Garver-Daniels}, Nate and {Gentile}, Peter A. and {Good}, Deborah C. and {Jones}, Megan L. and {Lorimer}, Duncan R. and {Luo}, Jing and {Lynch}, Ryan S. and {McLaughlin}, Maura A. and {Ng}, Cherry and {Nice}, David J. and {Pennucci}, Timothy T. and {Pol}, Nihan S. and {Ransom}, Scott M. and {Spiewak}, Ren{\'e}e and {Stairs}, Ingrid H. and {Stovall}, Kevin and {Swiggum}, Joseph K. and {Vigeland}, Sarah J.},
        title = "{The NANOGrav 12.5 yr Data Set: Probing Interstellar Turbulence and Precision Pulsar Timing with PSR J1903+0327}",
      journal = {\apj},
         year = 2025,
        month = jun,
       volume = {986},
       number = {2},
          eid = {191},
        pages = {191},
          doi = {10.3847/1538-4357/add0b6},
archivePrefix = {arXiv},
       eprint = {2411.08191},
 primaryClass = {astro-ph.HE},
       adsurl = {https://ui.adsabs.harvard.edu/abs/2025ApJ...986..191G}
}

@ARTICLE{2024A&A...689A.191Z,
       author = {{Zubieta}, E. and {Garc{\'\i}a}, F. and {del Palacio}, S. and {Araujo Furlan}, S.~B. and {Gancio}, G. and {Lousto}, C.~O. and {Combi}, J.~A. and {Espinoza}, C.~M.},
        title = "{Timing irregularities and glitches from the pulsar monitoring campaign at IAR}",
      journal = {\aap},
         year = 2024,
        month = sep,
       volume = {689},
          eid = {A191},
        pages = {A191},
          doi = {10.1051/0004-6361/202450441},
archivePrefix = {arXiv},
       eprint = {2406.17099},
 primaryClass = {astro-ph.HE},
       adsurl = {https://ui.adsabs.harvard.edu/abs/2024A&A...689A.191Z}
}

@article{Lam_2019,
doi = {10.3847/1538-4357/ab01cd},
url = {https://doi.org/10.3847/1538-4357/ab01cd},
year = {2019},
month = {feb},
publisher = {The American Astronomical Society},
volume = {872},
number = {2},
pages = {193},
author = {Lam, M. T. and McLaughlin, M. A. and Arzoumanian, Z. and Blumer, H. and Brook, P. R. and Cromartie, H. T. and Demorest, P. B. and DeCesar, M. E. and Dolch, T. and Ellis, J. A. and Ferdman, R. D. and Ferrara, E. C. and Fonseca, E. and Garver-Daniels, N. and Gentile, P. A. and Jones, M. L. and Lorimer, D. R. and Lynch, R. S. and Ng, C. and Nice, D. J. and Pennucci, T. T. and Ransom, S. M. and Spiewak, R. and Stairs, I. H. and Stovall, K. and Swiggum, J. K. and Vigeland, S. J. and Zhu, W. W.},
title = {The NANOGrav 12.5 yr Data Set: The Frequency Dependence of Pulse Jitter in Precision Millisecond Pulsars},
journal = {The Astrophysical Journal}
}

@article{10.1093/mnras/stu1420,
    author = {Liu, K. and Desvignes, G. and Cognard, I. and Stappers, B. W. and Verbiest, J. P. W. and Lee, K. J. and Champion, D. J. and Kramer, M. and Freire, P. C. C. and Karuppusamy, R.},
    title = {Measuring pulse times of arrival from broad-band pulsar observations},
    journal = {Monthly Notices of the Royal Astronomical Society},
    volume = {443},
    number = {4},
    pages = {3752-3760},
    year = {2014},
    month = {08},
    issn = {0035-8711},
    doi = {10.1093/mnras/stu1420},
    url = {https://doi.org/10.1093/mnras/stu1420},
    eprint = {https://academic.oup.com/mnras/article-pdf/443/4/3752/6297371/stu1420.pdf},
}

@ARTICLE{2013MNRAS.428.1147V,
       author = {{van Haasteren}, Rutger and {Levin}, Yuri},
        title = "{Understanding and analysing time-correlated stochastic signals in pulsar timing}",
      journal = {\mnras},
         year = 2013,
        month = jan,
       volume = {428},
       number = {2},
        pages = {1147-1159},
          doi = {10.1093/mnras/sts097},
archivePrefix = {arXiv},
       eprint = {1202.5932},
 primaryClass = {astro-ph.IM},
       adsurl = {https://ui.adsabs.harvard.edu/abs/2013MNRAS.428.1147V}
}

@ARTICLE{2023A&A...678A..48E,
       author = {{EPTA Collaboration} and {Antoniadis}, J. and {Babak}, S. and {Bak Nielsen}, A.-S. and {Bassa}, C.~G. and {Berthereau}, A. and {Bonetti}, M. and {Bortolas}, E. and {Brook}, P.~R. and {Burgay}, M. and {Caballero}, R.~N. and {Chalumeau}, A. and {Champion}, D.~J. and {Chanlaridis}, S. and {Chen}, S. and {Cognard}, I. and {Desvignes}, G. and {Falxa}, M. and {Ferdman}, R.~D. and {Franchini}, A. and {Gair}, J.~R. and {Goncharov}, B. and {Graikou}, E. and {Grie{\ss}meier}, J.-M. and {Guillemot}, L. and {Guo}, Y.~J. and {Hu}, H. and {Iraci}, F. and {Izquierdo-Villalba}, D. and {Jang}, J. and {Jawor}, J. and {Janssen}, G.~H. and {Jessner}, A. and {Karuppusamy}, R. and {Keane}, E.~F. and {Keith}, M.~J. and {Kramer}, M. and {Krishnakumar}, M.~A. and {Lackeos}, K. and {Lee}, K.~J. and {Liu}, K. and {Liu}, Y. and {Lyne}, A.~G. and {McKee}, J.~W. and {Main}, R.~A. and {Mickaliger}, M.~B. and {Ni{\c{t}}u}, I.~C. and {Parthasarathy}, A. and {Perera}, B.~B.~P. and {Perrodin}, D. and {Petiteau}, A. and {Porayko}, N.~K. and {Possenti}, A. and {Quelquejay Leclere}, H. and {Samajdar}, A. and {Sanidas}, S.~A. and {Sesana}, A. and {Shaifullah}, G. and {Speri}, L. and {Spiewak}, R. and {Stappers}, B.~W. and {Susarla}, S.~C. and {Theureau}, G. and {Tiburzi}, C. and {van der Wateren}, E. and {Vecchio}, A. and {Venkatraman Krishnan}, V. and {Verbiest}, J.~P.~W. and {Wang}, J. and {Wang}, L. and {Wu}, Z.},
        title = "{The second data release from the European Pulsar Timing Array. I. The dataset and timing analysis}",
      journal = {\aap},
         year = 2023,
        month = oct,
       volume = {678},
          eid = {A48},
        pages = {A48},
          doi = {10.1051/0004-6361/202346841},
archivePrefix = {arXiv},
       eprint = {2306.16224},
 primaryClass = {astro-ph.HE},
       adsurl = {https://ui.adsabs.harvard.edu/abs/2023A&A...678A..48E}
}

@ARTICLE{2023PASA...40...49Z,
       author = {{Zic}, Andrew and {Reardon}, Daniel J. and {Kapur}, Agastya and {Hobbs}, George and {Mandow}, Rami and {Cury{\l}o}, Ma{\l}gorzata and {Shannon}, Ryan M. and {Askew}, Jacob and {Bailes}, Matthew and {Bhat}, N.~D. Ramesh and {Cameron}, Andrew and {Chen}, Zu-Cheng and {Dai}, Shi and {Di Marco}, Valentina and {Feng}, Yi and {Kerr}, Matthew and {Kulkarni}, Atharva and {Lower}, Marcus E. and {Luo}, Rui and {Manchester}, Richard N. and {Miles}, Matthew T. and {Nathan}, Rowina S. and {Os{\l}owski}, Stefan and {Rogers}, Axl F. and {Russell}, Christopher J. and {Sarkissian}, John M. and {Shamohammadi}, Mohsen and {Spiewak}, Ren{\'e}e and {Thyagarajan}, Nithyanandan and {Toomey}, Lawrence and {Wang}, Shuangqiang and {Zhang}, Lei and {Zhang}, Songbo and {Zhu}, Xing-Jiang},
        title = "{The Parkes Pulsar Timing Array third data release}",
      journal = {\pasa},
         year = 2023,
        month = dec,
       volume = {40},
          eid = {e049},
        pages = {e049},
          doi = {10.1017/pasa.2023.36},
archivePrefix = {arXiv},
       eprint = {2306.16230},
 primaryClass = {astro-ph.HE},
       adsurl = {https://ui.adsabs.harvard.edu/abs/2023PASA...40...49Z}
}

@INPROCEEDINGS{2023AAS...24120907S,
       author = {{Starkman}, Taylor},
        title = "{Modeling Time of Arrival Delay for Pulsar Pulses with Hermite Polynomials}",
    booktitle = {American Astronomical Society Meeting Abstracts \#241},
         year = 2023,
       series = {American Astronomical Society Meeting Abstracts},
       volume = {241},
        month = jan,
          eid = {209.07},
        pages = {209.07},
       adsurl = {https://ui.adsabs.harvard.edu/abs/2023AAS...24120907S}
}

@article{10.1145/3447755,
author = {Gewers, Felipe L. and Ferreira, Gustavo R. and Arruda, Henrique F. De and Silva, Filipi N. and Comin, Cesar H. and Amancio, Diego R. and Costa, Luciano Da F.},
title = {Principal Component Analysis: A Natural Approach to Data Exploration},
year = {2021},
issue_date = {May 2022},
publisher = {Association for Computing Machinery},
address = {New York, NY, USA},
volume = {54},
number = {4},
issn = {0360-0300},
url = {https://doi.org/10.1145/3447755},
doi = {10.1145/3447755},
journal = {ACM Comput. Surv.},
month = may,
articleno = {70},
numpages = {34}
}

@book{AbramowitzStegun1965,
  author    = {Abramowitz, Milton and Stegun, Irene A.},
  title     = {Handbook of Mathematical Functions with Formulas, Graphs, and Mathematical Tables},
  publisher = {Dover Publications},
  address   = {New York},
  year      = {1965},
  edition   = {9th},
  series    = {Applied Mathematics Series},
  volume    = {55}
}

@ARTICLE{2020SciPy-NMeth,
  author  = {Virtanen, Pauli and Gommers, Ralf and Oliphant, Travis E. and
            Haberland, Matt and Reddy, Tyler and Cournapeau, David and
            Burovski, Evgeni and Peterson, Pearu and Weckesser, Warren and
            Bright, Jonathan and {van der Walt}, St{\'e}fan J. and
            Brett, Matthew and Wilson, Joshua and Millman, K. Jarrod and
            Mayorov, Nikolay and Nelson, Andrew R. J. and Jones, Eric and
            Kern, Robert and Larson, Eric and Carey, C J and
            Polat, {\.I}lhan and Feng, Yu and Moore, Eric W. and
            {VanderPlas}, Jake and Laxalde, Denis and Perktold, Josef and
            Cimrman, Robert and Henriksen, Ian and Quintero, E. A. and
            Harris, Charles R. and Archibald, Anne M. and
            Ribeiro, Ant{\^o}nio H. and Pedregosa, Fabian and
            {van Mulbregt}, Paul and {SciPy 1.0 Contributors}},
  title   = {{{SciPy} 1.0: Fundamental Algorithms for Scientific
            Computing in Python}},
  journal = {Nature Methods},
  year    = {2020},
  volume  = {17},
  pages   = {261--272},
  adsurl  = {https://rdcu.be/b08Wh},
  doi     = {10.1038/s41592-019-0686-2},
}

@MISC{2017ascl.soft06011L,
       author = {{Lam}, M.~T.},
        title = "{PyPulse: PSRFITS handler}",
 howpublished = {Astrophysics Source Code Library, record ascl:1706.011},
         year = 2017,
        month = jun,
          eid = {ascl:1706.011},
        pages = {ascl:1706.011},
archivePrefix = {ascl},
       eprint = {1706.011},
       adsurl = {https://ui.adsabs.harvard.edu/abs/2017ascl.soft06011L}
}

@incollection{brent1973algorithm,
  author    = {Brent, Richard P.},
  title     = {An Algorithm with Guaranteed Convergence for Finding a Zero of a Function},
  booktitle = {Algorithms for Minimization without Derivatives},
  chapter   = {4},
  publisher = {Prentice-Hall},
  address   = {Englewood Cliffs, NJ},
  year      = {1973},
  isbn      = {0-13-022335-2}
}

@article{aic,
author = {Kenneth P. Burnham and David R. Anderson},
title ={Multimodel Inference: Understanding AIC and BIC in Model Selection},
journal = {Sociological Methods \& Research},
volume = {33},
number = {2},
pages = {261-304},
year = {2004},
doi = {10.1177/0049124104268644},
URL = {https://doi.org/10.1177/0049124104268644},
eprint = {https://doi.org/10.1177/0049124104268644}
}

@article{Alam_2021,
doi = {10.3847/1538-4365/abc6a1},
url = {https://doi.org/10.3847/1538-4365/abc6a1},
year = {2020},
month = {dec},
publisher = {The American Astronomical Society},
volume = {252},
number = {1},
pages = {5},
author = {Alam, Md F. and Arzoumanian, Zaven and Baker, Paul T. and Blumer, Harsha and Bohler, Keith E. and Brazier, Adam and Brook, Paul R. and Burke-Spolaor, Sarah and Caballero, Keeisi and Camuccio, Richard S. and Chamberlain, Rachel L. and Chatterjee, Shami and Cordes, James M. and Cornish, Neil J. and Crawford, Fronefield and Cromartie, H. Thankful and DeCesar, Megan E. and Demorest, Paul B. and Dolch, Timothy and Ellis, Justin A. and Ferdman, Robert D. and Ferrara, Elizabeth C. and Fiore, William and Fonseca, Emmanuel and Garcia, Yhamil and Garver-Daniels, Nathan and Gentile, Peter A. and Good, Deborah C. and Gusdorff, Jordan A. and Halmrast, Daniel and Hazboun, Jeffrey S. and Islo, Kristina and Jennings, Ross J. and Jessup, Cody and Jones, Megan L. and Kaiser, Andrew R. and Kaplan, David L. and Kelley, Luke Zoltan and Key, Joey Shapiro and Lam, Michael T. and W. Lazio, T. Joseph and Lorimer, Duncan R. and Luo, Jing and Lynch, Ryan S. and Madison, Dustin R. and Maraccini, Kaleb and McLaughlin, Maura A. and Mingarelli, Chiara M. F. and Ng, Cherry and Nguyen, Benjamin M. X. and Nice, David J. and Pennucci, Timothy T. and Pol, Nihan S. and Ramette, Joshua and Ransom, Scott M. and Ray, Paul S. and Shapiro-Albert, Brent J. and Siemens, Xavier and Simon, Joseph and Spiewak, Renée and Stairs, Ingrid H. and Stinebring, Daniel R. and Stovall, Kevin and Swiggum, Joseph K. and Taylor, Stephen R. and Tripepi, Michael and Vallisneri, Michele and Vigeland, Sarah J. and Witt, Caitlin A. and Zhu, Weiwei and (The NANOGrav Collaboration)},
title = {The NANOGrav 12.5 yr Data Set: Wideband Timing of 47 Millisecond Pulsars},
journal = {The Astrophysical Journal Supplement Series}
}

@ARTICLE{2018ApJ...868..122B,
       author = {{Brook}, P.~R. and {Karastergiou}, A. and {McLaughlin}, M.~A. and {Lam}, M.~T. and {Arzoumanian}, Z. and {Chatterjee}, S. and {Cordes}, J.~M. and {Crowter}, K. and {DeCesar}, M. and {Demorest}, P.~B. and {Dolch}, T. and {Ellis}, J.~A. and {Ferdman}, R.~D. and {Ferrara}, E. and {Fonseca}, E. and {Gentile}, P.~A. and {Jones}, G. and {Jones}, M.~L. and {Lazio}, T.~J.~W. and {Levin}, L. and {Lorimer}, D.~R. and {Lynch}, R.~S. and {Ng}, C. and {Nice}, D.~J. and {Pennucci}, T.~T. and {Ransom}, S.~M. and {Ray}, P.~S. and {Spiewak}, R. and {Stairs}, I.~H. and {Stinebring}, D.~R. and {Stovall}, K. and {Swiggum}, J.~K. and {Zhu}, W.~W.},
        title = "{The NANOGrav 11-year Data Set: Pulse Profile Variability}",
      journal = {\apj},
         year = 2018,
        month = dec,
       volume = {868},
       number = {2},
          eid = {122},
        pages = {122},
          doi = {10.3847/1538-4357/aae9e3},
archivePrefix = {arXiv},
       eprint = {1810.08269},
 primaryClass = {astro-ph.HE},
       adsurl = {https://ui.adsabs.harvard.edu/abs/2018ApJ...868..122B}
}

@ARTICLE{2025ApJ...984..139F,
       author = {{Fiore}, William and {McLaughlin}, Maura A. and {Agazie}, Gabriella and {Anumarlapudi}, Akash and {Archibald}, Anne M. and {Arzoumanian}, Zaven and {Baker}, Paul T. and {Brook}, Paul R. and {Cromartie}, H. Thankful and {Crowter}, Kathryn and {DeCesar}, Megan E. and {Demorest}, Paul B. and {Dey}, Lankeswar and {Dolch}, Timothy and {Ferrara}, Elizabeth C. and {Fonseca}, Emmanuel and {Freedman}, Gabriel E. and {Garver-Daniels}, Nate and {Gentile}, Peter A. and {Glaser}, Joseph and {Good}, Deborah C. and {Hazboun}, Jeffrey S. and {Jennings}, Ross J. and {Jones}, Megan L. and {Kaplan}, David L. and {Kerr}, Matthew and {Lam}, Michael T. and {Lorimer}, Duncan R. and {Luo}, Jing and {Lynch}, Ryan S. and {McEwen}, Alexander and {McMann}, Natasha and {Meyers}, Bradley W. and {Ng}, Cherry and {Nice}, David J. and {Pennucci}, Timothy T. and {Perera}, Benetge B.~P. and {Pol}, Nihan S. and {Radovan}, Henri A. and {Ransom}, Scott M. and {Ray}, Paul S. and {Schmiedekamp}, Ann and {Schmiedekamp}, Carl and {Shapiro-Albert}, Brent J. and {Stairs}, Ingrid H. and {Stovall}, Kevin and {Susobhanan}, Abhimanyu and {Swiggum}, Joseph K. and {Wahl}, Haley M.},
        title = "{Pulse Profile Variability of PSR J1022+1001 in NANOGrav Data}",
      journal = {\apj},
         year = 2025,
        month = may,
       volume = {984},
       number = {2},
          eid = {139},
        pages = {139},
          doi = {10.3847/1538-4357/adc255},
archivePrefix = {arXiv},
       eprint = {2412.05452},
 primaryClass = {astro-ph.HE},
       adsurl = {https://ui.adsabs.harvard.edu/abs/2025ApJ...984..139F}
}

@article{Turner_2021,
doi = {10.3847/1538-4357/abfafe},
url = {https://doi.org/10.3847/1538-4357/abfafe},
year = {2021},
month = {aug},
publisher = {The American Astronomical Society},
volume = {917},
number = {1},
pages = {10},
author = {Turner, Jacob E. and McLaughlin, Maura A. and Cordes, James M. and Lam, Michael T. and Shapiro-Albert, Brent J. and Stinebring, Daniel R. and Arzoumanian, Zaven and Blumer, Harsha and Brook, Paul R. and Chatterjee, Shami and Cromartie, H. Thankful and DeCesar, Megan E. and Demorest, Paul B. and Dolch, Timothy and Ellis, Justin A. and Ferdman, Robert D. and Ferrara, Elizabeth C. and Fonseca, Emmanuel and Garver-Daniels, Nathan and Gentile, Peter A. and Good, Deborah C. and Jones, Megan L. and Lazio, T. Joseph W. and Lorimer, Duncan R. and Luo, Jing and Lynch, Ryan S. and Ng, Cherry and Nice, David J. and Pennucci, Timothy T. and Pol, Nihan S. and Ransom, Scott M. and Spiewak, Renée and Stairs, Ingrid H. and Stovall, Kevin and Swiggum, Joseph K. and Vigeland, Sarah J.},
title = {The NANOGrav 12.5 Year Data Set: Monitoring Interstellar Scattering Delays},
journal = {The Astrophysical Journal}
}

@ARTICLE{2025arXiv251122597H,
       author = {{Hazboun}, Jeffrey S. and {Simon}, Joseph and {Baier}, Jeremy and {Larsen}, Bjorn and {Oliver}, Daniel J. and {Baker}, Paul T. and {B{\'e}csy}, Bence and {Chen}, Siyuan and {Diaz Hernandez}, Alberto and {Ellis}, Justin A. and {Holgado}, A. Miguel and {Islo}, Kristina and {Johnson}, Aaron and {Kaiser}, Andrew R. and {Laal}, Nima and {McEwen}, Alexander and {Pol}, Nihan S. and {Shapiro Key}, Joey and {Kim}, Min Young and {Samson}, Matthew and {Shapiro-Albert}, Brent J. and {Sun}, Jerry P. and {Taylor}, Stephen R. and {Witt}, Caitlin A. and {Volpe}, Jeremy and {Ye}, Christine and {Blumer}, Harsha and {Brook}, Paul R. and {Chatterjee}, Shami and {Cordes}, James M. and {Crawford}, Fronefield and {Cromartie}, H. Thankful and {DeCesar}, Megan E. and {Demorest}, Paul B. and {Dolch}, Timothy and {Ferdman}, Robert D. and {Ferrara}, Elizabeth C. and {Fiore}, William and {Fonseca}, Emmanuel and {Garver-Daniels}, Nathan and {Gentile}, Peter A. and {Good}, Deborah C. and {Jennings}, Ross J. and {Jones}, Megan L. and {Kaplan}, David L. and {Lam}, Michael T. and {Lazio}, T. Joseph W. and {Lorimer}, Duncan R. and {Luo}, Jing and {Lynch}, Ryan S. and {Madison}, Dustin R. and {McLaughlin}, Maura A. and {Mingarelli}, Chiara M.~F. and {Ng}, Cherry and {Nice}, David J. and {Pennucci}, Timothy T. and {Ransom}, Scott M. and {Ray}, Paul S. and {Siemens}, Xavier and {Spiewak}, Ren{\'e}e and {Stairs}, Ingrid H. and {Stinebring}, Daniel R. and {Stovall}, Kevin and {Swiggum}, Joseph K. and {Turner}, Jacob E. and {Vallisneri}, Michele and {Vigeland}, Sarah J.},
        title = "{The NANOGrav 12.5-year Data Set: Chromatic Noise Characterization \& Mitigation with Time-Domain Kernels}",
      journal = {arXiv e-prints},
         year = 2025,
        month = nov,
          eid = {arXiv:2511.22597},
        pages = {arXiv:2511.22597},
          doi = {10.48550/arXiv.2511.22597},
archivePrefix = {arXiv},
       eprint = {2511.22597},
 primaryClass = {astro-ph.IM},
       adsurl = {https://ui.adsabs.harvard.edu/abs/2025arXiv251122597H}
}

@article{10.1093/mnras/staf886,
    author = {Keith, Michael J and Spiewak, Renée and Lyne, Andrew G and Weltevrede, Patrick and Antonopoulou, Danai and Stappers, Ben},
    title = {Measuring pulsar profile variations with 2D Gaussian process regression},
    journal = {Monthly Notices of the Royal Astronomical Society},
    volume = {540},
    number = {3},
    pages = {2486-2505},
    year = {2025},
    month = {05},
    issn = {0035-8711},
    doi = {10.1093/mnras/staf886},
    url = {https://doi.org/10.1093/mnras/staf886},
    eprint = {https://academic.oup.com/mnras/article-pdf/540/3/2486/63406512/staf886.pdf},
}

@article{10.1093/mnras/190.4.945,
    author = {Buschauer, Robert and Benford, Gregory},
    title = {Narrowband versus broadband emission processes in pulsars},
    journal = {Monthly Notices of the Royal Astronomical Society},
    volume = {190},
    number = {4},
    pages = {945-959},
    year = {1980},
    month = {04},
    issn = {0035-8711},
    doi = {10.1093/mnras/190.4.945},
    url = {https://doi.org/10.1093/mnras/190.4.945},
    eprint = {https://academic.oup.com/mnras/article-pdf/190/4/945/18194237/mnras190-0945.pdf},
}

@ARTICLE{1975ApJ...196...51R,
       author = {{Ruderman}, M.~A. and {Sutherland}, P.~G.},
        title = "{Theory of pulsars: polar gaps, sparks, and coherent microwave radiation.}",
      journal = {\apj},
         year = 1975,
        month = feb,
       volume = {196},
        pages = {51-72},
          doi = {10.1086/153393},
       adsurl = {https://ui.adsabs.harvard.edu/abs/1975ApJ...196...51R}
}

@ARTICLE{2000Natur.406..484S,
       author = {{Stairs}, I.~H. and {Lyne}, A.~G. and {Shemar}, S.~L.},
        title = "{Evidence for free precession in a pulsar}",
      journal = {\nat},
         year = 2000,
        month = aug,
       volume = {406},
       number = {6795},
        pages = {484-486},
          doi = {10.1038/35020010},
       adsurl = {https://ui.adsabs.harvard.edu/abs/2000Natur.406..484S}
}

@article{doi:10.1126/science.1243254,
author = {Andrew Lyne  and Francis Graham-Smith  and Patrick Weltevrede  and Christine Jordan  and Ben Stappers  and Cees Bassa  and Michael Kramer },
title = {Evolution of the Magnetic Field Structure of the Crab Pulsar},
journal = {Science},
volume = {342},
number = {6158},
pages = {598-601},
year = {2013},
doi = {10.1126/science.1243254},
URL = {https://www.science.org/doi/abs/10.1126/science.1243254},
eprint = {https://www.science.org/doi/pdf/10.1126/science.1243254}}

@article{Perera_2010,
doi = {10.1088/0004-637X/721/2/1193},
url = {https://doi.org/10.1088/0004-637X/721/2/1193},
year = {2010},
month = {sep},
publisher = {The American Astronomical Society},
volume = {721},
number = {2},
pages = {1193},
author = {Perera, B. B. P. and McLaughlin, M. A. and Kramer, M. and Stairs, I. H. and Ferdman, R. D. and Freire, P. C. C. and Possenti, A. and Breton, R. P. and Manchester, R. N. and Burgay, M. and Lyne, A. G. and Camilo, F.},
title = {THE EVOLUTION OF PSR J0737−3039B AND A MODEL FOR RELATIVISTIC SPIN PRECESSION},
journal = {The Astrophysical Journal}
}

@ARTICLE{2010Sci...329..408L,
       author = {{Lyne}, Andrew and {Hobbs}, George and {Kramer}, Michael and {Stairs}, Ingrid and {Stappers}, Ben},
        title = "{Switched Magnetospheric Regulation of Pulsar Spin-Down}",
      journal = {Science},
         year = 2010,
        month = jul,
       volume = {329},
       number = {5990},
        pages = {408},
          doi = {10.1126/science.1186683},
archivePrefix = {arXiv},
       eprint = {1006.5184},
 primaryClass = {astro-ph.GA},
       adsurl = {https://ui.adsabs.harvard.edu/abs/2010Sci...329..408L}
}

@article{Lazio:2017fos,
    author = "Lazio, T. Joseph W. and Bhaskaran, S. and Cutler, C. and Folkner, W. M. and Park, R. S. and Ellis, J. A. and Ely, T. and Taylor, S. R. and Vallisneri, M.",
    editor = "Sanidas, Sotiris and Weltevrede, Patrick and Preston, Lina Levin and Perera, Benetge B. P.",
    title = "{Solar System Ephemerides, Pulsar Timing, Gravitational Waves, \textbackslash{}\& Navigation}",
    eprint = "1801.02898",
    archivePrefix = "arXiv",
    primaryClass = "astro-ph.IM",
    doi = "10.1017/S1743921317009711",
    journal = "IAU Symp.",
    volume = "337",
    pages = "150--153",
    year = "2017"
}

@article{10.1093/mnras/stae2727,
    author = {Liu, K and Parthasarathy, A and Keith, M and Tiburzi, C and Susarla, S C and Antoniadis, J and Chalumeau, A and Chen, S and Cognard, I and Golden, A and Grießmeier, J -M and Guillemot, L and Janssen, G H and Keane, E F and Kramer, M and McKee, J W and Mickaliger, M B and Theureau, G and Wang, J},
    title = {The impact on astrometry by solar-wind effect in pulsar timing},
    journal = {Monthly Notices of the Royal Astronomical Society},
    volume = {536},
    number = {3},
    pages = {2603-2617},
    year = {2024},
    month = {12},
    issn = {0035-8711},
    doi = {10.1093/mnras/stae2727},
    url = {https://doi.org/10.1093/mnras/stae2727},
    eprint = {https://academic.oup.com/mnras/article-pdf/536/3/2603/61075114/stae2727.pdf},
}

@article{Dai_2024,
doi = {10.1088/1674-4527/ad484e},
url = {https://dx.doi.org/10.1088/1674-4527/ad484e},
year = {2024},
month = {jul},
publisher = {National Astromonical Observatories, CAS and IOP Publishing},
volume = {24},
number = {8},
pages = {085008},
author = {Dai, Jian-Peng and Han, Wei and Wang, Na},
title = {The Influence of Different Solar System Planetary Ephemerides on Pulsar Timing},
journal = {Research in Astronomy and Astrophysics}
}

@article{Vallisneri_2020,
doi = {10.3847/1538-4357/ab7b67},
url = {https://dx.doi.org/10.3847/1538-4357/ab7b67},
year = {2020},
month = {apr},
publisher = {The American Astronomical Society},
volume = {893},
number = {2},
pages = {112},
author = {Vallisneri, M. and Taylor, S. R. and Simon, J. and Folkner, W. M. and Park, R. S. and Cutler, C. and Ellis, J. A. and Lazio, T. J. W. and Vigeland, S. J. and Aggarwal, K. and Arzoumanian, Z. and Baker, P. T. and Brazier, A. and Brook, P. R. and Burke-Spolaor, S. and Chatterjee, S. and Cordes, J. M. and Cornish, N. J. and Crawford, F. and Cromartie, H. T. and Crowter, K. and DeCesar, M. and Demorest, P. B. and Dolch, T. and Ferdman, R. D. and Ferrara, E. C. and Fonseca, E. and Garver-Daniels, N. and Gentile, P. and Good, D. and Hazboun, J. S. and Holgado, A. M. and Huerta, E. A. and Islo, K. and Jennings, R. and Jones, G. and Jones, M. L. and Kaplan, D. L. and Kelley, L. Z. and Key, J. S. and Lam, M. T. and Levin, L. and Lorimer, D. R. and Luo, J. and Lynch, R. S. and Madison, D. R. and McLaughlin, M. A. and McWilliams, S. T. and Mingarelli, C. M. F. and Ng, C. and Nice, D. J. and Pennucci, T. T. and Pol, N. S. and Ransom, S. M. and Ray, P. S. and Siemens, X. and Spiewak, R. and Stairs, I. H. and Stinebring, D. R. and Stovall, K. and Swiggum, J. K. and van Haasteren, R. and Witt, C. A. and Zhu, W. W.},
title = {Modeling the Uncertainties of Solar System Ephemerides for Robust Gravitational-wave Searches with Pulsar-timing Arrays},
journal = {The Astrophysical Journal}
}

@ARTICLE{2021A&A...647A..84T,
       author = {{Tiburzi}, C. and {Shaifullah}, G.~M. and {Bassa}, C.~G. and {Zucca}, P. and {Verbiest}, J.~P.~W. and {Porayko}, N.~K. and {van der Wateren}, E. and {Fallows}, R.~A. and {Main}, R.~A. and {Janssen}, G.~H. and {Anderson}, J.~M. and {Bak Nielsen}, A. -S. and {Donner}, J.~Y. and {Keane}, E.~F. and {K{\"u}nsem{\"o}ller}, J. and {Os{\l}owski}, S. and {Grie{\ss}meier}, J. -M. and {Serylak}, M. and {Br{\"u}ggen}, M. and {Ciardi}, B. and {Dettmar}, R. -J. and {Hoeft}, M. and {Kramer}, M. and {Mann}, G. and {Vocks}, C.},
        title = "{The impact of solar wind variability on pulsar timing}",
      journal = {\aap},
         year = 2021,
        month = mar,
       volume = {647},
          eid = {A84},
        pages = {A84},
          doi = {10.1051/0004-6361/202039846},
archivePrefix = {arXiv},
       eprint = {2012.11726},
 primaryClass = {astro-ph.HE},
       adsurl = {https://ui.adsabs.harvard.edu/abs/2021A&A...647A..84T}
}

@ARTICLE{2006MNRAS.372.1549E,
       author = {{Edwards}, R.~T. and {Hobbs}, G.~B. and {Manchester}, R.~N.},
        title = "{TEMPO2, a new pulsar timing package - II. The timing model and precision estimates}",
      journal = {\mnras},
         year = 2006,
        month = nov,
       volume = {372},
       number = {4},
        pages = {1549-1574},
          doi = {10.1111/j.1365-2966.2006.10870.x},
archivePrefix = {arXiv},
       eprint = {astro-ph/0607664},
 primaryClass = {astro-ph},
       adsurl = {https://ui.adsabs.harvard.edu/abs/2006MNRAS.372.1549E}
}

@article{Luo_2021,
doi = {10.3847/1538-4357/abe62f},
url = {https://dx.doi.org/10.3847/1538-4357/abe62f},
year = {2021},
month = {apr},
publisher = {The American Astronomical Society},
volume = {911},
number = {1},
pages = {45},
author = {Luo, Jing and Ransom, Scott and Demorest, Paul and Ray, Paul S. and Archibald, Anne and Kerr, Matthew and Jennings, Ross J. and Bachetti, Matteo and van Haasteren, Rutger and Champagne, Chloe A. and Colen, Jonathan and Phillips, Camryn and Zimmerman, Josef and Stovall, Kevin and Lam, Michael T. and Jenet, Fredrick A.},
title = {PINT: A Modern Software Package for Pulsar Timing},
journal = {The Astrophysical Journal}
}

@ARTICLE{astropy:2022,
       author = {{Astropy Collaboration} and {Price-Whelan}, Adrian M. and {Lim}, Pey Lian and {Earl}, Nicholas and {Starkman}, Nathaniel and {Bradley}, Larry and {Shupe}, David L. and {Patil}, Aarya A. and {Corrales}, Lia and {Brasseur}, C.~E. and {N{\"o}the}, Maximilian and {Donath}, Axel and {Tollerud}, Erik and {Morris}, Brett M. and {Ginsburg}, Adam and {Vaher}, Eero and {Weaver}, Benjamin A. and {Tocknell}, James and {Jamieson}, William and {van Kerkwijk}, Marten H. and {Robitaille}, Thomas P. and {Merry}, Bruce and {Bachetti}, Matteo and {G{\"u}nther}, H. Moritz and {Aldcroft}, Thomas L. and {Alvarado-Montes}, Jaime A. and {Archibald}, Anne M. and {B{\'o}di}, Attila and {Bapat}, Shreyas and {Barentsen}, Geert and {Baz{\'a}n}, Juanjo and {Biswas}, Manish and {Boquien}, M{\'e}d{\'e}ric and {Burke}, D.~J. and {Cara}, Daria and {Cara}, Mihai and {Conroy}, Kyle E. and {Conseil}, Simon and {Craig}, Matthew W. and {Cross}, Robert M. and {Cruz}, Kelle L. and {D'Eugenio}, Francesco and {Dencheva}, Nadia and {Devillepoix}, Hadrien A.~R. and {Dietrich}, J{\"o}rg P. and {Eigenbrot}, Arthur Davis and {Erben}, Thomas and {Ferreira}, Leonardo and {Foreman-Mackey}, Daniel and {Fox}, Ryan and {Freij}, Nabil and {Garg}, Suyog and {Geda}, Robel and {Glattly}, Lauren and {Gondhalekar}, Yash and {Gordon}, Karl D. and {Grant}, David and {Greenfield}, Perry and {Groener}, Austen M. and {Guest}, Steve and {Gurovich}, Sebastian and {Handberg}, Rasmus and {Hart}, Akeem and {Hatfield-Dodds}, Zac and {Homeier}, Derek and {Hosseinzadeh}, Griffin and {Jenness}, Tim and {Jones}, Craig K. and {Joseph}, Prajwel and {Kalmbach}, J. Bryce and {Karamehmetoglu}, Emir and {Ka{\l}uszy{\'n}ski}, Miko{\l}aj and {Kelley}, Michael S.~P. and {Kern}, Nicholas and {Kerzendorf}, Wolfgang E. and {Koch}, Eric W. and {Kulumani}, Shankar and {Lee}, Antony and {Ly}, Chun and {Ma}, Zhiyuan and {MacBride}, Conor and {Maljaars}, Jakob M. and {Muna}, Demitri and {Murphy}, N.~A. and {Norman}, Henrik and {O'Steen}, Richard and {Oman}, Kyle A. and {Pacifici}, Camilla and {Pascual}, Sergio and {Pascual-Granado}, J. and {Patil}, Rohit R. and {Perren}, Gabriel I. and {Pickering}, Timothy E. and {Rastogi}, Tanuj and {Roulston}, Benjamin R. and {Ryan}, Daniel F. and {Rykoff}, Eli S. and {Sabater}, Jose and {Sakurikar}, Parikshit and {Salgado}, Jes{\'u}s and {Sanghi}, Aniket and {Saunders}, Nicholas and {Savchenko}, Volodymyr and {Schwardt}, Ludwig and {Seifert-Eckert}, Michael and {Shih}, Albert Y. and {Jain}, Anany Shrey and {Shukla}, Gyanendra and {Sick}, Jonathan and {Simpson}, Chris and {Singanamalla}, Sudheesh and {Singer}, Leo P. and {Singhal}, Jaladh and {Sinha}, Manodeep and {Sip{\H{o}}cz}, Brigitta M. and {Spitler}, Lee R. and {Stansby}, David and {Streicher}, Ole and {{\v{S}}umak}, Jani and {Swinbank}, John D. and {Taranu}, Dan S. and {Tewary}, Nikita and {Tremblay}, Grant R. and {de Val-Borro}, Miguel and {Van Kooten}, Samuel J. and {Vasovi{\'c}}, Zlatan and {Verma}, Shresth and {de Miranda Cardoso}, Jos{\'e} Vin{\'\i}cius and {Williams}, Peter K.~G. and {Wilson}, Tom J. and {Winkel}, Benjamin and {Wood-Vasey}, W.~M. and {Xue}, Rui and {Yoachim}, Peter and {Zhang}, Chen and {Zonca}, Andrea and {Astropy Project Contributors}},
        title = "{The Astropy Project: Sustaining and Growing a Community-oriented Open-source Project and the Latest Major Release (v5.0) of the Core Package}",
      journal = {\apj},
         year = 2022,
        month = aug,
       volume = {935},
       number = {2},
          eid = {167},
        pages = {167},
          doi = {10.3847/1538-4357/ac7c74},
archivePrefix = {arXiv},
       eprint = {2206.14220},
 primaryClass = {astro-ph.IM},
       adsurl = {https://ui.adsabs.harvard.edu/abs/2022ApJ...935..167A}
}

@ARTICLE{astropy:2018,
       author = {{Astropy Collaboration} and {Price-Whelan}, A.~M. and
         {Sip{\H{o}}cz}, B.~M. and {G{\"u}nther}, H.~M. and {Lim}, P.~L. and
         {Crawford}, S.~M. and {Conseil}, S. and {Shupe}, D.~L. and
         {Craig}, M.~W. and {Dencheva}, N. and {Ginsburg}, A. and {Vand
        erPlas}, J.~T. and {Bradley}, L.~D. and {P{\'e}rez-Su{\'a}rez}, D. and
         {de Val-Borro}, M. and {Aldcroft}, T.~L. and {Cruz}, K.~L. and
         {Robitaille}, T.~P. and {Tollerud}, E.~J. and {Ardelean}, C. and
         {Babej}, T. and {Bach}, Y.~P. and {Bachetti}, M. and {Bakanov}, A.~V. and
         {Bamford}, S.~P. and {Barentsen}, G. and {Barmby}, P. and
         {Baumbach}, A. and {Berry}, K.~L. and {Biscani}, F. and {Boquien}, M. and
         {Bostroem}, K.~A. and {Bouma}, L.~G. and {Brammer}, G.~B. and
         {Bray}, E.~M. and {Breytenbach}, H. and {Buddelmeijer}, H. and
         {Burke}, D.~J. and {Calderone}, G. and {Cano Rodr{\'\i}guez}, J.~L. and
         {Cara}, M. and {Cardoso}, J.~V.~M. and {Cheedella}, S. and {Copin}, Y. and
         {Corrales}, L. and {Crichton}, D. and {D'Avella}, D. and {Deil}, C. and
         {Depagne}, {\'E}. and {Dietrich}, J.~P. and {Donath}, A. and
         {Droettboom}, M. and {Earl}, N. and {Erben}, T. and {Fabbro}, S. and
         {Ferreira}, L.~A. and {Finethy}, T. and {Fox}, R.~T. and
         {Garrison}, L.~H. and {Gibbons}, S.~L.~J. and {Goldstein}, D.~A. and
         {Gommers}, R. and {Greco}, J.~P. and {Greenfield}, P. and
         {Groener}, A.~M. and {Grollier}, F. and {Hagen}, A. and {Hirst}, P. and
         {Homeier}, D. and {Horton}, A.~J. and {Hosseinzadeh}, G. and {Hu}, L. and
         {Hunkeler}, J.~S. and {Ivezi{\'c}}, {\v{Z}}. and {Jain}, A. and
         {Jenness}, T. and {Kanarek}, G. and {Kendrew}, S. and {Kern}, N.~S. and
         {Kerzendorf}, W.~E. and {Khvalko}, A. and {King}, J. and {Kirkby}, D. and
         {Kulkarni}, A.~M. and {Kumar}, A. and {Lee}, A. and {Lenz}, D. and
         {Littlefair}, S.~P. and {Ma}, Z. and {Macleod}, D.~M. and
         {Mastropietro}, M. and {McCully}, C. and {Montagnac}, S. and
         {Morris}, B.~M. and {Mueller}, M. and {Mumford}, S.~J. and {Muna}, D. and
         {Murphy}, N.~A. and {Nelson}, S. and {Nguyen}, G.~H. and
         {Ninan}, J.~P. and {N{\"o}the}, M. and {Ogaz}, S. and {Oh}, S. and
         {Parejko}, J.~K. and {Parley}, N. and {Pascual}, S. and {Patil}, R. and
         {Patil}, A.~A. and {Plunkett}, A.~L. and {Prochaska}, J.~X. and
         {Rastogi}, T. and {Reddy Janga}, V. and {Sabater}, J. and
         {Sakurikar}, P. and {Seifert}, M. and {Sherbert}, L.~E. and
         {Sherwood-Taylor}, H. and {Shih}, A.~Y. and {Sick}, J. and
         {Silbiger}, M.~T. and {Singanamalla}, S. and {Singer}, L.~P. and
         {Sladen}, P.~H. and {Sooley}, K.~A. and {Sornarajah}, S. and
         {Streicher}, O. and {Teuben}, P. and {Thomas}, S.~W. and
         {Tremblay}, G.~R. and {Turner}, J.~E.~H. and {Terr{\'o}n}, V. and
         {van Kerkwijk}, M.~H. and {de la Vega}, A. and {Watkins}, L.~L. and
         {Weaver}, B.~A. and {Whitmore}, J.~B. and {Woillez}, J. and
         {Zabalza}, V. and {Astropy Contributors}},
        title = "{The Astropy Project: Building an Open-science Project and Status of the v2.0 Core Package}",
      journal = {\aj},
         year = 2018,
        month = sep,
       volume = {156},
       number = {3},
          eid = {123},
        pages = {123},
          doi = {10.3847/1538-3881/aabc4f},
archivePrefix = {arXiv},
       eprint = {1801.02634},
 primaryClass = {astro-ph.IM},
       adsurl = {https://ui.adsabs.harvard.edu/abs/2018AJ....156..123A}
}

@ARTICLE{astropy:2013,
       author = {{Astropy Collaboration} and {Robitaille}, Thomas P. and {Tollerud}, Erik J. and {Greenfield}, Perry and {Droettboom}, Michael and {Bray}, Erik and {Aldcroft}, Tom and {Davis}, Matt and {Ginsburg}, Adam and {Price-Whelan}, Adrian M. and {Kerzendorf}, Wolfgang E. and {Conley}, Alexander and {Crighton}, Neil and {Barbary}, Kyle and {Muna}, Demitri and {Ferguson}, Henry and {Grollier}, Fr{\'e}d{\'e}ric and {Parikh}, Madhura M. and {Nair}, Prasanth H. and {Unther}, Hans M. and {Deil}, Christoph and {Woillez}, Julien and {Conseil}, Simon and {Kramer}, Roban and {Turner}, James E.~H. and {Singer}, Leo and {Fox}, Ryan and {Weaver}, Benjamin A. and {Zabalza}, Victor and {Edwards}, Zachary I. and {Azalee Bostroem}, K. and {Burke}, D.~J. and {Casey}, Andrew R. and {Crawford}, Steven M. and {Dencheva}, Nadia and {Ely}, Justin and {Jenness}, Tim and {Labrie}, Kathleen and {Lim}, Pey Lian and {Pierfederici}, Francesco and {Pontzen}, Andrew and {Ptak}, Andy and {Refsdal}, Brian and {Servillat}, Mathieu and {Streicher}, Ole},
        title = "{Astropy: A community Python package for astronomy}",
      journal = {\aap},
         year = 2013,
        month = oct,
       volume = {558},
          eid = {A33},
        pages = {A33},
          doi = {10.1051/0004-6361/201322068},
archivePrefix = {arXiv},
       eprint = {1307.6212},
 primaryClass = {astro-ph.IM},
       adsurl = {https://ui.adsabs.harvard.edu/abs/2013A&A...558A..33A}
}
